\documentclass[a4paper,12pt]{article}

\usepackage{graphics}
\usepackage{graphicx}
\usepackage{amsmath}
\usepackage{amssymb}
\usepackage[sort&compress,numbers]{natbib}

\evensidemargin \oddsidemargin
\newcommand{\be}{\begin{equation}}
\newcommand{\ee}{\end{equation}}
\newcommand{\bea}{\begin{eqnarray}}
\newcommand{\eea}{\end{eqnarray}}

\begin{document}

\title{\vspace{3cm}Dark matter and the early Universe: a review\vspace{1cm}}

\author{A.\ Arbey and F. Mahmoudi\\[1.cm]
\normalsize Univ Lyon, Univ Claude Bernard Lyon 1, CNRS/IN2P3,\\
\normalsize Institut de Physique des 2 Infinis de Lyon, UMR 5822, 69622 Villeurbanne, France\\[0.2cm]
\normalsize Theoretical Physics Department, CERN, CH-1211 Geneva 23, Switzerland\\[0.2cm]
\normalsize Institut Universitaire de France, 103 boulevard Saint-Michel, 75005 Paris, France\\[0.5cm]}
 
\date{}

\maketitle
 
\vspace*{1.cm}

\begin{abstract}
Dark matter represents currently an outstanding problem in both cosmology and particle physics. In this review we discuss the possible explanations for dark matter and the experimental observables which can eventually lead to the discovery of dark matter and its nature, and demonstrate the close interplay between the cosmological properties of the early Universe and the observables used to constrain dark matter models in the context of new physics beyond the Standard Model.
\end{abstract}

\newpage

\tableofcontents

\section{Introduction}

The nature of dark matter (DM) is currently one of the most intriguing questions of fundamental physics. Even if the question of dark matter finds its roots in astrophysics and cosmology, it is also actively searched for in particle physics experiments, at colliders as well as in dark matter detection experiments, which aim at discovering dark matter particles. The main experimental challenge comes of course from the elusive nature of dark matter. In addition to dark matter, several other cosmological questions remain unanswered, such as the nature of dark energy, the properties of the inflationary period, the existence of phase transitions in the early Universe, and the origin of baryon asymmetry in the Universe.

In the context of particle physics, dark matter can be made of one or several new particles, which are expected to be electrically neutral, uncoloured, weakly-interacting and stable. Since the Standard Model (SM) fails at providing a dark matter candidate, it is necessary to consider scenarios beyond the Standard Model, which may in addition have a broad phenomenology at colliders. New physics scenarios generally rely on new symmetries which are broken at high energies or extra-dimensions, and as such they can impact the properties of the early Universe, either by the presence of new particles in the primordial thermal bath, or via phase transitions.

One of the most important observables connecting dark matter, cosmology and particle physics is the cosmological density of dark matter, which is called in this context the relic density. Dark matter particles originate from the very early Universe, they have been in interaction with the thermal bath before decoupling from it, they have then decayed or annihilated and we observe today the particles which have survived until now, the relics. There exist many possible modifications to this model, but this simple scenario shows how the dark matter relic density is connected to the primordial Universe. Since dark matter density has been very precisely measured by cosmological observations, it can be used to set constraints not only on particle physics models, but also on early Universe scenarios.

In the following we will first briefly present the standard cosmological scenario in Section 2, which is based on the Friedmann-Lemaître-Robertson-Walker (FLRW) model, and present the relevant observational constraints. We will discuss in Section 3 the problem of dark matter in cosmology, and in Section 4 we will briefly review cosmological scenarios beyond the standard cosmological model which could modify the early Universe properties. In Section 5 we describe the problem of dark matter in the context of particle physics, before presenting the calculation of the relic density in Section 6. In Section 7 we demonstrate the interplay between cosmology and particle physics via the example of supersymmetric dark matter. Finally in Section 8 we discuss the interplay between dark matter and the early Universe, and conclude in Section 9.

%%%%%%%%%%%%%%%%%%%%%%

\section{Standard Cosmological Model}

In this section, we describe the standard cosmological scenario, which is based on the FLRW model. In the standard cosmological model it is assumed that dark matter and dark energy drive the expansion of the Universe today, and at the beginning of the Universe radiation was the dominant energy. We will consider natural units with $c = \hbar = k_B = 1$.

\subsection{Friedmann-Lemaître-Robertson-Walker model}

The cosmological principle states that the energy distribution in the Universe is homogeneous and isotropic. Observations of the Universe have revealed that this principle is valid at large scales, whereas it is obviously not valid locally. Yet, this principle is the basis of the standard cosmological model. In terms of geometry, this is equivalent of considering that space is globally homogeneous and isotropic. The most general metric $g_{\mu\nu}$ which corresponds to these assumptions is the Robertson and Walker metric, which is written in terms of proper time as
\begin{equation}
 d\tau^2 = g_ {\mu\nu} dx^\mu dx^\nu = dt^2 - a^2(t) \left\{\displaystyle\frac{dr^2}{1-k r^2} + r^2 (d\theta^2 + \sin^2\theta d\phi^2) \right\}\,,
\end{equation}
in a spherical coordinate system $x^\mu=(t,r,\theta,\phi)$, where $a(t)$ is the scale factor of the Universe which characterizes its expansion and is a function of the cosmological time $t$, and $k$ is a global curvature which is the same at any point of the spacetime. The cosmological principle implies that all quantities in the Universe are only dependent on time $t$ and do not depend on position.

It is assumed that the different forms of energies in the Universe can be described as perfect fluids in adiabatic expansions with a total density $\rho = \sum \rho_{\rm fluid}$ and pressure $P = \sum P_{\rm fluid}$, leading to the stress-energy momentum tensor
\begin{equation}
 T^{\mu\nu} = (P+\rho) U^\mu U^\nu - P g^{\mu\nu} \,,
\end{equation}
where $U^\mu$ is the 4-velocity. In the rest frame of the cosmological fluids the nonzero components are $T^{00} = \rho$ and $T^{ij} = -P g^{ij}$.

For each of the cosmic fluids the energy conservation, which can be expressed as
\begin{equation}
 D_\mu T^{\mu\nu}_{\rm fluid} = 0\,,
\end{equation}
where $D_\mu$ is the covariant derivative, leads to
\begin{equation}
 \dot{\rho}_{\rm fluid} + 3 H (\rho_{\rm fluid} + P_{\rm fluid}) = 0\,. \label{eq:energy_conservation}
\end{equation}
The dot in the equation above denotes a time derivative and the Hubble parameter $H$ is
\begin{equation}
 H = \displaystyle\frac{\dot{a}}{a}\,.
\end{equation}

The Einstein equation leads to two independent equations:
\begin{eqnarray}
 H^2 &=& \frac{8\pi G}{3} \rho - \displaystyle\frac{k}{a^2}\,, \label{eq:friedmann1}\\
 \frac{\ddot{a}}{a} &=& -\frac{4\pi G}{3} (\rho + 3 P)\,, \label{eq:friedmann2}
\end{eqnarray}
where $G$ is the Newton gravitational constant. These equations are referred to as the first and second Friedmann equations, respectively. It is important to note that the energy conservation equation (\ref{eq:energy_conservation}) is not independent from the two Friedmann equations.

The values of the parameters at present time are denoted with a $0$ index. In particular $a_0$ is the current value of the scale factor and we have:
\begin{equation}
 H_0^2 = \frac{8\pi G}{3} \rho_0 - \displaystyle\frac{k}{a_0^2} = \frac{8\pi G}{3} \rho^c_0\,,
\end{equation}
where $\rho^c_0$ is called ``critical density'', and $H_0$ is the ``Hubble constant''. The densities of different fluids are generally normalized by the critical density
\begin{equation}
 \Omega_{\rm fluid} = \displaystyle\frac{\rho_0^{\rm fluid}}{\rho^c_0}\,,
\end{equation}
and the $\Omega_{\rm fluid}$ densities are called ``cosmological parameters''.

The curvature term can be absorbed into a curvature energy density:
\begin{equation}
 \rho_k = \displaystyle - \frac{3}{8\pi G}\frac{k}{a^2}\,,
\end{equation}
and has a pressure $P_k = -\rho_k/3$.

The different fluids in the Universe are matter, made of a mixture of dark matter and baryonic matter; radiation, composed of relativistic particle species such as photons and neutrinos; and dark energy, which is an unidentified component of negative pressure. In the standard cosmological model, dark matter is considered as cold, i.e. with small velocities, and dark energy is considered to be a ``cosmological constant'' $\Lambda$ with a constant density and pressure such as $\rho_\Lambda = - P_\Lambda$, forming the $\Lambda$CDM paradigm.

The Planck Collaboration has provided the observational values of the Hubble constant and the cosmological parameters of (total) matter, cold dark matter, baryonic matter, cosmological constant and curvature, respectively \cite{Aghanim:2018eyx}:
\begin{eqnarray}
 H_0 &=& 67.66 \pm 0.42 \;{\rm km/s/Mpc}\,,\nonumber\\
 \Omega_m &=& 0.3111 \pm 0.0056\,,\nonumber\\
 \Omega_c h^2 &=& 0.11933 \pm 0.00091\,,\nonumber\\
 \Omega_b h^2 &=& 0.02242 \pm 0.00014\,,\label{eq:planck}\\
 \Omega_\Lambda &=& 0.6889 \pm 0.0056\,,\nonumber\\
 \Omega_k &=& 0.0007 \pm 0.0037\,,\nonumber
\end{eqnarray}
where $h = H_0 / ({100\;{\rm km/s/Mpc}})$ is the reduced Hubble parameter. The radiation density is currently negligible and because of the very small value of the curvature the Universe is generally considered to be flat with $k=\Omega_k=0$.

\subsection{A quick story of the Universe}

By convention, the beginning of the history of the Universe is set at $t=0$, which corresponds to a scale factor $a(0) = 0$. It is however often considered that our physical description of Nature is not valid before the Planck time $t_P = 5.39\times10^{-44}\,$s where a quantum gravity description would be necessary. One therefore generally considers the Planck time as the effective start time, when the scale factor was negligible.

From the properties of the different components, namely that matter has a pressure negligible in comparison to its density, radiation pressure is $P_r = \rho_r/3$ (obtained from statistical physics considerations) and $P_\Lambda = - \rho_\Lambda$. Using the energy conservation equation (\ref{eq:energy_conservation}), it is straightforward to show that
\begin{eqnarray}
 \rho_m &=& \displaystyle\rho_{m,0}\left(\frac{a_0}{a}\right)^{-3}\,,\\
 \rho_r &=& \displaystyle\rho_{r,0}\left(\frac{a_0}{a}\right)^{-4}\,,\\
 \rho_\Lambda &=& \rho_{\Lambda,0}\,.
\end{eqnarray}

The total energy density is currently dominated by the cosmological constant, but from the above equations and observational values (\ref{eq:planck}), it is clear that matter was dominating in the recent past, whereas at the beginning of the Universe radiation was dominating.

Instead of referring to the scale factor $a$ to describe the past Universe, one generally refers to the redshift $z$ such as
\begin{equation}
 z = \frac{a_0}{a} - 1\,,
\end{equation}
so that today corresponds to a zero redshift and the very beginning of the Universe to an infinite redshift. The Planck Collaboration deduced from the observational data the value of the redshift at which matter and radiation had equal densities \cite{Aghanim:2018eyx}:
\begin{equation}
 z_{eq} = 3387 \pm 21\,,
\end{equation}
which gives for radiation $\Omega_r \approx 9.2 \times 10^{-5}$. In terms of temperature, from statistical physics we have $\rho_r \propto T^4$, as long as the expansion is adiabatic. The temperature thus simply evolves as the inverse of the scale factor:
\begin{equation}
 T \propto a^{-1}\,.
\end{equation}

In the standard cosmological model there are therefore three main phases:
\begin{itemize}
 \item a radiation domination era until $z \sim 3400$,
 \item a matter domination era until $z \approx 0.3$,
 \item a cosmological constant (co-)domination era at low redshift.
\end{itemize}
Structure formation occurred mainly during the matter domination era, and the cosmological constant thus started playing a role only very recently.

\begin{figure}[!ht]
 \begin{center}
  \includegraphics[width=8.cm,angle=90]{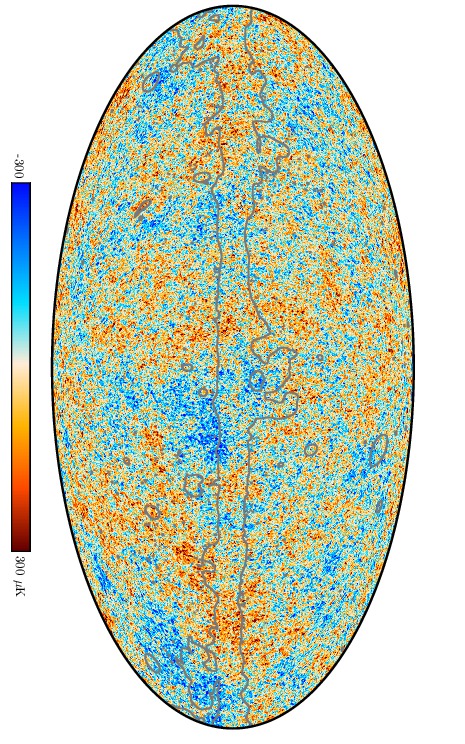}
\caption{Full-sky map of the tiny temperature anisotropies over an average temperature of 2.725 K of the Cosmic Microwave Background. The grey contours show the position of the microwave emissions in the Milky Way and local structures, which have been removed from the data. From \cite{Akrami:2018vks}.\label{fig:cmb}}
 \end{center}
\end{figure}

The standard cosmological model offers a simple framework to study the evolution of the Universe, but it does not describe the other phenomena which may have occurred in the early Universe, for which specific models are required. The physics of the early Universe is indeed complicated, since the Universe went through different phases, in particular an inflation period during which the scale factor increased exponentially and also several phase transitions such as the electroweak phase transition. These periods are however not observable directly, because the photons emitted during these phases were interacting too strongly with the plasma to carry information, and only after the formation of neutral atoms during the ``recombination'' period at a temperature of about 1 eV, or equivalently $10^5\,$K, the photons can be considered as messengers which carry information. The photons emitted at the recombination time can still be observed today in the cosmic microwave background (CMB), on which most of the results of the Planck Collaboration are based. The energetic photons emitted during recombination have had their wavelength redshifted and reduced by the expansion, and are today observed in the microwave range corresponding to a temperature of 2.725 K. The CMB is an important vector of information, since its anisotropies, as shown in Fig.~\ref{fig:cmb}, are imprints of prior phenomena and their careful examination can provide constraints on scenarios describing the beginning of the Universe. %
Precise analyses of the CMB have shown that the standard cosmological model is in agreement with the data \cite{Aghanim:2018eyx}, and that no large deviation from new phenomena is possible at the time of recombination.

Another indirect information about the early Universe comes from the abundance of the chemical elements. Indeed the nuclei have been formed during the so-called Big-Bang nucleosynthesis (BBN), by the combination of the nucleons, and this epoch is well described theoretically. In spite of inconsistencies with the predictions of the lithium abundance, models of BBN are considered as successful and important tests of the Big-Bang model, and their predictions are in agreement with the observed abundances. We will discuss more thoroughly the BBN physics in the following, since BBN is the most ancient period from which one can derive constraints on the cosmological models.

%%%

\subsection{Big-Bang nucleosynthesis}

During Big-Bang nucleosynthesis baryons combine to form complex nuclei, involving nuclear reactions of the type:
\begin{equation}
N_i\,\, ^{A_i}Z_i + N_j\,\, ^{A_j}Z_j + N_k\,\, ^{A_k}Z_k \longleftrightarrow N_l\,\, ^{A_l}Z_l + N_m\,\, ^{A_m}Z_m + N_n\,\, ^{A_n} Z_n \,,
\end{equation}
where $N_i$ is the number of nuclei $Z_i$ involved in the reactions and $A_i$ their atomic number. The abundance $Y_i$ of a given nucleus $Z_i$ is defined as the ratio of the nucleus density $\rho_{Z_i}$ to the baryon density $\rho_b$. The evolution of the abundance $Y_i$ is then given by the Boltzmann equation:
\begin{equation}
\frac{{d}Y_i}{{d}t} = N_i \sum_{j,k,l,m,n} \left( -\frac{Y_i^{N_i} Y_j^{N_j}Y_k^{N_k}}{N_i!N_j!N_k!}\Gamma_{ijk\rightarrow lmn} + \frac{Y_l^{N_l}Y_m^{N_m}Y_n^{N_n}}{N_l!N_m!N_n!}\Gamma_{lmn\rightarrow ijk} \right)\,, \label{eq_Y}
\end{equation}
where $\Gamma_{ijk\rightarrow lmn}$ and $\Gamma_{lmn\rightarrow ijk}$ are the forward and reverse nuclear reaction rates, respectively, which can be measured or computed in the context of nuclear physics. In this equation, $t$ is the cosmological time of the FLRW Universe, which can be obtained by solving the Friedmann equations.

At the beginning of BBN, i.e. a few seconds after the beginning of the Universe at a redshift $z_{\rm BBN}\sim 10^{11}$, the temperature was of the order of 100 MeV, or equivalently $10^{13}\,$K, and the cosmic plasma was made of photons, neutrinos, electrons and positrons in thermal equilibrium, and of baryons. The densities of photons, neutrinos, electrons and positrons can be obtained from statistical physics considerations. The baryons had non-relativistic velocities and were not in thermal equilibrium. Their density is therefore unknown, but the ratio of baryon number density to photon number density has been obtained from the CMB to be $6.09\times10^{-10}$ \cite{Tanabashi:2018oca}. During BBN, the electrons and positrons annihilated (mainly into photons), and the equations of conservation of electron, positron and photon numbers are thus not independent. In addition the chemical potential of the electrons and positrons have to be taken into account.

To solve the system of Eqs. (\ref{eq_Y}), it is assumed that the original nuclei present in the Universe are protons and neutrons in thermal equilibrium:
\begin{equation}
Y_{\rm p}(T) = \displaystyle \frac{1}{1+e^{-q/T}}\;, \qquad\qquad Y_{\rm n}(T) = \frac{1}{1+e^{q/T}}\;.
\end{equation}
As the expansion proceeded, the temperature decreased, thermal equilibrium was broken, and the nuclei of the plasma interacted and formed heavier nuclei. BBN ended about 3 minutes after the beginning of the Universe at a temperature of about $10^9\,$K.

Using the public code {\tt AlterBBN} \cite{Arbey:2011nf,Arbey:2018zfh}, which is dedicated to the calculation of the abundance of the elements, one shows that the main abundances are\footnote{The notation X/Y denotes the ratio of the abundances of X over Y.}:
\begin{eqnarray}
Y_{\rm p} &=& 0.2473 \pm 0.0003\,,\nonumber\\
^2{\rm H/H} &=& (2.463\pm0.038) \times 10^{-5}\,,\\
^3{\rm He/H} &=& (1.034\pm 0.016) \times 10^{-5}\,,\nonumber\\
^7{\rm Li/H} &=& (5.376 \pm 0.352) \times 10^{-10}\,,\nonumber
\end{eqnarray}
which can be compared to the observational values \cite{Tanabashi:2018oca}:
\begin{eqnarray}
Y_{\rm p} &=& 0.2450 \pm 0.0030\,,\nonumber\\
^2{\rm H/H} &=& (2.569\pm0.027) \times 10^{-5}\,,\\
^3{\rm He/H} &=& (1.1\pm 0.2) \times 10^{-5}\,,\nonumber\\
^7{\rm Li/H} &=& (1.6 \pm 0.3) \times 10^{-10}\,.\nonumber
\end{eqnarray}
There is an excellent agreement for hydrogen, deuterium and helium, whereas there is a large discrepancy for lithium-7: this is the well-known lithium problem \cite{Cyburt:2008kw,Fields:2011zzb}, which may be a sign of either new physics or the existence of a process which destroys lithium-7 after its formation at BBN time. 

%%%%%%%%%%%%%%%%%%%%%%

\section{Dark matter(s)}

The nature of dark matter is presently a major question for both cosmology and particle physics. Yet, its existence is still debated, and relies on the observations of gravitational effects in large-scale structures and in cosmology.

\subsection{Observational evidences}

There are evidences for the existence of dark matter at different scales.

\subsubsection{Galaxies}

One of the most striking evidences of unexpected gravitational effects come from galaxies, and in particular spiral galaxies. In spiral galaxies, most of the visible mass is gathered in the budge and the disc. Using the Gauss' theorem, the velocity of stars $v$ at the distance $R$ from the galactic centre reads:
\begin{equation}
 v(R) = \sqrt{\frac{G M(R)}{R}}\,,
\end{equation}
where $M(R)$ is the total mass contained in a sphere of radius $R$. Far from the centre, the disc fades away, so that the star density decreases and the total mass inside the radius $R$ becomes constant. As a consequence the velocity is expected to decrease as $v(R) \propto R^{-1/2}$. For most of spiral galaxies however the observed velocity far from the centre is approximately constant, leading to the so-called flat rotation curves of spiral galaxies (see for example \cite{Persic:1995ru}). Considering a constant velocity $v_0$ within a radius $R$ the mass inside this radius is given by:
\begin{equation}
 M(R) = \frac{R v_0^2}{G} \,,
\end{equation}
and therefore the mass continues increasing far beyond the fading of the visible disc. In addition globular clusters around spiral galaxies are spread within spherical distributions around the galactic centres and extend well beyond the visible disc, constituting a spherical halo. Using a spherically symmetric distribution it is possible to show that the density leading to flat rotation curves reads:
\begin{equation}
 \rho_M(R) = \frac{v_0^2}{4\pi G R^2}\,,
\end{equation}
corresponding to a decrease as $R^{-2}$, whereas the visible disc density decreases exponentially. Since most of the matter in the halo is invisible, its mass is considered to mainly come from dark matter.

\begin{figure}[!ht]
\begin{center}
\includegraphics[width=0.9\textwidth]{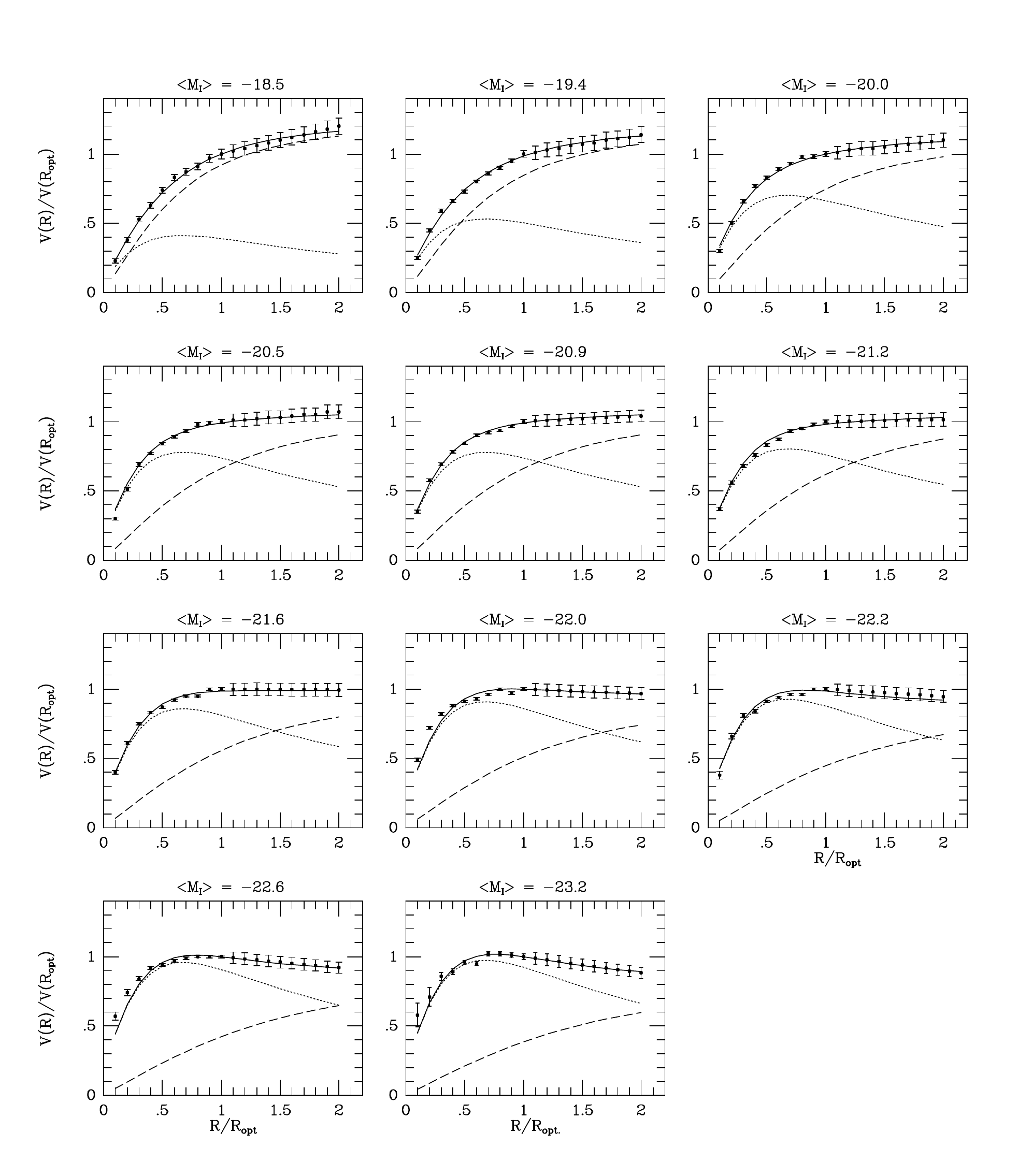}
\caption{Average rotation velocity of spiral galaxies as a function of the reduced radius for different luminosities. The plain lines correspond to the best fits to the observational data, the dotted lines to the contribution from the disc and the dashed lines to the contribution from the halo. From \cite{Persic:1995ru}.\label{fig:PPS}} 
\end{center}
\end{figure}

Examples of average rotation curves are shown in Fig.~\ref{fig:PPS}, where a large number of galaxies have been studied and gathered into luminosity groups, and average rotation curves are plotted in addition to the contributions to the velocity from the disc and the halo. For the faintest galaxies, the rotation curves are slightly increasing for large radii, whereas the brightest ones have slightly decreasing velocities at large radii.

Similar results can be obtained when studying the dynamics of elliptical galaxies, and globally dark matter seems to represent about 80-90\% of the total mass of galaxies.

\subsubsection{Galaxy clusters}

Galaxy clusters are very massive objects which contain large quantities of gases in the intergalactic medium. These gases accelerated by the ambient gravity can reach high velocities and emit x-ray via thermal bremsstrahlung. During decades the analysis of the emitted x-ray together with a modelling of the galaxy clusters were used to weigh the total mass of the clusters, which was in turn compared to the visible mass. The results of such analyses were generally that a large fraction of the mass is invisible, but were often criticized for their approximate modelling.

\begin{figure}[!ht]
\begin{center}
\hspace*{-0.4cm} \includegraphics[height=6.5cm]{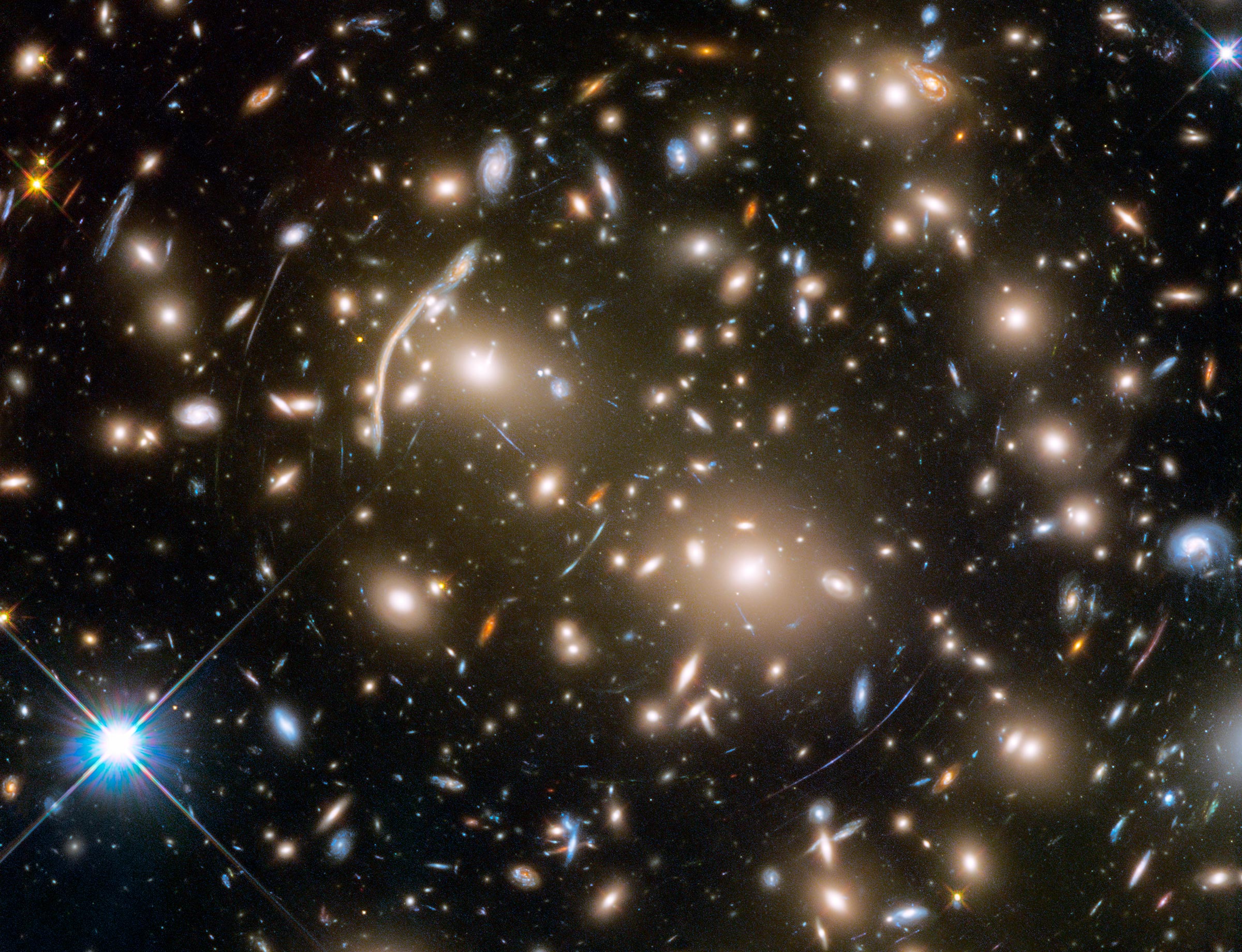}~
 \includegraphics[height=6.5cm]{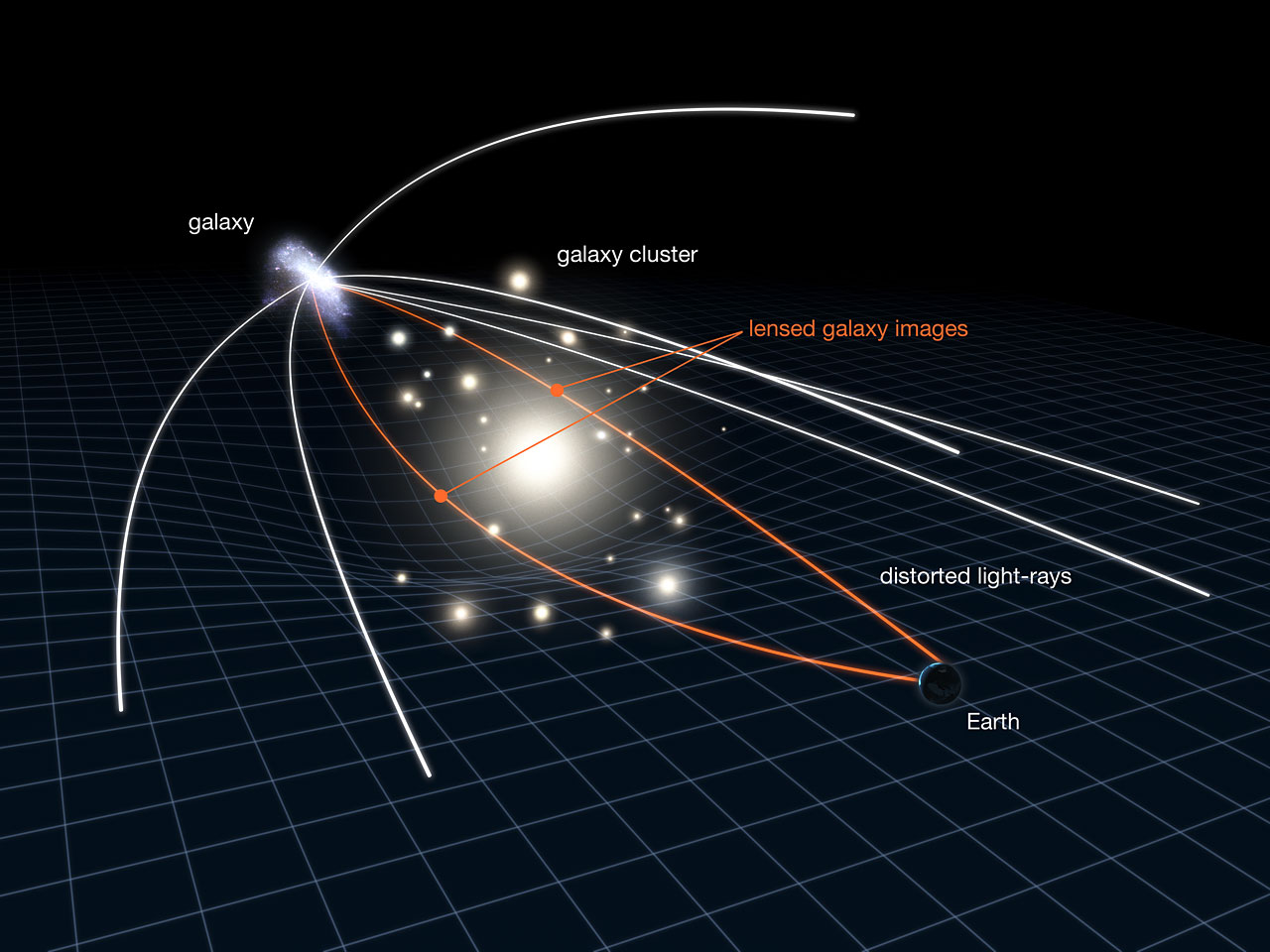}
\caption{(Left) Galaxy cluster Abell 370 composed of several hundreds of galaxies. The blue arcs are distorted images of remote galaxies behind the cluster. Credit: NASA, ESA, and J. Lotz and the HFF Team. (Right) Schematic explanation of strong gravitational lensing by a cluster. Credit: NASA, ESA and L. Calçada.\label{fig:lensing}}
\end{center}
\end{figure}

Such methods to determine the total mass of galaxy clusters have been superseded by gravitational lensing methods consisting in observing the deformed images of distant objects and reconstructing the path of the light which has been modified by the presence of masses affecting geometry locally. The most striking cases are the ones of strong gravitational lensing of distant galaxies behind a galaxy cluster, which appear under the form of several distorted images spread on the so-called Einstein circle. This is shown in Fig.~\ref{fig:lensing}.

The angular radius of the Einstein circle is related to the mass which gave rise to light deflection through the formula \cite{Einstein:1956zz}:
\begin{equation}
 \theta_E=\sqrt{\displaystyle\frac{4 G M}{c^2} \frac{(D_S-D_L)}{D_S D_L}}\,,
\end{equation}
where $\theta_E$ is the Einstein angular radius, $M$ the mass of the lens, $D_L$ the distance to the lens and $D_S$ the distance to the source.

Gravitational lensing is therefore often used to weigh galaxy clusters, and numerous studies are consistent and tend to demonstrate that the visible mass represents only 10-20\% of the total mass, similarly to the commonly found result in galaxies.

\begin{figure}[!ht]
\begin{center}
 \includegraphics[width=10cm]{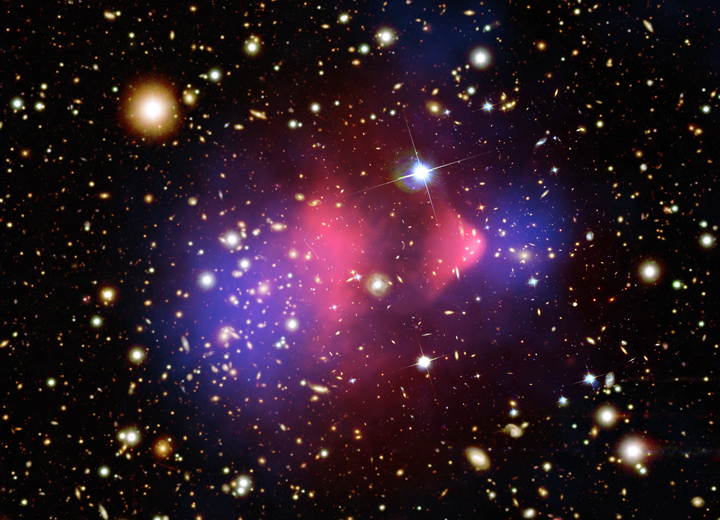}
\caption{Composite image of the galaxy cluster 1E0657-56. Red: hot gas observed by the Chandra X-Ray Observatory. Blue: major part of the mass deduced from gravitational lensing. Credit: NASA/CXC/CfA/M. Markevitch et al.; NASA/STScI; Magellan/U.Arizona/D. Clowe et al.; NASA/STScI; ESO WFI.\label{fig:bullet}}
\end{center}
\end{figure}

In the context of gravitational lensing and x-ray emission of clusters the analysis in Ref.~\cite{Clowe:2006eq} of the Bullet Cluster 1E0657-558 can be considered as a breakthrough for understanding the nature of dark matter. In Fig.~\ref{fig:bullet} an image of the bullet cluster is shown, which is in fact composed of two galaxy clusters having crossed each other. The bullet-shaped red region corresponds to the hot gas observed by the Chandra X-Ray Observatory, and the blue regions to the major part of the mass deduced from gravitational lensing. 

The hot gas has experienced a drag force due to interactions during the collision, giving the characteristic bullet shape. The visible mass is composed of the galaxies and the hot gas. The most important result is that the major part of the mass is separated from the gas. This proves that baryonic mass and dark matter are separate entities.

This result can be considered as a disproof of theories which assume that dark matter does not exist and that what is observed in galaxies and clusters comes from new gravitational effects. One of such theories is MOdified Newtonian Dynamics (MOND) \cite{Milgrom:1983ca}, which fails to explain the Bullet Cluster results. It can be noted however that modified versions of the original MOND model such as the covariant theory TeVeS \cite{Bekenstein:2004ne} may explain the observations \cite{Angus:2006qy}.

\subsubsection{Large and cosmological scales}

The problem of dark matter extends to larger and cosmological scales. After recombination and emission of the cosmic microwave background, atoms are formed and experience gravity as the main interaction. Matter starts gathering via gravitational collapse in the denser regions, forming large-scale structures in which galaxies also emerge. Two main effects affect structure formation: gravitational interaction attracts matter in the centres of masses, and the expansion of the Universe drives structures away from each other. Simulations of structure formation have been performed on supercomputers to study the interplay between expansion and gravity and study the role of dark matter in this process. In particular dark matter is generally considered as collisionless contrary to baryonic matter, so that it is possible to distinguish the role of dark matter from the one of baryonic matter in the simulations.

\begin{figure}[!t]
\begin{center}
 \includegraphics[width=8.cm]{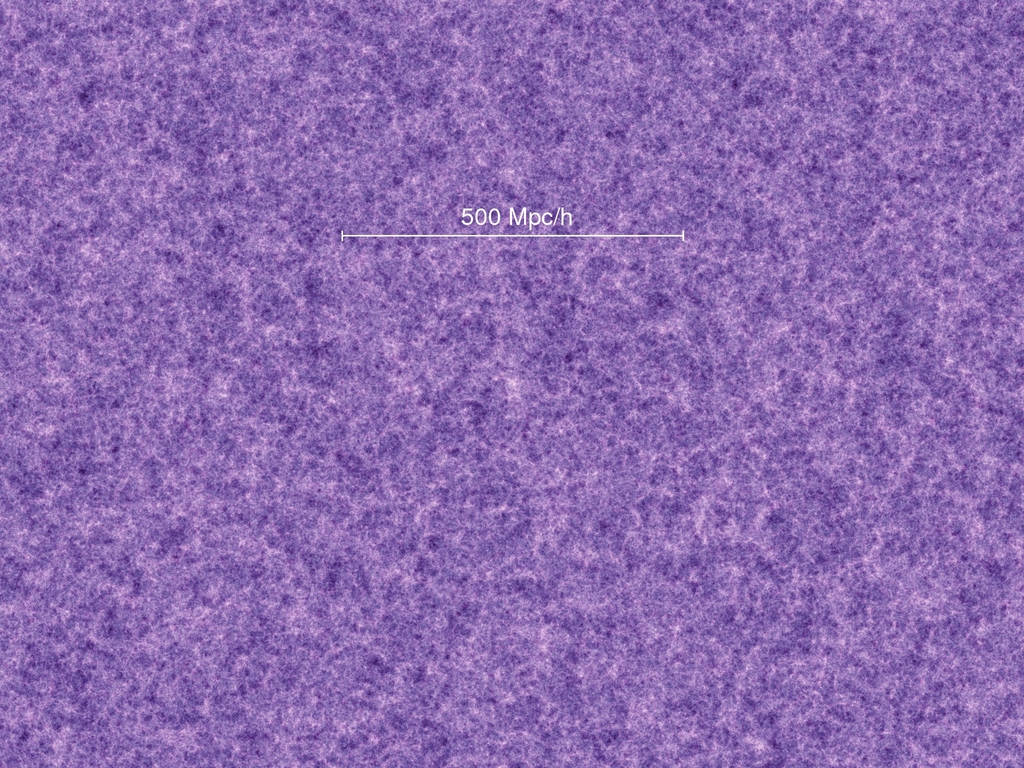}\includegraphics[width=8.cm]{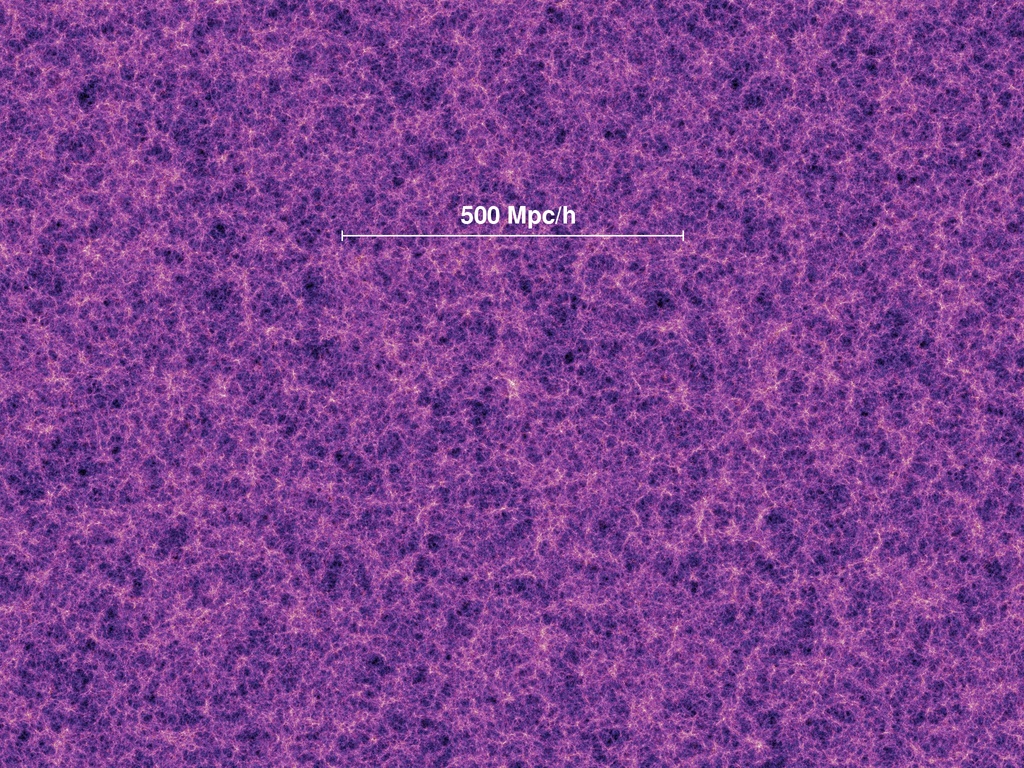}\\[-0.1cm]
 \includegraphics[width=8.cm]{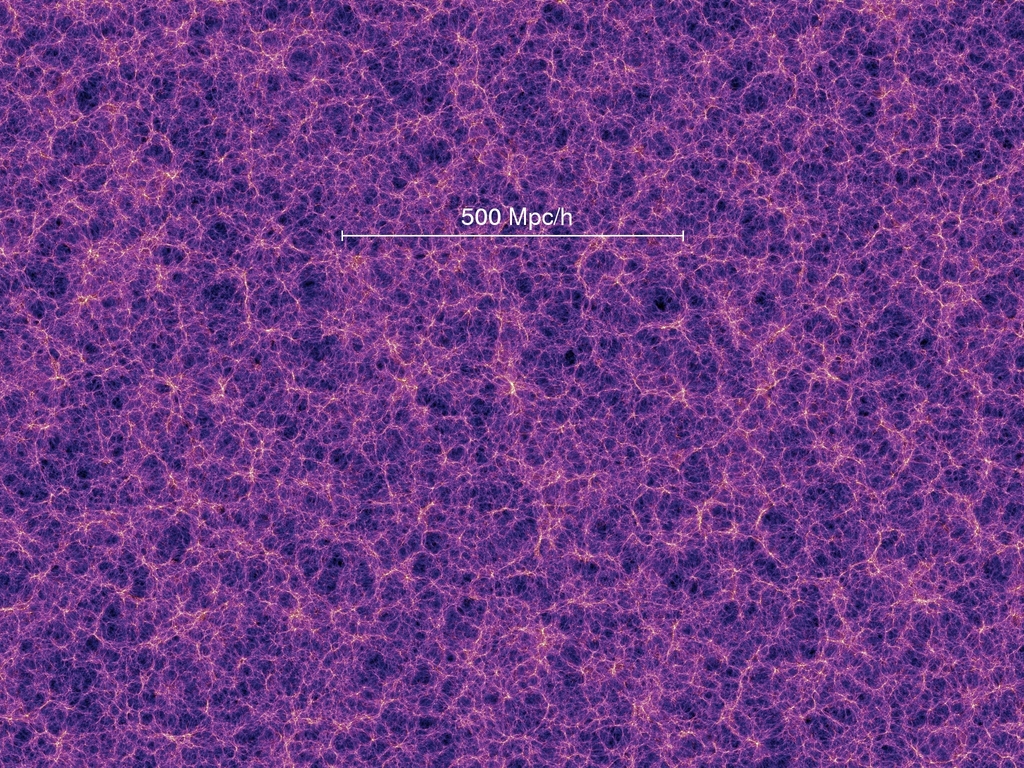}\includegraphics[width=8.cm]{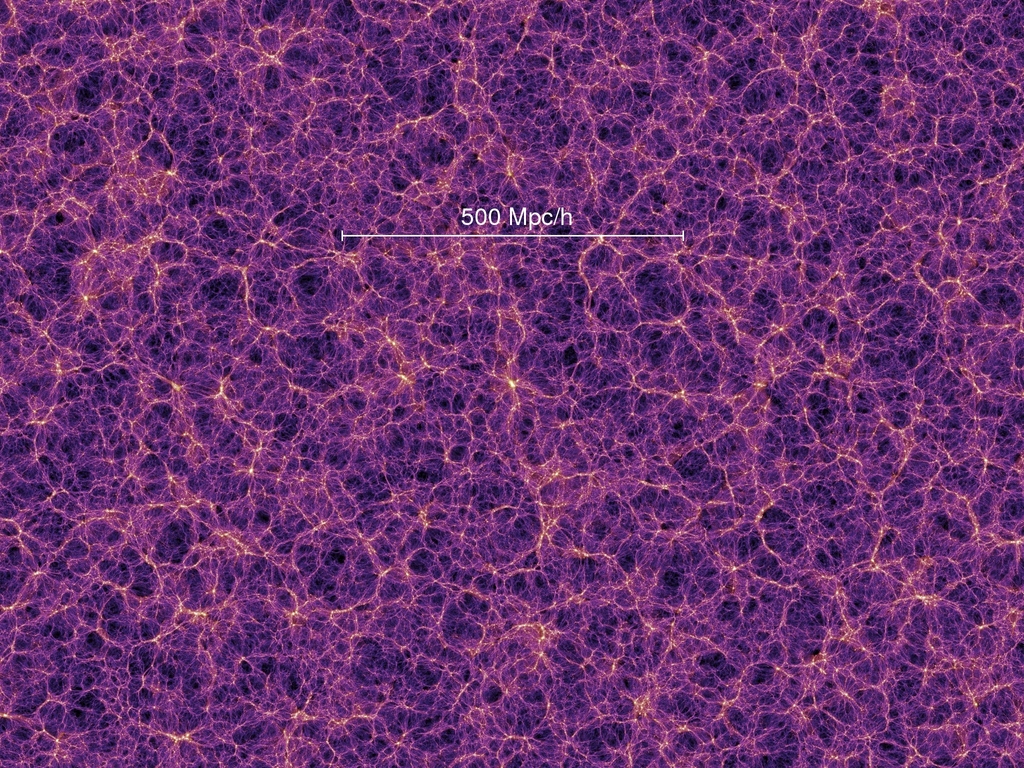}
\caption{Simulated dark matter distribution in the Millenium simulation \cite{Springel:2005nw} at redshifts (cosmological times) from upper left to lower right $z=18.3$ ($t = 0.21$ Gyr), $z=5.7$ ($t = 1.0$ Gyr), $z=1.4$ ($t = 4.7$ Gyr) and $z=0$ ($t = 13.6$ Gyr). From \cite{Croton:2006my}.\label{fig:millenium}}
\end{center}
\end{figure}

In Fig.~\ref{fig:millenium} the distribution of dark matter computed in the Millenium simulation \cite{Springel:2005nw} is shown as a function of cosmological time and redshift. This simulation reveals a hierarchical scenario: the initial distribution of the cosmological densities is relatively uniform at early times, then gravity makes the denser centres-of-mass attract more mass, thus increasing their masses and generating heavier centres-of-mass. In this process filaments of matter are formed on which the number of structures is much larger.

\begin{figure}[!ht]
\begin{center}
 \includegraphics[width=15.cm]{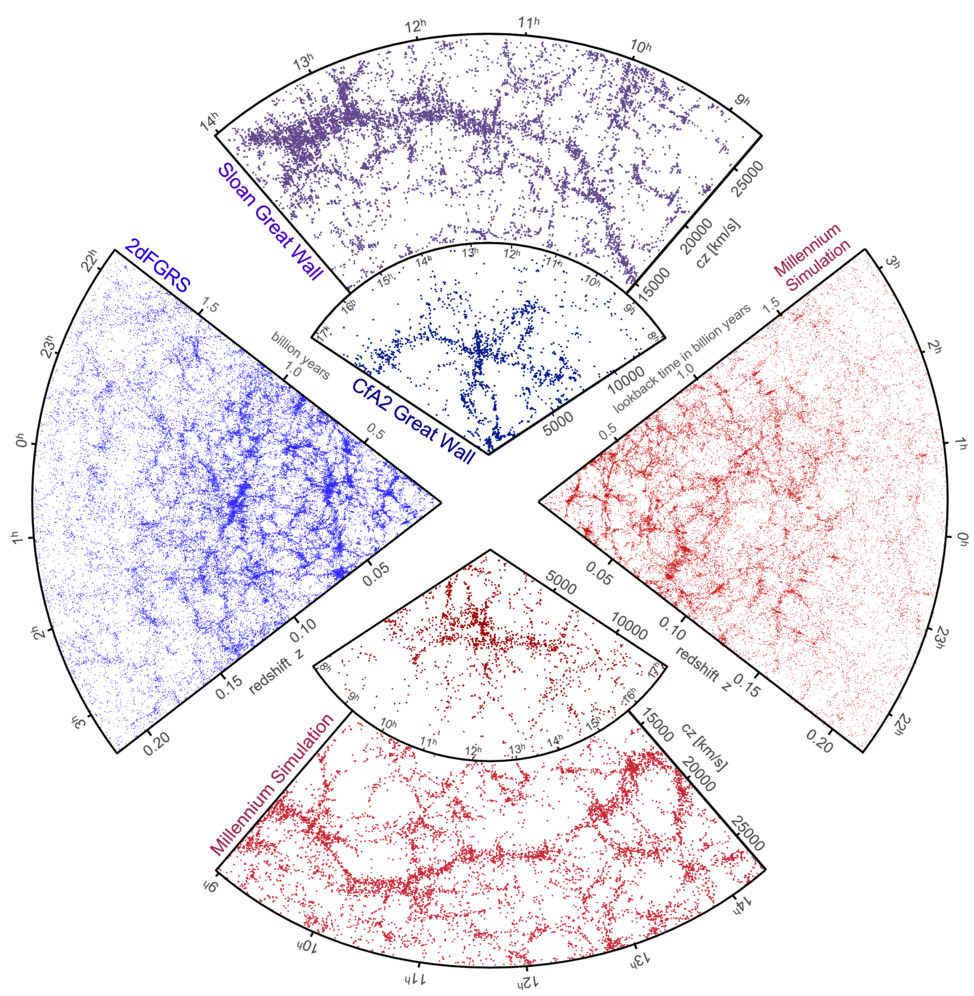}
\caption{Simulated distribution of structures from the Millenium simulation (in red) to be compared to the observed distributions by CfA2 \cite{Geller:1989da}, 2dFGRS \cite{Colless:2001gk} and SDSS \cite{York:2000gk}. From \cite{Springel:2006vs}.\label{fig:millenium_obs}}
\end{center}
\end{figure}

It is possible to compare the simulations with observational data. In Fig.~\ref{fig:millenium_obs}, Millenium simulated maps are shown together with observational maps of CfA2 \cite{Geller:1989da}, 2dFGRS \cite{Colless:2001gk} and SDSS \cite{York:2000gk}.

\begin{figure}[!ht]
\begin{center}
 \includegraphics[width=10.cm]{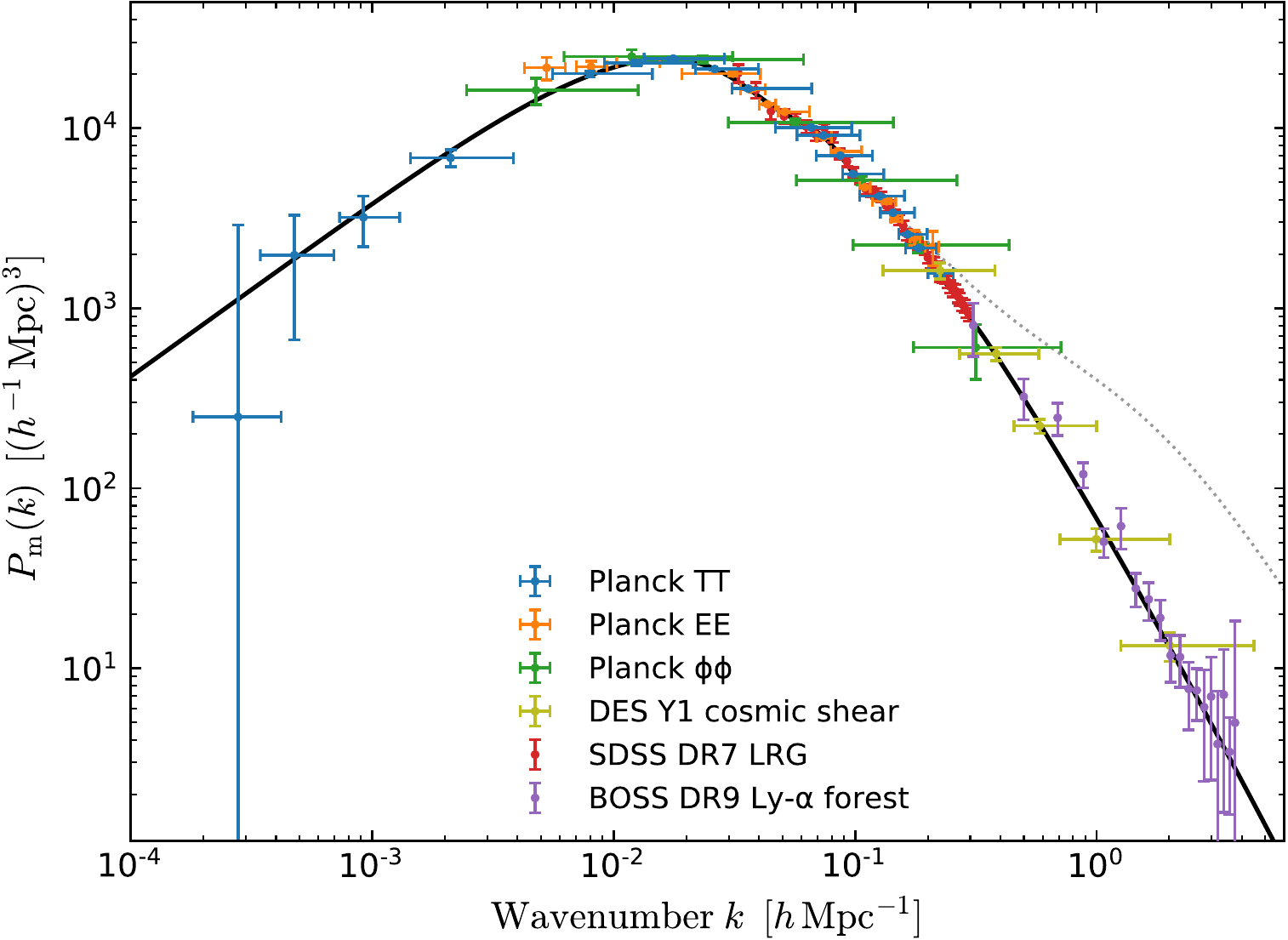} 
\caption{Matter power spectrum from different cosmological probes. The black line is the prediction of the $\Lambda$CDM standard cosmological model with the parameters given in Eq.~(\ref{eq:planck}). From~\cite{Akrami:2018vks}.\label{fig:PS}}
\end{center}
\end{figure}

For a careful comparison, a precise statistical analysis is needed, because simulations do not reproduce the actual data, but a similar set of data. To assess the similarity, the so-called power spectrum is used, which corresponds to the mass density contrast as a function of scale and is the Fourier transform of the matter correlation function (see for example \cite{Dodelson:2003ft}). Since the temperature fluctuations in the CMB can be seen as markers of small-size matter density contrast, the CMB data can also be used to extend the power spectrum to lower scales. The Planck Collaboration has released such a power spectrum \cite{Akrami:2018vks}, which is shown in Fig.~\ref{fig:PS}.%
The standard cosmological model $\Lambda$CDM shows a remarkable agreement with all the observations from disparate scales.

Finally the Planck combined results of Eq.~(\ref{eq:planck}) are consistent with the large-scale structure data, and reveal that the ratio of baryonic mass over total matter mass is close to 15\%, which is in agreement with the results at galaxy and galaxy cluster scales.

%%%

\subsection{Generic types of dark matter}

With so many observations supporting the existence of dark matter, the determination of its nature becomes an important challenge. In particular the observations reveal that dark matter is massive, stable over billions of years, very likely collisionless, interacting mostly gravitationally and separate from baryonic matter. 

If dark matter is made of new particles or objects, the first question is related to its thermal velocity. The terms ``cold'' and ``hot'' are used to characterize nonrelativistic and relativistic speeds, respectively. Beside baryons, the only electromagnetically neutral, stable and massive particles that exist in the Standard Model are the neutrinos, which are so light that they have relativistic speeds and constitute a good example of hot dark matter. The important fact about hot dark matter is that, being composed of massive and high velocity components, it tends to spread structures such as galaxies, and if constituting the totality of dark matter, structures would be shaped very differently from what we observe, and the CMB anisotropies would have larger sizes. For this reason hot dark matter can constitute only a small part of total dark matter, and it is believed that most of dark matter is cold. Under the $\Lambda$CDM assumption it is considered that the whole dark matter is cold.

Other models are possible, such as warm dark matter with intermediate properties between hot and cold, fuzzy dark matter which considers that dark matter is made of an ultra-light scalar field condensating under the form of a galaxy-sized Bose-Einstein condensate \cite{Matos:1998vk,Hu:2000ke,Arbey:2001qi,Arbey:2003sj,Hui:2016ltb,Irsic:2017yje}, dark fluid which hypothesizes that dark matter and dark energy are the same entity \cite{Boyle:2001du,Bilic:2001cg,Arbey:2006it,Guo:2005qy}, self-interacting dark matter which has a self-interaction and is not collisionless \cite{Carlson:1992fn,Spergel:1999mh}, etc. Even cold dark matter is a broad term which can go from weakly-interacting massive particles (WIMPs) of new physics models to primordial black holes. Basically, the masses of dark matter candidates span the whole range between $10^{-24}\,$eV and millions of solar masses. Primordial black holes \cite{Carr:1975qj,Carr:2009jm,Carr:2016drx} for example have attracted interest after the discovery of gravitational waves by LIGO and Virgo \cite{Abbott:2016blz}, especially because they can be used as probes for gravity \cite{Eardley:1974nw,AmelinoCamelia:1998ax,Arbey:2019zsx,Carson:2019kkh}, for quantum effects through their evaporation \cite{Unruh:1976db,Alexeyev:2002tg,Arbey:2019mbc,Ashtekar:2020ifw} and for primordial Universe models \cite{Carr:1975qj,Jedamzik:1999am,Sasaki:2016jop,Arbey:2019jmj,Inomata:2020lmk}. In the following, in order to discuss the relation between early Universe and dark matter, we will focus on particle physics candidates.

%%%%%%%%%%%%%%%%%%%%%%

\section{Beyond the standard cosmological model}

The cosmological standard model is a simple and useful scenario with a basic description of the early Universe, but does not provide an explanation for the natures of dark matter and dark energy in the present Universe. Studying possible specific cosmological models is therefore of importance. We discuss in this section some models which go beyond the standard cosmological scenario.

\subsection{Dark energy}

The cosmological constant $\Lambda$ in the standard cosmological model is a simple addition which enters naturally Einstein's equation. However it remains an assumption and it is possible that the cosmological constant is in fact the manifestation of a dynamical component such as a scalar field dominated by its potential, or a vacuum energy. The generic term ``dark energy'' encompasses a large variety of models, in which the main feature is that the pressure and density have approximately the same value with an opposite sign in the present time.

One of the simplest classes of models is the so-called quintessence model \cite{Zlatev:1998tr,Tsujikawa:2013fta,Arbey:2019cpf}, in which a real scalar field $\phi$ with a specific potential $V(\phi)$ can act as dark energy. The scalar field Lagrangian reads:
\begin{equation}
 \mathcal{L}_\phi = g^{\mu\nu}\partial_\mu\phi \partial_\nu\phi - V(\phi)\,, \label{eq:quint_lag}
\end{equation}
which gives in terms of density and pressure in the Friedmann-Lemaître Universe:
\begin{eqnarray}
 \rho_\phi &=& \frac12 \dot{\phi}^2 + V(\phi)\,, \label{eq:quint_rho}\\
 P_\phi &=& \frac12 \dot{\phi}^2 - V(\phi)\,. \label{eq:quint_P}
\end{eqnarray}
In addition, the scalar field obeys the Klein-Gordon equation:
\begin{equation}
 \ddot{\phi} + 3 H \dot{\phi} + \frac{dV}{d\phi}\,, \label{eq:quint_KG}
\end{equation}
which is equivalent to the energy conservation equation (\ref{eq:energy_conservation}) for the scalar field. The choice of the potential is arbitrary, and several potentials have been studied in the literature (see Ref.~\cite{Tsujikawa:2013fta} for a review). As an example we consider the following inverse power-law potential \cite{Linder:2006uf}:
\begin{equation}
 V(\phi) = M^4 \left(\frac{M_P^4}{\phi}\right)^p\,,
\end{equation}
where $M$ is a parameter with a mass unit, $M_P$ the Planck mass and $p$ a constant exponent. The value of $M$ together with the initial conditions have to be chosen in order to find the equation of state $w_\phi = P_\phi / \rho_\phi$ close to $-1$ at present time and a density $\rho_\phi$ in agreement with the Planck's results given in~(\ref{eq:planck}).

\begin{figure}[!ht]
 \begin{center}
  \includegraphics[width=15.cm]{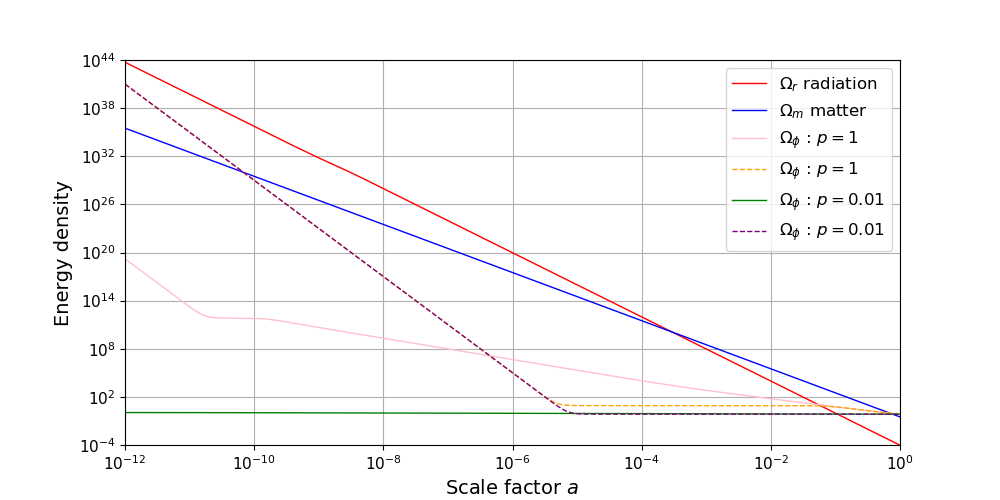}\\\includegraphics[width=15cm]{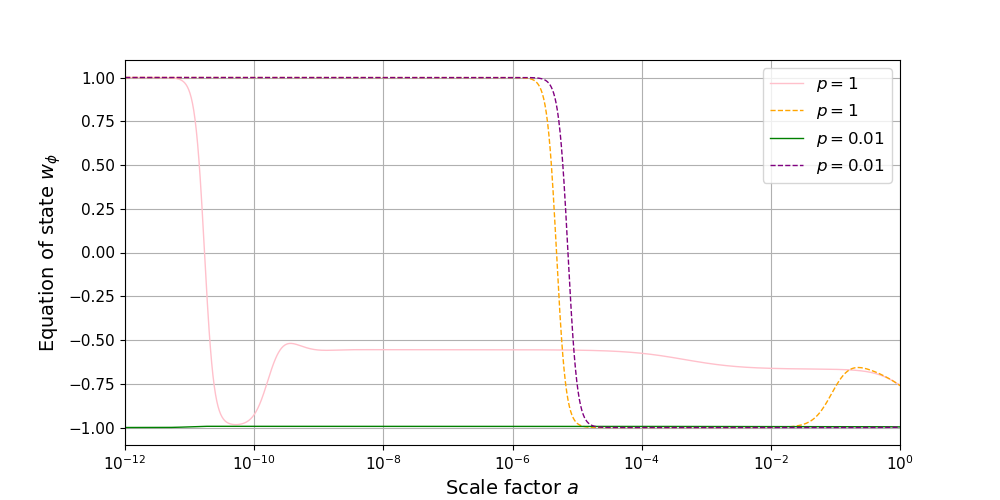}
\caption{Behaviour of a quintessence scalar field with an inverse power-law potential $V(\phi) = M^4 (M_P^4/\phi)^p$ for two different choices of $p$ and for two sets of initial conditions. The upper (lower) plot shows the evolution of the cosmological densities (equations of state of quintessence) as a function of the scale factor.\label{fig:quintessence} From~\cite{Arbey:2019cpf}.}
 \end{center}
\end{figure}

In Fig.~\ref{fig:quintessence}, the evolution of the quintessence density and its equation of state are shown as a function of the scale factor, for two values of $p$ and two different sets of initial conditions. $M$ has been chosen to obtain a quintessence density today equal to the observed cosmological constant density. The radiation and matter densities are shown for comparison. It is interesting to notice the sensitivity to the initial conditions and exponent. Also, contrary to the cosmological constant, for some choices of the parameters, the quintessence density can become dominant at very high redshifts, for a scale factor smaller that $10^{-12}$. The field is at this time in a regime where its kinetic term dominates, and its density decreases as $a^{-6}$.

Many other models for dark energy have been developed, some well-known with esoteric names such as phantom energy \cite{Caldwell:1999ew} or chameleon fields \cite{Khoury:2003rn}. They all lead to specific behaviours of dark energy in the recent Universe, but show that dark energy could also play a role in the early Universe.

%%%

\subsection{Inflation and reheating}

Three major problems have been identified regarding the early Universe. The first one is about the apparent flatness of the Universe. We observe that the curvature $k$ is close to zero, but its exact value is unknown. The curvature density may represent a small fraction of the critical density and we know that it evolves as $a^{-2}$, whereas the matter and radiation densities evolve as $a^{-3}$ and $a^{-4}$, respectively. Therefore curvature has always had a negligible effect in comparison to matter and radiation. Yet curvature is related to gravity, thus one can suppose that there existed a small non-negligible curvature at the beginning of the Universe. In such a case, because of the evolution of the curvature density, curvature would dominate the expansion today, contrary to the observations.

The second problem is about the isotropy of the CMB. Indeed, the expansion of the Universe is a global dilatation, so that the apparent velocity of a fixed object at a distance $r$ from us is given by $v = H r$. If this velocity becomes larger than the speed of light, a horizon forms, delimiting causally disconnected regions. The corresponding radius $r_H = c / H$ is called Hubble radius. Integrating from the beginning of the Universe until recombination, it is possible to show that the Hubble radius at the beginning of the Universe corresponds to regions of about 2$^{\circ}$ in the CMB sky. If so, we should observe anisotropies of angular sizes close to this value. This is in contradiction with the excellent isotropy of the CMB which is of the order of $10^{-5}$.

The third problem concerns the CMB anisotropies themselves. Starting from a purely homogeneous and isotropic Universe, it is not possible to generate inhomogeneities. It has been shown that quantum fluctuations cannot account for the size of the inhomegeneities which seeded large scale structures and imprinted the CMB.

These three problems have found a common explanation in the scenario of inflation \cite{Starobinsky:1980te,Starobinsky:1982ee}: about $10^{-36}$ seconds after the beginning of the Universe, the Universe underwent an exponential expansion called inflation. Inflation could be driven by the presence of a scalar field, the inflaton, which decayed into more standard particles at the end of inflation during the so-called reheating period. Because of this exponential expansion, regions originally causally connected before inflation have been causally disconnected, and the apparent isotropy of the CMB results from the causal connection between the different regions of the sky in the pre-inflation period. Also to generate an exponential expansion the inflaton had to be dominant with a nearly constant density, so that any curvature existing in the early Universe would be flattened by the expansion and the effective curvature density has rapidly become negligible. Finally during the reheating quantum fluctuations have also been enlarged by the rapid expansion resulting in large enough inhomogeneities to seed the formation of large scale structures.

The inflation mechanism is very similar to the one of quintessence, with a real scalar field following Eqs. (\ref{eq:quint_lag})--(\ref{eq:quint_KG}). In order to reach an exponential expansion it is necessary for the field to have a slowly-varying density resulting in a negative pressure, an equation of state $w_\phi \sim - 1$, a negligible kinetic term and a dominating potential. In other words the scalar field is slowly rolling down the potential, and such a model of inflation is generally called slow-roll inflation \cite{Linde:1981mu,Albrecht:1982wi}. This implies two conditions: $\epsilon,|\eta| \ll 1$ where
\begin{equation}
\epsilon=\frac{M_P^2}{16\pi}\left(\frac{V'}{V}\right)^2 \;,\;\;\; \eta=\frac{M_P^2}{8\pi}\left(\frac{V''}{V}\right)\,.
\end{equation}
At the end of inflation, the field reaches the minimum of the potential and starts oscillating around the minimum. If coupled to Standard Model particles, the field then decays into them, following the modified Klein-Gordon equation:
\begin{equation}
 \ddot{\phi}+3H\dot{\phi}+\frac{dV}{d\phi}= - \Gamma_\phi \rho_\phi \,,
\end{equation}
where $\Gamma_\phi$ is the decay width of the scalar field. If the field decays into radiation, radiation receives an entropy injection:
\begin{equation}
\frac{ds_\text{rad}}{dt} = -3 H s_\text{rad} + \frac{\Gamma_\phi \rho_\phi}{T}\,,
\end{equation}
resulting in generation and (re)heating of the primordial plasma. The reheating temperature $T_\text{RH}$ of the scalar field reads \cite{Gelmini:2006pw,Arbey:2018uho}:
\begin{equation}
 \Gamma_\phi = \sqrt{\frac{4\pi^3 g_\text{eff}(T_\text{RH})}{45}} \, \frac{T^2_\text{RH}}{M_P}\,, \label{eq:TRH}
\end{equation}
where $g_\text{eff}$ is the number of effective energy degrees of freedom of radiation.

Several models of inflation exist depending on the form of the potential, and they can be constrained via the analysis of the shape of the anisotropies of the CMB \cite{Akrami:2018odb}.

%%%

\subsection{Other models}

More fundamental properties of our Universe may also affect cosmology, in particular in the Planck epoch, where a quantum gravity description would be required. Beside quantum gravity other modifications of Einstein's gravity are possible.

Generically one can postulate that the geometric Lagrangian of gravity, which reduces to the scalar curvature $R$ in Einstein's gravity, is an unknown function $f(R)$ \cite{Buchdahl:1983zz,Sotiriou:2008rp}. Since the curvature $R$ is small in the Solar System, $f(R)\approx R$ for small values of $R$, so that Einstein's gravity is retrieved close to the Earth and in recent cosmology. In the early Universe when the densities were extremely large, the scalar curvature was large and new gravitational effects could occur. It is remarkable that for
\begin{equation}
 f(R) = R + \frac{R^2}{6 M}\,,
\end{equation}
where $M$ is a parameter with the dimension of a mass, Starobinsky's inflation \cite{Starobinsky:1980te} is retrieved, since the gravity modification is equivalent to the presence of a scalar field through a conformal transformation.

Another question concerns the existence of extra-dimensions, which are predicted by string theories but are today invisible to us. To hide the extra-dimensions, one generally considers that they are compactified with very small radii much below the scales we are sensitive to. It is often assumed that Standard Model particles are confined outside the extra-dimensions, but gravity may enter them, leading for example to the modification of the speed of gravitational waves \cite{Andriot:2019hay}. The existence of such extra-dimensions could give birth to effective scalar fields in our 4-dimensional world, such as dilatons \cite{deCarlos:1993wie,Gasperini:1994xg}, or other new particles \cite{ArkaniHamed:1998nn,Randall:1999ee}.

Generally speaking, our limited knowledge of high energy physics and quantum gravity does not allow us to rule out new phenomena which could modify the early Universe properties.

%%%

\subsection{Phase transitions}

Other phenomena that certainly occurred in the early Universe are phase transitions, at temperatures much larger than the ones of Big-Bang nucleosynthesis and recombination. We are aware of two such phase transitions: hadronization of the quark-gluon plasma around 200 MeV, and spontaneous symmetry breaking of the SU(2)$\times$U(1) gauge theory of electroweak interactions around 200 GeV. Both phase transitions are indirectly studied at colliders and in particular at the LHC, but their exact mechanisms are still unclear.

Moreover Grand Unified Theories which postulate the unification at a high energy of the three Standard Model interactions require the existence of a phase transition at a temperature of about $10^{15}\,$GeV during which the electroweak interaction decouples from the strong interaction.

In addition our Universe presents a deficit of antibaryons in comparison to baryons. Antibaryons are extremely rare, and yet the Standard Model does not provide mechanisms to explain the baryon asymmetry. Sakharov stated three necessary conditions for such a mechanism \cite{Sakharov:1967dj}:\vspace*{-0.1cm}
\begin{itemize}
 \item baryon number violation,\vspace*{-0.1cm}
 \item C-symmetry and CP-symmetry violation,\vspace*{-0.1cm}
 \item interactions out of thermal equilibrium,\vspace*{-0.1cm}
\end{itemize}
which if satisfied may enable an over-production of baryons. The Standard Model has too small baryon number violation and CP-violation, and the first two conditions necessitate new physics beyond the Standard Model, while the third condition implies the existence of a phase transition where thermal equilibrium is broken via for example entropy injection. This phase transition is called baryogenesis. Two main classes of models exist: on the one hand baryogenesis could be related to the symmetry breaking of Grand Unification \cite{Kuzmin:1987wn,Kolb:1996jt}; on the other hand it may be related to the electroweak symmetry \cite{Fukugita:1986hr,Anderson:1991zb,Cohen:1993nk,Trodden:1998ym}. It is of course possible that other phase transitions existed at intermediate energies. In any case these phase transitions involve new fields and entropy injection which are likely to modify strongly the properties of the early Universe.

%%%%%%%%%%%%%%%%%%%%%%

\section{Dark matter in particle physics}

We describe in this chapter the connections between dark matter and particle physics.

\subsection{Dark matter and new physics}

As discussed in the previous sections, the Standard Model of particle physics does not provide a candidate for cold dark matter. The main requirements for an adequate candidate are that it has to be massive, nonrelativistic, weakly-interacting and stable over tens of billions of years. This rules out in general coloured particles, charged particles and very light particles. Stability is the main condition for dark matter and an important requirement when building new physics models.

In order to guarantee the stability, it is necessary to consider either a new sector secluded from the Standard Model, or to have a discrete symmetry to protect particles, or to have composite particles with a strong binding. The second method, i.e. a protective discrete symmetry, is the most common technique to obtain dark matter particles. For example, under a $Z_2$ symmetry, new physics particles have a multiplicative charge $-1$, and the Standard Model particles have a charge $+1$. The main consequences are first that new particles can only decay into an odd number of new particles in addition to standard particles, and second two new particles can (co-)annihilate into Standard Model particles. With such a set-up, the lightest new particles are stable, can be produced by collisions of Standard Model particles and can annihilate inside dark matter halos. In such a case, the surviving dark matter particles are called relics.

Dark matter is not the only motivation for new physics models. Extensions of the Standard Model are based on new symmetries ({\it e.g.} supersymmetry), new symmetry groups ({\it e.g.} SO(10)), new hypotheses ({\it e.g.} extra-dimensions), new particles ({\it e.g.} vector-like quarks), ... which can solve theoretical problems affecting the Standard Model ({\it e.g.} hierarchy problem, baryon asymmetry, ...).

In a complete scenario it is possible to compute the relic abundance of dark matter. For this purpose, two cases can be considered: thermal relics and non-thermal relics.

\subsubsection{Thermal relics}

The name ``thermal relics'' encompasses models in which new particles are in thermal equilibrium in the early Universe. During this equilibrium the distribution functions of the new particles can be of Fermi-Dirac for fermions and of Bose-Einstein for bosons. For particle species $i$, the distribution of their energy $E_i$ at a temperature $T$ is given by:
\begin{equation}
 f_i(E_i,\mu,T) = \frac{g_i}{\exp\left(\displaystyle\frac{E_i-\mu}{k_B T}\right) \pm 1}\,,
\end{equation}
where the $+$ sign is for fermions and $-$ for bosons, $g_i$ is the degeneracy,  $\mu$ the chemical potential and $k_B$ the Boltzmann constant. These distributions in the early Universe plasma where the temperature is extremely high reduce to the Maxwell-Boltzmann distribution for the cases where the interaction between particles is weak and the chemical potential is negligible:
\begin{equation}
 f_i(E_i,T) = g_i \, \exp\left(-\frac{E_i}{k_B T}\right)\,.
\end{equation}

As long as the particles are in interaction with the plasma with a common temperature, particles are in thermal equilibrium  and the distributions are driven only by the temperature. However, since the Universe is in expansion, in turn the temperature decreases and the average distance between particles in the plasma increases, which limit the interaction rate. The equilibrium can therefore be broken, the unstable particles decay and the stable ones undergo a ``freeze-out''. Since the stable particles do not decay and the interactions are heavily suppressed, their number is frozen and nearly constant. Consequently the number density of stable particles decreases as $a^{-3}$.

This scenario is the most common one. A detailed description of the relic density calculation of thermal relics is given in Section~\ref{sec:thermal}.

\subsubsection{Non-thermal relics}

Non-thermal relics are particles which were not in thermal equilibrium in the plasma of the early Universe. To have non-thermal relics, it is necessary that either dark matter particles are produced at low temperatures so that they never reached thermal equilibrium, or that these particles were so weakly interacting with the plasma that they could be considered as decoupled from the primordial plasma. There exist several models of non-thermal relics, and we will discuss here three classes: axion-like particles, freeze-in models and decay products of primordial fields.

\paragraph{Axion-like particles (ALPs)}

Axion-like particles are Weakly Interacting Sub-eV Particles (WISPs). The original axion model is related to the strong CP-violation of the Standard Model. The QCD Lagrangian can be written as
\begin{equation}
 \mathcal{L}_{\rm QCD} = \mathcal{L}_{\rm QCD,\,pert} + \bar{\Theta} \frac{g^2}{32\pi^2} G^a_{\mu\nu} \bar{G}^a_{\mu\nu}\,,
\end{equation}
where $G^a_{\mu\nu}$ is the gluon tensor, $\bar{G}^a_{\mu\nu}$ its dual and $\bar{\Theta}$ a constant. $\mathcal{L}_{\rm QCD,\,pert}$ is the usual perturbative QCD Lagrangian. The second term is an effective term coming from the topological properties of vacuum in non-Abelian gauge theories and from the diagonalization of the quark mass matrix. It is a nonperturbative term which does not affect the pertubative QCD calculations and can be written as a divergence of a current. However this term violates P- and CP-symmetries. Measurements of the neutron electric dipole moment reveal that $|\bar{\Theta}| \le 10^{-10}$, showing that CP-symmetry can be only slightly violated. This is the strong CP-problem: why is this CP-violation constrained to be tiny?

A solution was proposed by Peccei and Quinn \cite{Peccei:1977hh,Peccei:1977ur}, in which $\bar{\Theta}$ is replaced by a dynamical term. For this, they added to the Standard Model a new chiral global symmetry $U(1)_{PQ}$ which is spontaneously broken and gives birth to a new Goldstone (massless real scalar) boson, the instanton $\phi$. Under a $U(1)_{PQ}$ transformation the instanton scalar field transforms as
\begin{equation}
 \phi \longrightarrow \phi + f_a\, \epsilon\,,
\end{equation}
where $f_a$ is a constant with the dimension of an energy, of the order of the symmetry breaking scale. However this symmetry suffers from an Adler-Bell-Jackiw anomaly \cite{Adler:1969gk,Bell:1969ts} which generates an effective term in the Lagrangian. The Peccei-Quinn additional term of the Lagrangian therefore reads:
\begin{equation}
 \mathcal{L}_\phi = \frac12 \partial_\mu\phi \partial^\mu\phi + \frac{g^2}{32\pi^2 f_a} \phi G^a_{\mu\nu} \bar{G}^a_{\mu\nu}\,, 
\end{equation}
and by choosing the average value of the field to be
\begin{equation}
 \langle\phi\rangle = - f_a \bar{\Theta}\,,
\end{equation}
the P- and CP-violating terms are eliminated. The fluctuation of the field $\delta\phi = \phi - \langle\phi\rangle$ represents the axion field, which because of the anomaly acquires a mass
\begin{equation}
 m_\phi \sim \frac{\Lambda_{\rm QCD}^2}{f_a}\,,
\end{equation}
where $\Lambda_{\rm QCD} \approx 200\,$MeV.
Axions are expected to have masses in the $\mu$eV--eV range, and are only nonpertubatively directly coupled to the QCD sector. As a consequence, axions are stable in the vacuum and interact only very weakly with the pertubative Standard Model.

In cosmology most of the relic axions are generated via the decay of primordial cosmological scalar fields, and because of their very weak couplings to standard particles they in general do not thermalize with the thermal plasma.

More generally axion-like particles refer to very light particles which are produced nonthermally in the early Universe. Their relic abundance is related to the details of the cosmological production mechanism, but can be compatible with the observed dark matter density \cite{Abbott:1982af,Marsh:2015xka}.

An important question is related to the very tiny mass of the axions: can such particles have nonrelativistic velocities and constitute dark matter? Indeed, standard thermal relics are expected to leave thermal equilibrium and decouple from the plasma at high temperatures because of their very low couplings to standard particles. If their mass is much smaller than the temperature, they would be relativistic. This is the case for example for the Standard Model neutrinos, which constitute a hot matter. A way out would be to produce these particles with nonrelativistic velocities, which is possible as long as they do not thermalize with the plasma. Another mechanism drives the dark matter behaviour of ultralight particles: because of their small mass, the associated Compton wavelengths can be large. The particles have therefore wave functions spread over a large distance, and they can condense into Bose-Einstein condensates \cite{Jetzer:1991jr,Boehmer:2007um}, whose size and mass are determined by the scalar field density and mass. For example, for a scalar field with a simple quadratic potential, the Compton wavelength is given by
\begin{equation}
 l_{\rm Compton} = \frac{\hbar}{m\,c} \,,
\end{equation}
where $\hbar$ is the reduced Planck constant and $m$ the mass of the quadratic potential. The size of the condensate is related to the value of the field at its centre $\phi_0$ \cite{Arbey:2001jj}:
\begin{equation} 
 L \sim \sqrt{\frac{M_P}{\phi_0}} \frac{\hbar}{m\,c}\,,
\end{equation}
and since the condensate is gravitationally bound, the velocities $v$ inside the condensate are related to the scalar field central value by:
\begin{equation} 
 \frac{v}{c} \sim \frac{\phi_0}{M_P}\,.
\end{equation}
The observed sizes of galactic halos and their velocities fix an upper limit on the Bose-Einstein condensate size: for a size $l \sim 100$ kpc and velocity $v \sim 100\,$km/s, one finds $m\sim 10^{-24}$\,eV, which is the typical mass of the so-called fuzzy dark matter \cite{Matos:1998vk,Hu:2000ke,Arbey:2001qi,Arbey:2003sj,Hui:2016ltb,Irsic:2017yje}. This value is of course a minimum mass $m$ and maximum size $L$ for Bose-Einstein condensation. It is indeed possible to have small-size Bose-Einstein condensates with larger masses $m$ acting as dark matter particles \cite{Sikivie:2009qn}.

\paragraph{Freeze-in models}

The freeze-in scenarios concern Feebly Interacting Dark Matter Particles (FIMPs) \cite{Hall:2009bx}: these particles have so feeble couplings to the standard particles that they are initially decoupled from the primordial thermal plasma, and could have been produced in small quantities by the decay of primordial fields. But even if feeble, interactions with the bath still lead to production of FIMPs and increase their density, up to the point that the temperature of the plasma is too low for such a production. This scenario is somehow reversed in comparison to the freeze-out scenario in which the new particles that are initially thermalized quit equilibrium when the temperature decreases.

Whereas freeze-out scenarios make no assumption on the initial production of the new particles since they are considered as thermalized, freeze-in scenarios need a description of the production mechanisms. Therefore freeze-in scenarios are predictive only when they derive from a high-energy theory which can predict their original density. Such models are nevertheless actively considered by the particle physics community, since the feeble couplings may constitute a reason why no new particles have been discovered so far, neither at colliders nor in dark matter experiments.

\paragraph{Decay products}

Massive and very weakly-interacting particles can also be generated non-thermally in the early Universe by the decay of primordial fields during phase transitions. If they are too weakly coupled to Standard Model particles, they do not interact with the thermal plasma and their relic density is determined by the production mechanism. However, in most new physics scenarios, other unstable new particles can also exist, which can thermalize with the plasma and subsequently decay into dark matter particles. In such scenarios the next-to-lightest new particle is in general stable over a long time because of its very weak coupling to the dark matter particles, before decaying into dark matter. At colliders the next-to-lightest particle can then be produced and be subject to searches for charged, coloured and/or long-lived particles.

\paragraph{General case}

In the most general case, dark matter can be composed of several new particles, which can come from one or several of the scenarios discussed earlier. In supersymmetry for example, the lightest supersymmetric particle, in general a neutralino or a gravitino, can be a dark matter candidate, but it is possible to have in addition an axion and its supersymmetric partner, the axino \cite{Chun:1992zk,Brandenburg:2004du}. Also in presence of several extra-dimensions, each extra-dimension can give birth to one dark matter candidate or more \cite{Arbey:2012ke}. In addition, it is possible to have dark matter in the form of primordial black holes.

%%%

\subsection{Experimental constraints on dark matter scenarios}

\begin{figure}[!ht]
\begin{center}
 \includegraphics[width=6.cm]{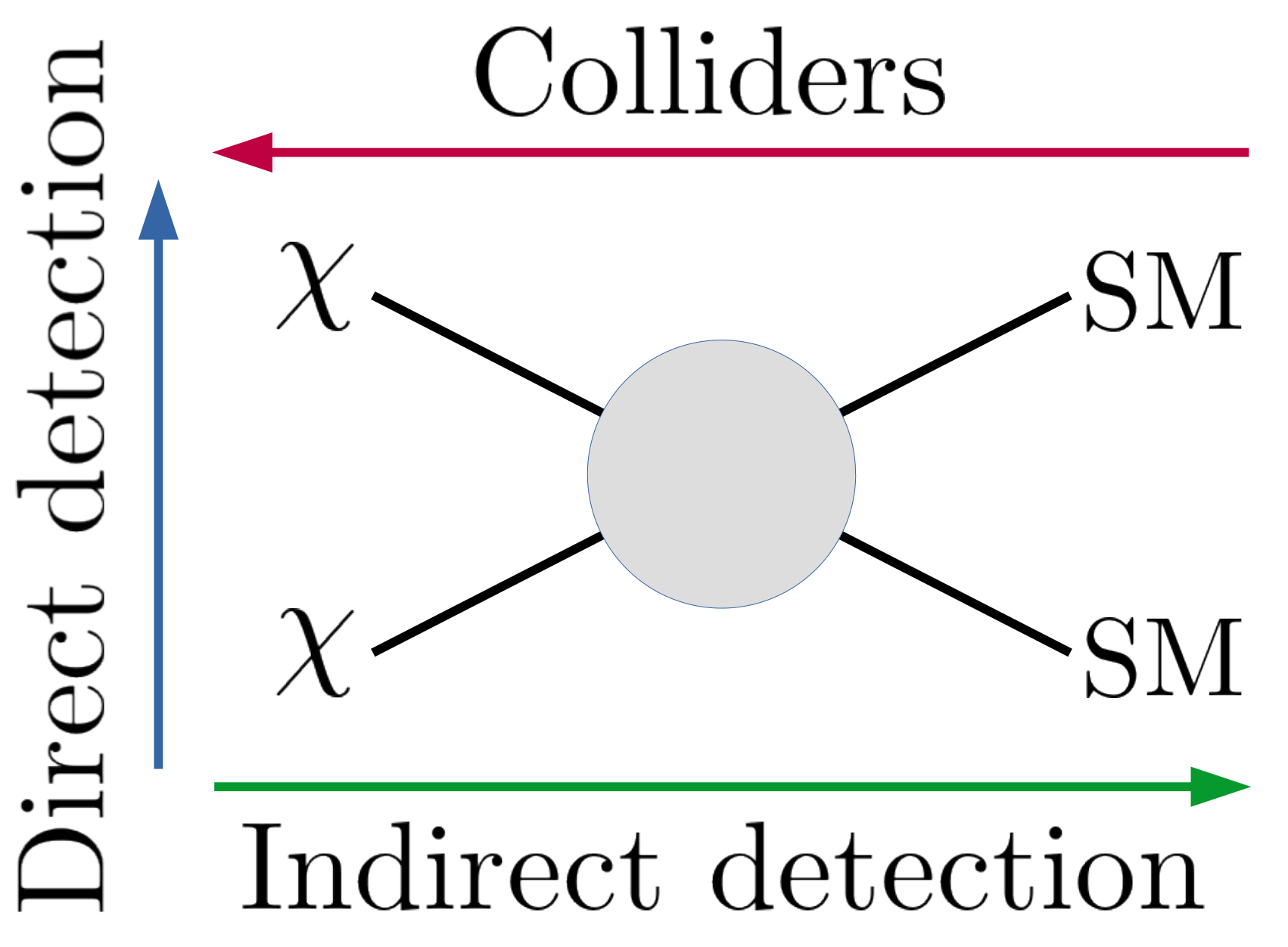}
\caption{Schematic representation of the three different types of dark matter particle searches and their interplay.\label{fig:DMsearches}}
\end{center}
\end{figure}

Dark matter particles are currently actively searched for at colliders and in dedicated dark matter experiments. In addition the relic density imposes strong constraints on dark matter scenarios. We discuss here the main experiment types: dark matter direct and indirect detections, and collider searches. A schematic and simplified view of the interplay between the three types of searches is shown in Fig.~\ref{fig:DMsearches}.

\subsubsection{Dark matter direct detection}

The local density of dark matter in the Solar System is of the order of 0.4 GeV/cm$^3$ \cite{Catena:2009mf,Nesti:2013uwa,Sivertsson:2017rkp}. If dark matter is composed of particles of masses around 100 GeV, this results in about 4 particles per cubic meter. Galactic halos are considered to be fixed in comparison to the rotating disc of the galaxy, thus dark matter has a relative velocity of about 200 km/s. During one year each cubic meter on Earth would therefore be crossed by $\sim10^{13}$ dark matter particles. It is thus possible to look for interactions of dark matter particles on Earth, by building large tanks and detectors of specific materials which maximize the probability of interaction. 

\begin{figure}[!ht]
\begin{center}
 \includegraphics[width=15.cm]{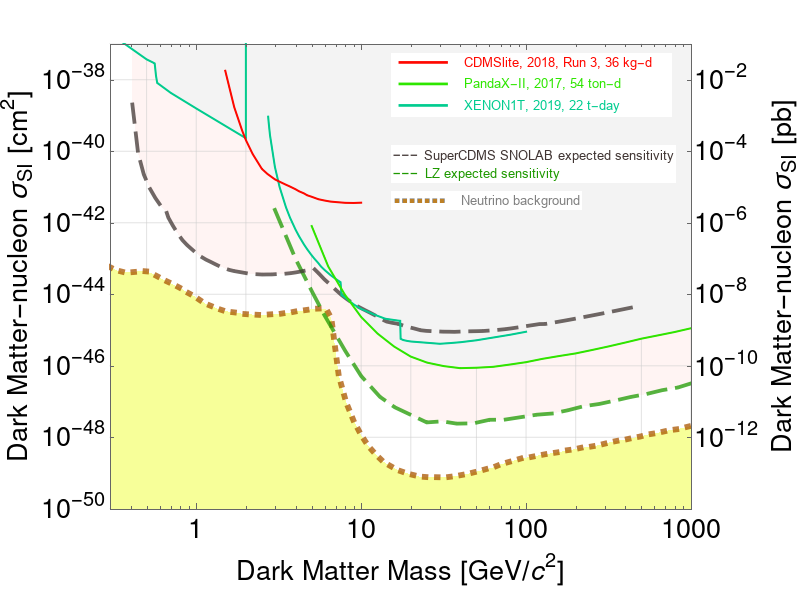}
\caption{Exclusion limits (solid lines) of dark matter detection experiments in the scattering cross section vs. dark matter mass plane, for CDMSLite \cite{Agnese:2018gze}, PANDAX-II \cite{Cui:2017nnn} and XENON1T \cite{Aprile:2019xxb,Aprile:2019jmx}. The dashed lines correspond to prospective limits for the future SuperCDMS+SNOLAB and LZ experiments. The yellow region corresponds to the cosmic neutrino background \cite{Ruppin:2014bra}.\label{fig:directdetectionexp} Obtained using the {\tt Dark Matter Limit Plotter} tool\protect\footnotemark.}
\end{center}
\end{figure}

Dark matter interacts with standard matter via a scattering with the nucleons (or electrons) of the atoms present in the experiments. The idea is to measure the recoil energy of nuclei in order to detect their interactions with matter, and to estimate the dark matter mass and the scattering cross section with nucleons, $\sigma_N$. Several direct detection experiments are currently running. A summary of the present and prospective results is shown in Fig.~\ref{fig:directdetectionexp}, based on the assumption that dark matter is composed of a single type of weakly-interacting massive particles.

\footnotetext{From {\tt https://supercdms.slac.stanford.edu/dark-matter-limit-plotter}.}

\begin{figure}[!t]
\begin{center}
 \includegraphics[width=12.cm]{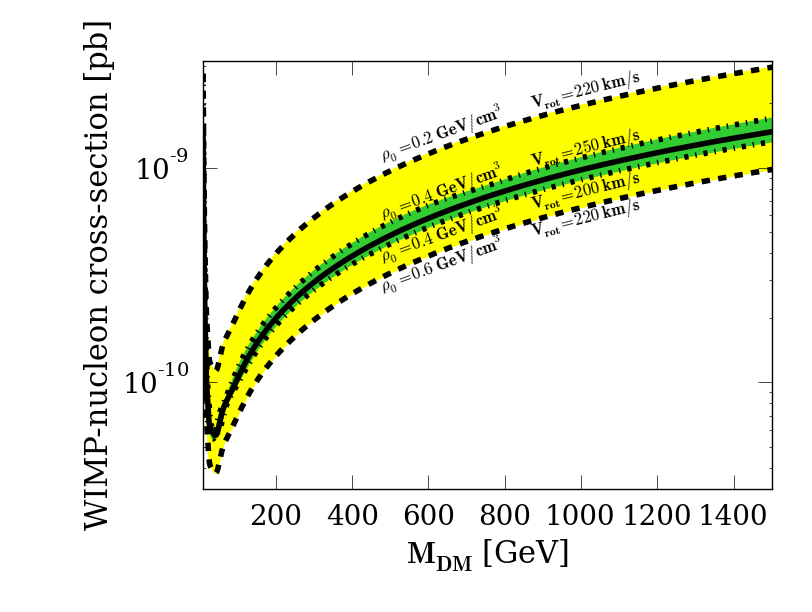}
\caption{XENON1T WIMP-nucleon cross section upper limit as a function of the dark matter particle mass, for different values of the local density and velocity.\label{fig:dduncert} From \cite{Arbey:2017eos}.}
\end{center}
\end{figure}

Beside uncertainties related to nucleon models such analyses are still limited by our poor knowledge of the local dark matter density and velocity, as can be seen in Fig.~\ref{fig:dduncert}, where the XENON1T experiment \cite{Aprile:2017iyp} limits are displayed for different choices of the local dark matter density and velocity.

The sensitivity of direct detection experiments is improving generation after generation, and will reach the limit where the scattering of the ambient neutrinos with nucleons will be detectable. Even if interesting, this background will constitute a difficult limit to bypass in order to search for more weakly-interacting dark matter particles. The scattering cross sections can be computed in a specific new physics model and the limits can be used to set constraints.

For very light particles like ALPs, such detection methods are not possible, because the recoil of nucleons would be too tiny. There are however specific experiments to directly search for ALPs through their interaction with electromagnetic fields. Indeed axions have couplings with photons, and in axion search experiments \cite{Graham:2015ouw} closed electromagnetic cavities are used in which the interaction of axions with the electromagnetic field can produce detectable photons.

\subsubsection{Dark matter indirect detection}

Indirect detection of dark matter is based on the production of Standard Model particles through dark matter interaction or decay, which are then detected on Earth. The different types of indirect detection experiments depend on the kind of particles which are searched for. The main types of decay or annihilation products are photons, charged particles (protons, antiprotons, electrons, positrons) and neutrinos. Photons are the easiest to detect, and their propagation is mostly unaffected by the interstellar and intergalactic media, but their background can be large. The neutrinos have also an unaffected propagation with a reduced background, but are difficult to detect. Charged particles are on the other hand strongly affected by the interstellar medium and magnetic fields and their propagation is complex to model.

\begin{figure}[!t]
\begin{center}
 \includegraphics[width=10.cm]{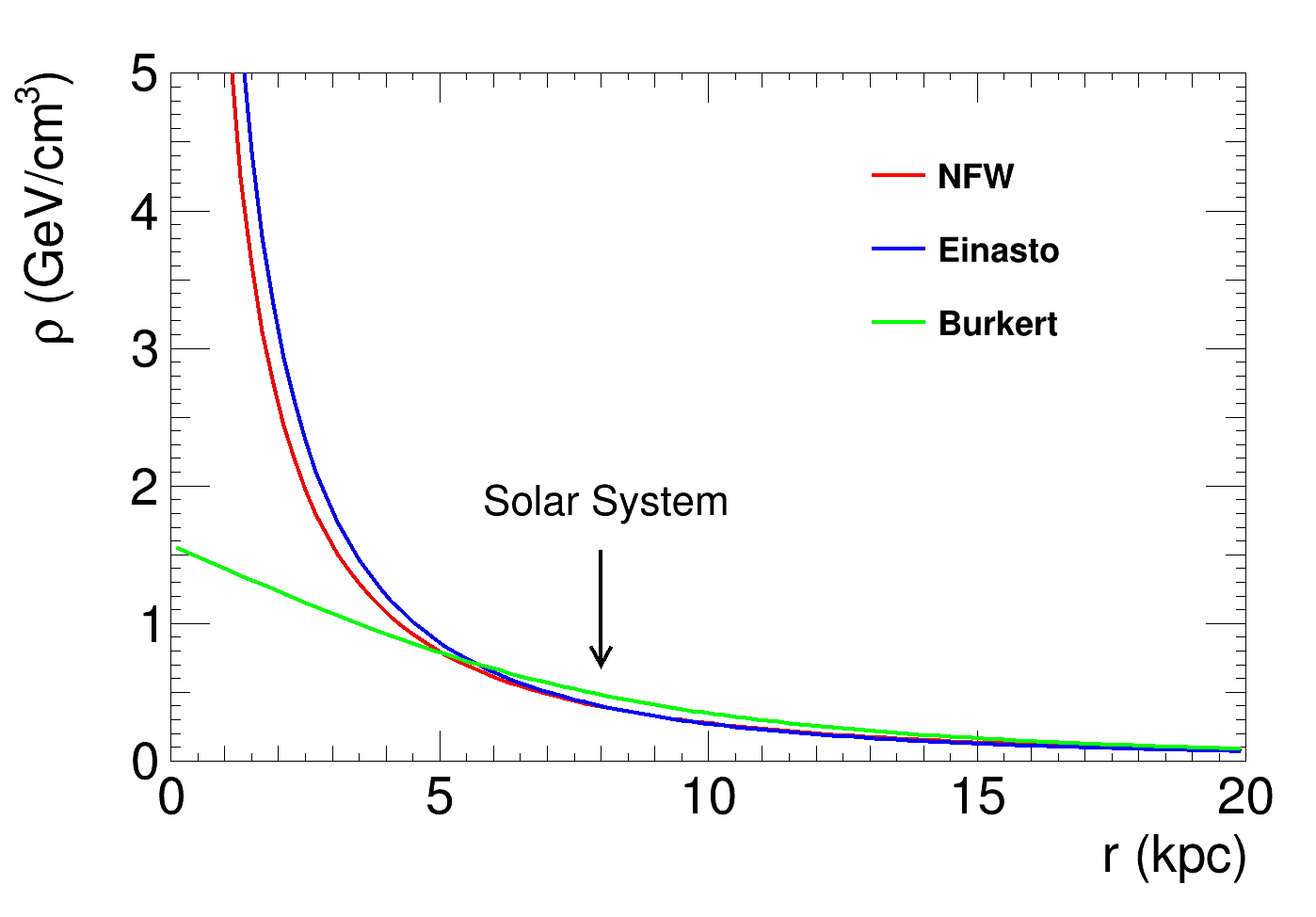}
\caption{Dark matter density as a function of the radius from the galactic centre in the Milky Way, for the NFW, Einasto and Burkert profiles.\label{fig:profiles}}
\end{center}
\end{figure}

It is important to look for these indirectly produced particles in the regions where dark matter is denser. There is currently a global agreement about the possible distributions of dark matter in galaxies. Three main galactic density profiles that are generally considered are:
\begin{itemize}
 \item Navarro, Frenk and White (NFW) profile~\cite{Navarro:1995iw} given by
\begin{equation}
\rho_{\rm NFW}(r)=\displaystyle\rho_s \frac{r_s}{r}\left(1+\frac{r}{r_s}\right)^{-2}\,, \label{eq:NFW}
\end{equation}
where in the Milky Way $r_s=19.6$ kpc and $\rho_s=0.32$ GeV/cm$^3$ \cite{McMillan:2011wd}.

\item Einasto profile~\cite{Graham:2005xx} given by
\begin{equation}
\rho_{\rm Ein}(r)=\displaystyle\rho_s \exp\left\lbrace -\frac{2}{\alpha} \left[\left(\frac{r}{r_s}\right)^{\alpha}-1\right] \right\rbrace \,,
\end{equation}
where in the Milky Way $\alpha=0.22$, $r_s=16.07$ kpc and $\rho_s=0.11$ GeV/cm$^3$~\cite{Catena:2009mf}. This profile is in better agreement with the latest simulations~\cite{Navarro:2008kc}.

\item Burkert profile~\cite{Burkert:1995yz} given by
\begin{equation}
\rho_{\rm Bur}(r)=\frac{\rho_s}{\displaystyle\left(1+\frac{r}{r_s}\right)\left(1+{\left(\frac{r}{r_s}\right)}^2\right)}\,,
\end{equation}
where in the Milky Way $r_s=9.26$ kpc and $\rho_s=1.57$ GeV/cm$^3$~\cite{Nesti:2013uwa}. With this profile, the dark matter density in the galactic centre is smoother.
\end{itemize}%
The densities of the different profiles are shown in Fig.~\ref{fig:profiles}.
The fact that different profiles are possible constitutes a challenge for interpreting indirect detection results, because even if the different profiles agree with the local density of dark matter, they also show that the dark matter density in the galactic centre is subject to large uncertainties.

To study indirect detection, it is necessary to understand how dark matter generates indirectly Standard Model particles. We consider here the case of annihilations, but decaying dark matter would be studied in a similar way. Two dark matter particles, when close to each other can interact and annihilate into SM particles, provided there exist small couplings between the dark matter sector and the SM sector. The probability of interaction depends on the relative velocity between dark matter particles and on the annihilation cross section into SM particles. However it is often the case that the largest cross sections are into unstable or coloured particles which will subsequently decay or hadronize. If the final stable annihilation products are charged it is then necessary to model their propagation.

\begin{figure}[!t]
\begin{center}
 \includegraphics[width=8.5cm]{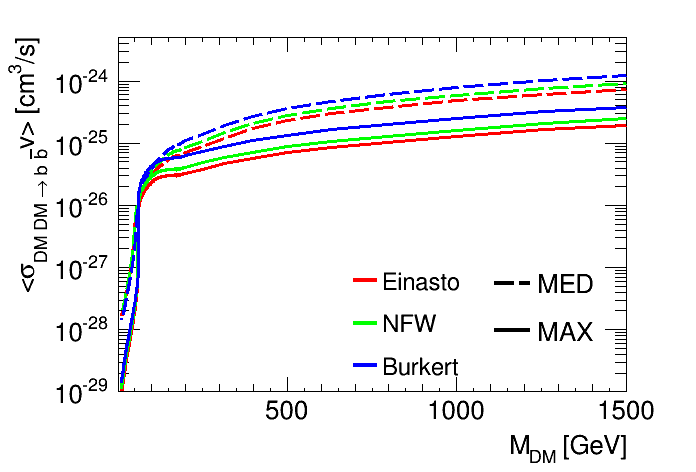}\includegraphics[width=8.5cm]{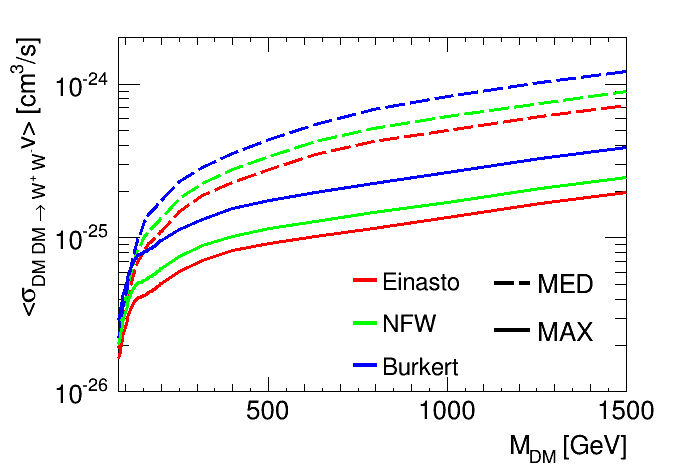}
\caption{95\% C.L. upper limit of dark matter annihilation cross section into $b\bar{b}$ (left) and $W^+ W^-$ (right) as a function of the dark matter mass, derived from AMS-02 antiproton data, for different choices of dark matter profiles (Einasto, Burkert, NFW) and propagation models (MED, MAX) \cite{Maurin:2001sj,Donato:2003xg,Boudaud:2016jvj}. From~\cite{Arbey:2017eos}.\label{fig:ams2_prop}}
\end{center}
\end{figure}

The AMS-02 experiment has set very strong constraints on the dark matter annihilation cross sections based on antiproton flux measurements \cite{Aguilar:2016kjl}. We show in Fig.~\ref{fig:ams2_prop} the dependence of the AMS-02 constraints on the choice of dark matter profiles and antiproton propagation models. We see that differences of one order of magnitude in the constraints are expected, depending on the choice of the dark matter profiles and propagation models. It is therefore particularly important to use the same profile and propagation model when comparing results from different experiments.

Since annihilation and scattering cross sections are related within a given particle physics model, the combination of direct and indirect detection results can set complementary constraints on new physics scenarios. However, as discussed above, such constraints are subject to large astrophysical uncertainties.

\subsubsection{Collider searches for dark matter}
\label{sec:collider}

New physics scenarios can be directly searched for at particle colliders, and in particular at the LHC. In a very general manner colliders probe new physics scenarios by aiming at producing new particles in high energy collisions of protons or electrons and positrons, or at generating heavy particles which decay into new particles. 

Most of the Standard Model particle properties have been thoroughly probed experimentally, and their decays are well constrained. Nevertheless the $Z^0$ boson, top quark or $B$-mesons or hadrons can still have small decay rates into new particles, provided the new particle mass is smaller than half of the decaying particle mass. The Higgs boson in this sense is of particular interest as it can be considered as a possible portal to new physics: The decay width of the Higgs boson into invisible particles is still weakly constrained and the branching fraction to invisible particles can be as high as 26\%~\cite{Tanabashi:2018oca}. Therefore the Higgs can still have non-negligible couplings to invisible new particles lighter than about 60 GeV. This also shows that, even if the measured Higgs couplings are mostly Standard Model-like~\cite{Tanabashi:2018oca}, there can still exist large couplings between the Higgs boson and the heavier new physics particles, and the precise measurements of the Higgs couplings may constitute a way to probe new physics, hence the name of ``Higgs portal'' to new physics \cite{Patt:2006fw,MarchRussell:2008yu,Djouadi:2011aa}.

The LHC is a hadron collider which collides protons at energies up to 14 TeV in the centre-of-mass. This allows us to probe new physics sectors with particle masses of a few TeV and with a high luminosity, in spite of the intense QCD background. Lepton colliders such as LEP or the future FCC-ee \cite{Abada:2019zxq} are on the other hand very well suited to study the electroweak sector. They have however lower luminosities and energies in the centre-of-mass, giving them a limited reach but more capabilities for precision measurements. As such, both kinds of colliders are complementary. The LHC is currently running and offers an unprecedented environment to study heavy sectors of new physics, especially the strongly interacting sector.

Concerning dark matter searches, the main challenge is that dark matter particles are expected to be neutral, uncoloured and weakly interacting, they can therefore have only small couplings to the QCD sector of the Standard Model. The high LHC luminosity is well-suited to probe such small couplings. Searches for new physics are generally focused on final states with missing transverse energy, which is generated by the decay of heavy New Physics particles into dark matter particles that remain undetected and constitute missing energy. The associated Standard Model particles in such processes act as markers for new physics phenomena. Such searches depend strongly on the characteristics of new physics, and the existence of heavier unknown particles. There are more specific searches generally (but not correctly) called ``dark matter searches'', or in somewhat clearer terms ``mono-X searches''. Indeed if dark matter is produced in a collider, it will remain undetected since it interacts too weakly. The only way to reveal that it has been produced is to observe the accompanying particles. At the LHC the prototypical channel is the monojet search \cite{Aaboud:2017phn}, where missing energy (i.e. dark matter) is accompanied by a high energy jet which has generally been produced as initial state radiation. At LEP the corresponding channels were the monophoton searches, i.e. missing energy and a hard photon, processes which are also studied at the LHC \cite{Sirunyan:2017ewk}. These channels are generally referred to as mono-X, where X can be a jet, a photon, a top quark, a gauge boson, a Higgs boson, or a lepton. In a simple scenario where only a dark matter and a mediator coupling to SM particle exist it is obvious that mono-X searches can be considered as dark matter collider searches. However in more realistic models involving many different new particles the interpretation is more complicated and it is difficult to relate the measured missing energy to dark matter particles \cite{Arbey:2013iza,Arbey:2015hca}. In the case of the discovery of new physics through mono-X searches it would in any case be very difficult to reinterpret the results in terms of dark matter in any realistic scenario, unless new physics is discovered in other channels or in dark matter detection experiments.

\subsection{Simplified dark matter models}

\begin{figure}[!t]
\begin{center}
\hspace*{-1.cm}\includegraphics[width=6.cm]{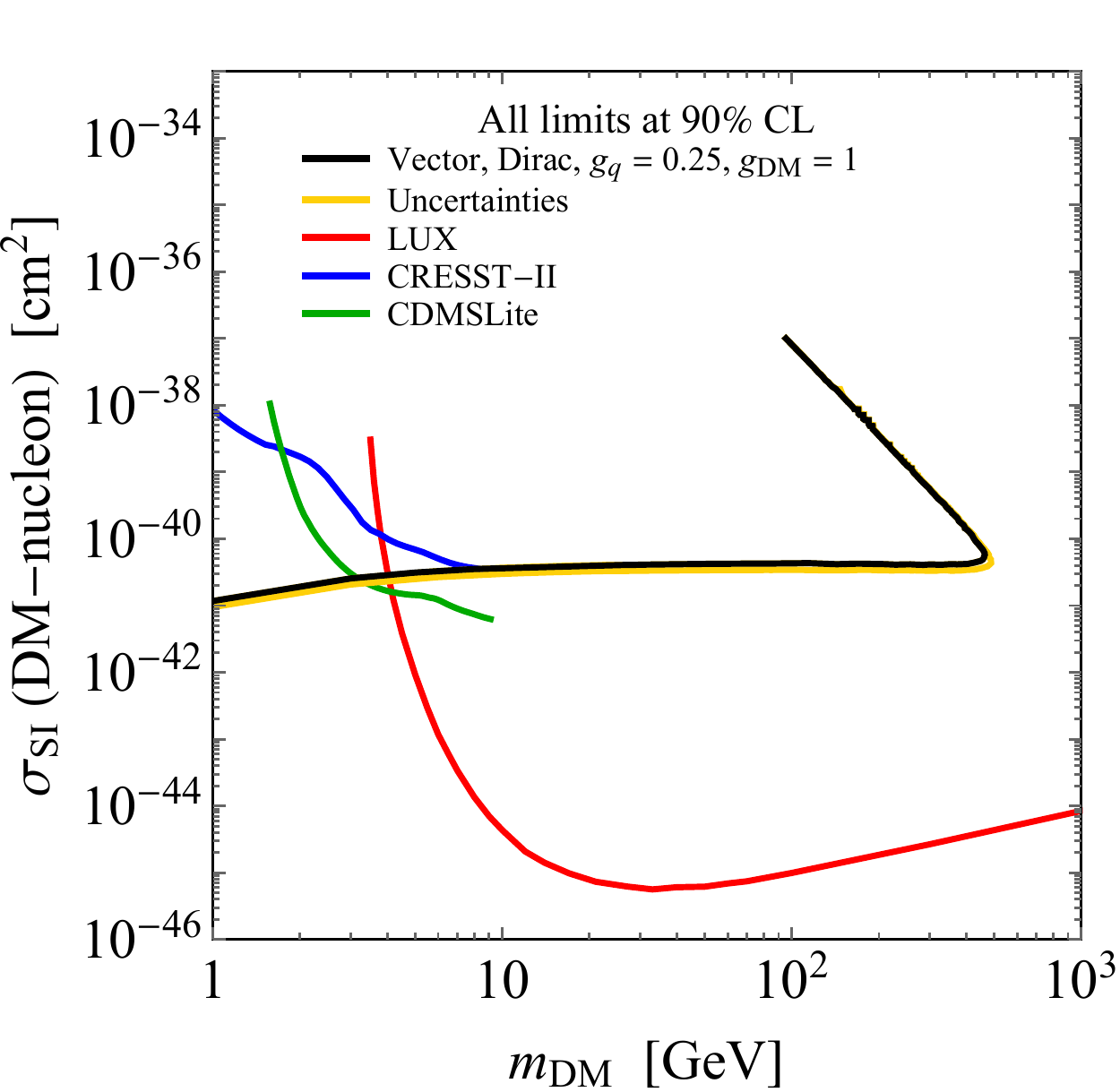}\includegraphics[width=6.cm]{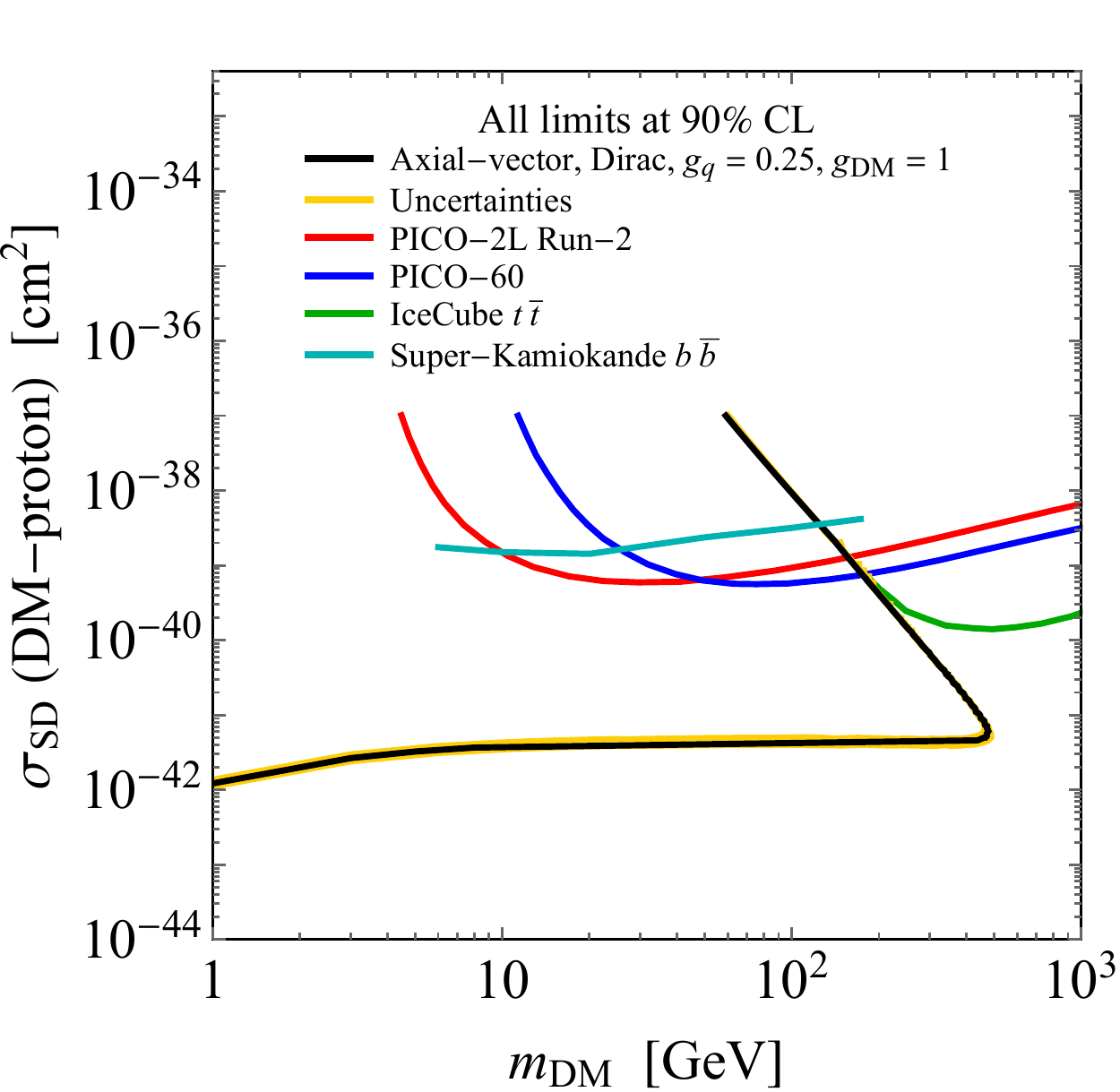}\includegraphics[width=6.cm]{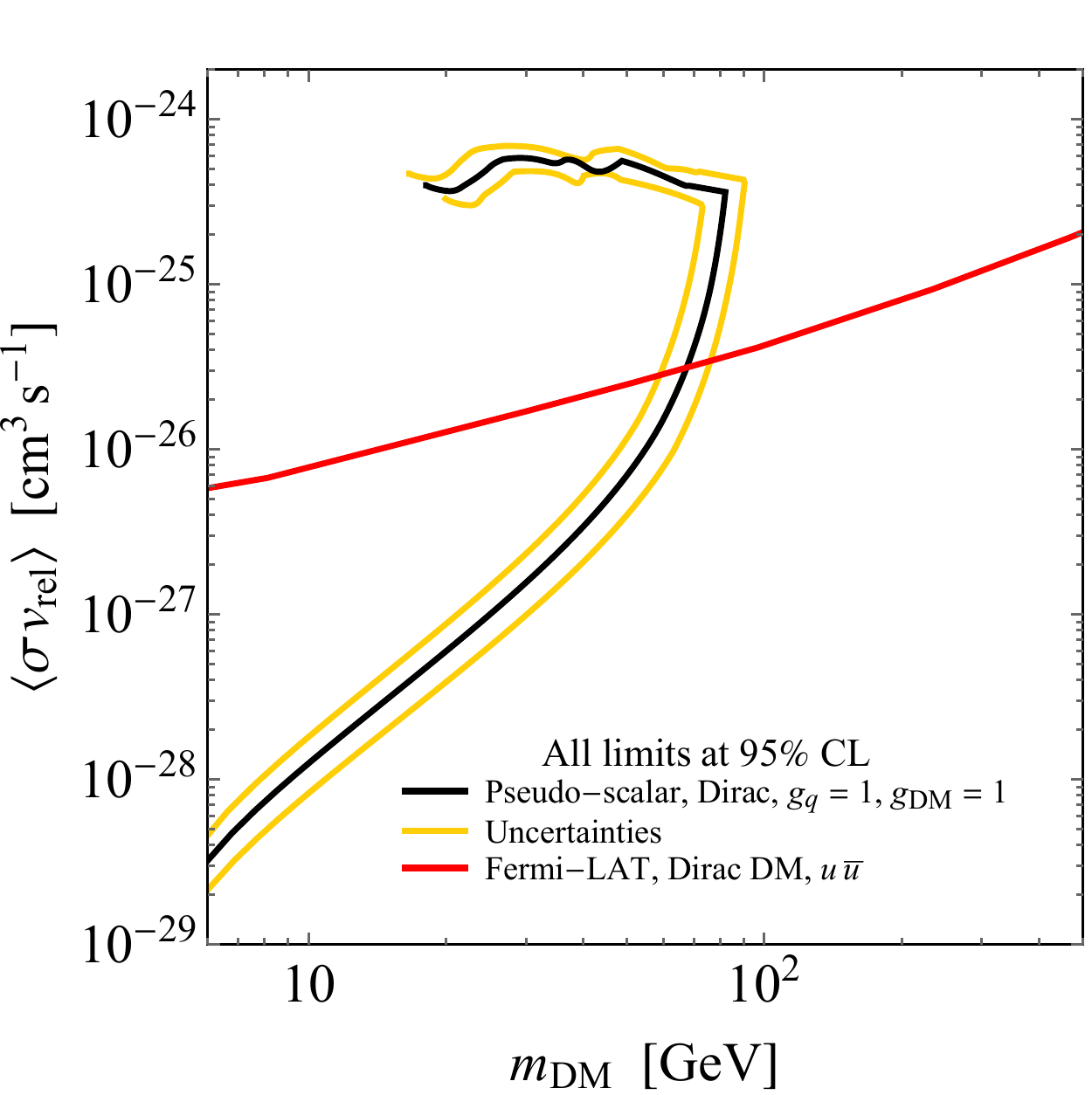}
\caption{90\% C.L. upper limit on the dark matter spin-independent scattering cross section with nucleon (left), spin-dependent scattering cross section with nucleon (centre) and annihilation cross section times velocity (right) for a Dirac fermion dark matter with different types of mediators. The black lines show the LHC constraints, the coloured lines correspond to direct detection results from LUX \cite{Akerib:2015rjg}, CRESST-II \cite{Angloher:2015ewa}, CDMSLite \cite{Agnese:2015nto}, PICO-2L \cite{Amole:2016pye} and PICO-60 \cite{Amole:2015pla}, to neutrino indirect detection results from IceCube \cite{Aartsen:2016exj} and Super-Kamiokande \cite{Choi:2015ara}, and to gamma-ray indirect detection from Fermi-LAT \cite{Ackermann:2015zua}. From~\cite{Boveia:2016mrp}.\label{fig:simplified}}
\end{center}
\end{figure}

We consider here a very simple scenario in which one dark matter particle and one mediator is added to the Standard Model spectrum. In such a scenario the effective 4-coupling (two dark matter particles and two Standard Model particles) can be considered to characterize the different searches: two dark matter particles annihilating into two Standard Model particles correspond to indirect detection, one dark matter particle scattering with one Standard Model particle correspond to direct detection, and two Standard Model particles colliding to form two dark matter particles to collider searches (see Fig.~\ref{fig:DMsearches}). The main interest of such simplified models is to be able to compare directly results from collider searches, dark matter direct detection and dark matter indirect detection, and check their consistencies \cite{Abdallah:2015ter,Abercrombie:2015wmb}. Such comparisons are exemplified in Fig.~\ref{fig:simplified}.

While such analyses are useful for testing compatibilities of different constraints, it is important to keep in mind that the direct and indirect detection constraints (as shown in the previous sections) are dependent on the dark matter astrophysical properties, which can strongly alter the limits, and that in more realistic scenarios the collider results are also impacted by the decay of heavy unstable new particles which breaks the direct relation between dark matter particles and missing transverse energy.

%%%%%%%%%%%%%%%%%%%%%%

\section{Relic density of thermal dark matter particles}
\label{sec:thermal}

An important theoretical observable to constrain dark matter scenarios is the relic density computed within a given model, which can be compared to the observed cosmological dark matter density. Contrary to direct and indirect detections which are mainly related to the dark matter properties, the value of the relic density depends in addition on the new physics scenario, and in particular on the properties of unstable new particles which are close in mass to the dark matter particle, as well as on the production mechanisms of the dark matter particles. In this section we assume that dark matter is produced thermally and was in thermal equilibrium in the early Universe.

\subsection{General case}

In a given new physics scenario, different particle species can exist and be in thermal equilibrium. When leaving thermal equilibrium, the number density of the species $a$ is given by the Boltzmann equation:
{\small \begin{align}
&\dfrac{dn_a}{dt} + 3 H n_a =\\
&\displaystyle - \sum_{bc\to de\cdots} \Bigg\{ \dfrac{1}{1+\delta_{bc}} \, (N_a^{in} - N_a^{out})\int (2\pi)^4\, \delta^4(p_b+p_c-p_d-p_e-\cdots) \, d\vec\Pi_b d\vec\Pi_c d\vec\Pi_d \cdots \, |\mathcal{M}(bc\to def\cdots)|^2 \, f_b f_c
 \Bigg\}\nonumber\\
&\displaystyle - \sum_{b\to cd\cdots} \Bigg\{ (N_a^{in} - N_a^{out}) \int (2\pi)^4\, \delta^4(p_b-p_c-p_d-\cdots) \, d\vec\Pi_b d\vec\Pi_c d\vec\Pi_d d\vec\Pi_e \cdots \, |\mathcal{M}(b\to cde\cdots)|^2 \, f_b
 \Bigg\} \,,\nonumber
\end{align}}%
where $N_a^{in (out)}$ is the number of particles $a$ in the initial (final) state, $n_b$ the density number of species $b$, $n_b^{eq}$ the density number of species $b$ at equilibrium, and 
\begin{equation}
 d\vec\Pi_i = g_i \dfrac{d^3\vec{p}_i}{2\,E_i\,(2\pi)^3}\,.
\end{equation}
The $f_i$ are the distribution functions. One assumes kinetic equilibrium so that
\begin{equation}
 f_i(E,t) = \dfrac{n_i(t)}{n_i^{eq}(t)} f_i^{eq}(E,t)\,,
\end{equation}
with the equilibrium number density
\begin{equation}
 n_i^{eq}=g_i \int \dfrac{d^3\vec{p}}{(2\pi)^3} f_i^{eq}(\vec{p})\,.
\end{equation}
To simplify we consider the distribution functions in the Maxwell-Boltzmann approximation, so that
\begin{equation}
 f_i^{eq}(\vec{p}) = \exp(-E_i/T) \,,
\end{equation}
and
\begin{equation}
 n_i^{eq} = \frac{T}{2\pi^2} g_i m_i^2 K_2\left(\frac{m_i}{T}\right)\,,
\end{equation}
where $K_2$ is the modified Bessel function of the second kind. Also assuming that the Standard Model particles are still at equilibrium, $f_i^{SM}=f_i^{eq}$ and $n_i^{SM}=n_i^{eq}$.

Considering only the processes with 1 or 2 particles in the initial and final states, assuming CP symmetry of the amplitudes, and at equilibrium, the Boltzmann equation becomes:
\begin{align}
\nonumber\dfrac{dn_a}{dt} + 3 H n_a =&\displaystyle - \sum_{bc \leftrightarrow de} \Bigg\{(N_a^{in} - N_a^{out}) \, \langle \sigma v\rangle_{bc\to de} \, \Bigl(n_b n_c - n_b^{eq} n_c^{eq}\frac{n_d n_e}{n_d^{eq} n_e^{eq}}\Bigr)  \Bigg\}\\
&\displaystyle - \sum_{b \leftrightarrow cd} \Bigg\{ (N_a^{in} - N_a^{out}) \, \langle \Gamma \rangle_{b\to cd} \, \Bigl(n_b-\frac{n_c n_d}{n_c^{eq} n_d^{eq}} n_b^{eq}\Bigr) \Bigg\}\,,
\end{align}%
with
\begin{equation}
\langle \sigma v\rangle_{bc\to de} = \frac{1}{n_b^{eq} n_c^{eq}} \int (2\pi)^4\, \delta^4(p_b+p_c-p_d-p_e) \, d\vec\Pi_b d\vec\Pi_c d\vec\Pi_d d\vec\Pi_e \, |\mathcal{M}(bc\to de)|^2 \, e^{-(E_b+E_c)/T}\,,
\end{equation}
and
\begin{equation}
\langle \Gamma \rangle_{b\to cd} = \frac{1}{n_b^{eq}} \int (2\pi)^4\, \delta^4(p_b-p_c-p_d) \, d\vec\Pi_b d\vec\Pi_c d\vec\Pi_d \, |\mathcal{M}(b\to cd)|^2 \, e^{-E_b/T}\,.
\end{equation}

The link between entropy and time is given by the adiabaticity condition:
\begin{equation}
 \frac{d s_{\rm rad}}{dt} = - 3 H s_{\rm rad}\,, \label{eq:radiation_entropy}
\end{equation}
where $s_{\rm rad}$ is the radiation entropy density.

Solving this set of differential equations together with the Friedmann equations, it is possible to obtain the densities of each type of new particles. The new particles that remain can constitute dark matter candidates.

\subsection{Standard case: single dark matter component with thermal freeze-out}
\label{sec:relic_standard}

In many new physics scenarios only one particle is stable and can form dark matter. We consider here this standard case: dark matter is made of one single type of particles, and the new physics particles are protected by a discrete symmetry so that new particles are produced in pair. To simplify we also assume that the new particles follow the Maxwell-Boltzmann distribution.

The standard hypothesis states that the early Universe was dominated by a radiation density, so that the expansion rate $H$ of the Universe is determined by the Friedmann equation:
\begin{equation}
H^2=\frac{8 \pi G}{3} \rho_{rad}\,,\label{eq:friedmann_stand}
\end{equation}
where
\begin{equation}
\rho_{rad}(T)=g_{\mbox{eff}}(T) \frac{\pi^2}{30} T^4
\end{equation}
is the radiation density and $g_{\mbox{eff}}$ is the effective number of degrees of freedom of radiation. The computation of the relic density is based on solving the Boltzmann evolution equation \cite{Gondolo:1990dk,Edsjo:1997bg}:
\begin{equation}
dn/dt=-3Hn-\langle \sigma_{\mbox{eff}} v\rangle (n^2 - n_{eq}^2)\,, \label{eq:evol_eq}
\end{equation}
where $n = \displaystyle\sum_i n_i$ is the number density of all new physics particles, $n_{eq}$ their equilibrium density given by
\begin{equation}
n_{eq} = \frac{T}{2\pi^2} \sum_i g_i m_i^2 K_2\left(\frac{m_i}{T}\right)\,,
\end{equation}
and $\langle \sigma_{\mbox{eff}} v\rangle$ is the thermal average of the annihilation rate of the new physics particles to the Standard Model particles. By solving Eq.~(\ref{eq:evol_eq}), the number density of new physics particles in the present Universe and consequently the relic density can be determined.

The annihilation rate of two new physics particles $i$ and $j$ into two Standard Model particles $k$ and $l$ can be defined as \cite{Gondolo:1990dk,Edsjo:1997bg}:
\begin{equation}
W_{ij\to kl} = \frac{p_{kl}}{16\pi^2 g_i g_j S_{kl} \sqrt{s}} \sum_{\rm{internal~d.o.f.}} \int \left| \mathcal{M}(ij\to kl) \right|^2 d\Omega \,,
\end{equation}
where $\mathcal{M}$ is the transition amplitude, $s$ is the centre-of-mass energy, $g_i$ is the number of degrees of freedom of particle $i$, $S_{kl}$ is a symmetry factor equal to 2 for identical final particles and to 1 otherwise, and $p_{kl}$ is the final centre-of-mass momentum such as
\begin{equation}
p_{kl} = \frac{\left[s-(m_k+m_l)^2\right]^{1/2} \left[s-(m_k-m_l)^2\right]^{1/2}}{2\sqrt{s}}\,.
\end{equation}
The integration is over the outgoing directions of one of the final particles. Moreover, a sum over the initial state internal degrees of freedom is performed.

The effective annihilation rate $W_{\rm eff}$ is defined as:
\begin{equation}
g_{DM}^2 p_{\rm{eff}} W_{\rm{eff}} \equiv \sum_{ij} g_i g_j p_{ij} W_{ij}\,,
\end{equation}
with
\begin{equation}
p_{\rm{eff}}(\sqrt{s}) = \frac{1}{2} \sqrt{(\sqrt{s})^2 -4 m_{DM}^2} \,.
\end{equation}

Practically, one computes
\begin{equation}
\frac{d W_{\rm eff}}{d \cos\theta} = \sum_{ijkl} \frac{p_{ij} p_{kl}}{ 8 \pi g_{DM}^2 p_{\rm eff} S_{kl} \sqrt{s} }
\sum_{\rm helicities} \left| \sum_{\rm diagrams}  \mathcal{M}(ij \to kl) \right|^2 \,,\label{dWeff}
\end{equation}
where $\theta$ is the angle between particles $i$ and $k$.

The thermal average of the effective cross section is then obtained with
\begin{equation}
\langle \sigma_{\rm{eff}}v \rangle = \dfrac{\displaystyle\int_0^\infty dp_{\rm{eff}} p_{\rm{eff}}^2 W_{\rm{eff}}(\sqrt{s}) K_1 \left(\dfrac{\sqrt{s}}{T} \right) } { m_{DM}^4 T \left[ \displaystyle\sum_i \dfrac{g_i}{g_{DM}} \dfrac{m_i^2}{m_1^2} K_2 \left(\dfrac{m_i}{T}\right) \right]^2}\,,\label{sigmaeffv}
\end{equation}
where $K_1$ and $K_2$ are the modified Bessel functions of the second kind of order 1 and 2 respectively.

The Boltzmann equation is rewritten in terms of the ratio of the number density to the radiation entropy density, $Y(T)=n(T)/s(T)$, where
\begin{equation}
s(T)=h_{\mbox{eff}}(T) \frac{2 \pi^2}{45} T^3 \,.
\end{equation}
$h_{\mbox{eff}}$ is the effective number of entropic degrees of freedom of radiation. 

Combining Eqs. (\ref{eq:friedmann_stand}) and (\ref{eq:evol_eq}) together with (\ref{eq:radiation_entropy}), the Boltzmann equation takes the form
\begin{equation}
\frac{dY}{dx}=-\sqrt{\frac{\pi}{45 G}}\frac{g_*^{1/2} m_{\mbox{\small DM}}}{x^2} \langle \sigma_{\mbox{eff}} v\rangle (Y^2 - Y^2_{eq}) \,, \label{eq:main}
\end{equation}
where $x=m_{\mbox{\small DM}}/T$ is the ratio of the DM mass over temperature, $Y_{eq}=n_{eq}/s$, and
\begin{equation}
g_*^{1/2}=\frac{h_{\mbox{eff}}}{\sqrt{g_{\mbox{eff}}}}\left(1+\frac{T}{3 h_{\mbox{eff}}}\frac{dh_{\mbox{eff}}}{dT}\right) \,.
\end{equation}

\begin{figure}[!t]
\begin{center}
 \includegraphics[width=12.cm]{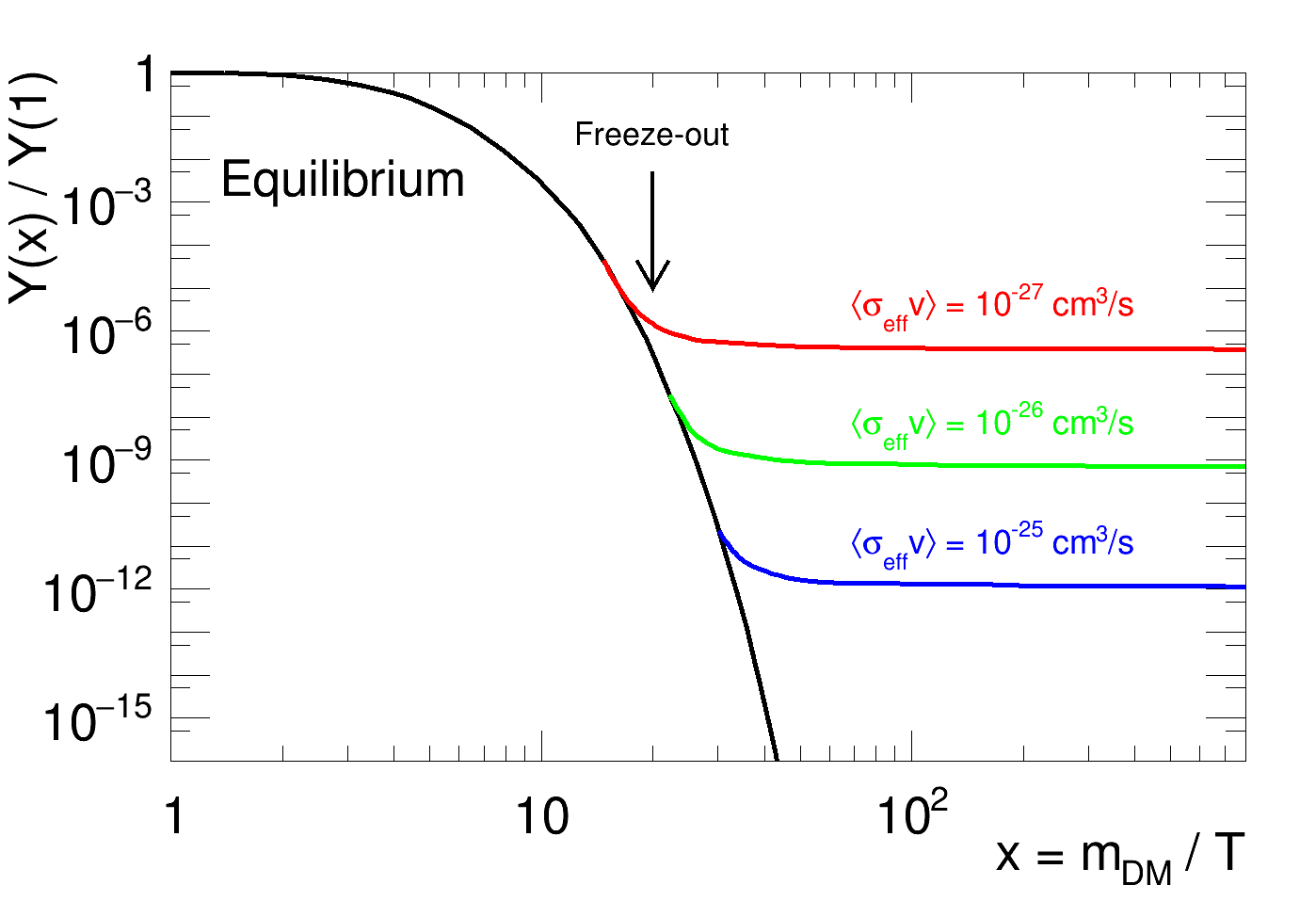}
\caption{Dark matter number density relative to the entropy density $Y=n/s$ normalized to its value at $T=m_{\mbox{\small DM}}$ as a function of $x=m_{\mbox{\small DM}}/T$, for different choices of the averaged annihilation cross sections. The black line corresponds to thermal equilibrium and the coloured lines to the values of $Y$ after freeze-out.\label{fig:freezeout}}
\end{center}
\end{figure}

The freeze-out temperature $T_f$ is the temperature at which the dark matter particles leave thermal equilibrium, corresponding to $Y (T_f) = (1 + \delta) Y_{\mbox{eq}}(T_f)$, with $\delta \simeq 1.5$. The evolution of $Y$ as a function of the temperature and the relation between freeze-out and cross section are shown in Fig.~\ref{fig:freezeout}, obtained with the public code {\tt SuperIso Relic} \cite{Arbey:2009gu,Arbey:2011zz,Arbey:2018msw}. The dark matter particles are originally in thermal equilibrium that they leave at the freeze-out temperature which depends on the value of the average annihilation cross section. Larger cross sections lead to smaller freeze-out temperatures.

The relic density is obtained by integrating Eq. (\ref{eq:main}) from $x=0$ to $m_{\mbox{\small DM}}/T_0$, where $T_0=2.726$ K is the temperature of the Universe today \cite{Gondolo:1990dk,Edsjo:1997bg}:
\begin{equation}
\Omega_{\mbox{\small DM}} h^2 = \frac{m_{\mbox{\small DM}} s(T_0) Y(T_0) h^2}{\rho_c^0} \approx 2.755\times 10^8 \frac{m_{\mbox{\small DM}}}{1 \mbox{ GeV}} Y(T_0)\,,
\end{equation}
with $\rho_c^0$ the critical density of the Universe. The obtained value can then be directly compared to the observed value $\Omega_c h^2$ (see Eq.~(\ref{eq:planck})) to set constraints on the new physics scenarios.

%%%%%%%%%%%%%%%%%%%%%%

\section{An example: supersymmetric dark matter}

In order to demonstrate the complementarity of the different dark matter searches, we consider in this section the minimal supersymmetric extension of the Standard Model (MSSM) \cite{Fayet:1977yc}, which is one of the most studied dark matter and new physics scenarios.

\subsection{Minimal supersymmetric extension of the Standard Model}

Supersymmetry is a global symmetry which relates the two classes of elementary particles: To each fermion corresponds a boson and vice versa. In absence of supersymmetry breaking the supersymmetric partners of the Standard Model bosons and fermions would have the same masses as the Standard Model particles, which is obviously not the case since they would have been discovered otherwise. Supersymmetry is therefore expected to be softly broken at a high energy scale, leading to different masses for the Standard Model and supersymmetric particles. The supersymmetric partners of the Standard Model particles have not been discovered up to now. 

The simplest supersymmetric model is the Minimal Supersymmetric extension of the Standard Model (MSSM), in which to each Standard Model particle corresponds one or two superpartners, and the Higgs sector is extended by the addition of a second Higgs doublet. Phenomenologically, the particle content of the MSSM, in addition to the SM particles (except the Higgs boson), is:\vspace*{-0.1cm}
\begin{itemize}
 \item two CP-even neutral Higgs bosons $h^0$ and $H^0$, one CP-odd neutral Higgs boson $A^0$ and two charged Higgs bosons $H^\pm$\vspace*{-0.1cm}
 \item twelve scalar quarks (or squarks) $\tilde u_{L,R}$, $\tilde d_{L,R}$, $\tilde c_{L,R}$, $\tilde s_{L,R}$, $\tilde t_{L,R}$, $\tilde b_{L,R}$\vspace*{-0.1cm}
 \item six charged scalar leptons (or sleptons) $\tilde e_{L,R}$, $\tilde \mu_{L,R}$, $\tilde \tau_{L,R}$\vspace*{-0.1cm}
 \item three scalar neutrinos (or sneutrinos) $\tilde \nu_e$, $\tilde \nu_\mu$, $\tilde \nu_\tau$\vspace*{-0.1cm}
 \item four neutral Majorana fermions (2 Higgsinos, zino, photino), superpartners of the Higgs bosons, $Z^0$ boson and photon\vspace*{-0.1cm}
 \item two charged Majorana fermions (charged Higgsino, wino), superpartners of the charged Higgs and $W^+$ bosons\vspace*{-0.1cm}
 \item one neutral and coloured Majorana fermion (gluino), superpartner of the gluon\vspace*{-0.1cm}
 \item a spin-3/2 particle called gravitino, superpartner of the spin-2 graviton\vspace*{-0.1cm}
\end{itemize}
and their antiparticles. The Higgsino, zino and photino states are mixed to form four neutralinos, and the charged Higgsino and wino mix into two charginos. The twelve squarks can also be mixed, as well as the sleptons and the sneutrinos.

In order for the MSSM to have a dark matter candidate, it is necessary that it contains a neutral, uncoloured and stable superpartner. The dark matter particle can be a neutralino, a sneutrino or a gravitino. If the dark matter particle is very heavy, it decays into Standard Model particles. In order to ensure the stability of a supersymmetric particle, a $Z_2$ symmetry called $R$-parity acting on the MSSM is generally assumed, such that all supersymmetric particles have a charge $-1$ and the Standard Model particles a charge $+1$. When $R$-parity is conserved, this multiplicative charge implies that supersymmetric particles can only be produced in pair and that during the decay of a supersymmetric particle there is always an odd number of supersymmetric particles in the final state. A direct consequence is that the lightest supersymmetric particle is stable, and can only annihilate into Standard Model particles.

In the following we will consider the MSSM with $R$-parity conservation, together with two additional assumptions: CP-conservation and minimal flavour violation. This implies that there is no extra source of CP-violation and flavour violation beyond the one in the Standard Model. This is a rather general assumption which leads to the reduction of the number of parameters of the MSSM from about one hundred to 20, within the so-called phenomenological MSSM (pMSSM) scenario:\vspace*{-0.1cm}
\begin{itemize}
 \item CP-odd Higgs mass $M_A$\vspace*{-0.1cm}
 \item ratio of the vacuum expectation values of the two Higgs doublets $\tan\beta= v_2/v_1$\vspace*{-0.1cm}
 \item bino mass parameter $M_1$\vspace*{-0.1cm}
 \item Higgsino mass parameter $\mu$\vspace*{-0.1cm}
 \item wino mass parameter $M_2$\vspace*{-0.1cm}
 \item gluino mass parameter $M_3$\vspace*{-0.1cm}
 \item left-handed third family mass parameter $M_{\tilde{q}_{3L}}$\vspace*{-0.1cm}
 \item right-handed stop mass parameter $M_{\tilde{t}_R}$\vspace*{-0.1cm}
 \item right-handed sbottom mass parameter $M_{\tilde{b}_R}$\vspace*{-0.1cm}
 \item top trilinar coupling $A_t$\vspace*{-0.1cm}
 \item bottom trilinar coupling $A_b$\vspace*{-0.1cm}
 \item left-handed first and second family mass parameter $M_{\tilde{q}_{L}}$\vspace*{-0.1cm}
 \item right-handed up or charm squark mass parameter $M_{\tilde{U}_R}$\vspace*{-0.1cm}
 \item right-handed down or strange squark mass parameter $M_{\tilde{D}_R}$\vspace*{-0.1cm}
 \item left-handed stau mass parameter $M_{\tilde{\tau}_L}$\vspace*{-0.1cm}
 \item right-handed stau mass parameter $M_{\tilde{\tau}_R}$\vspace*{-0.1cm}
 \item tau trilinar coupling $A_\tau$\vspace*{-0.1cm}
 \item left-handed electron or muon slepton mass parameter $M_{\tilde{\ell}_L}$\vspace*{-0.1cm}
 \item right-handed electron or muon slepton mass parameter $M_{\tilde{\ell}_R}$\vspace*{-0.1cm}
 \item gravitino mass $M_{\tilde{G}}$ \vspace*{-0.1cm}
\end{itemize}

The neutralino masses and couplings rely mostly on $M_1, M_2, \mu$ and the chargino ones on $M_2, \mu$. The neutralinos (charginos) mix into $\tilde\chi_1^0,\tilde\chi_2^0,\tilde\chi_3^0,\tilde\chi_4^0$ ($\tilde\chi_1^\pm,\tilde\chi_2^\pm$) where the index corresponds to the mass ordering from lighter to heavier. The squarks of different flavour do not mix, but the stops, sbottoms and staus mix into $\tilde{t}_1,\tilde{t}_2$, $\tilde{b}_1,\tilde{b}_2$ and $\tilde{\tau}_1,\tilde{\tau}_2$, respectively. The other particles do not mix, but their masses and couplings can receive higher order corrections.

The lightest Higgs boson, $h^0$, is generally considered as the Higgs which has been discovered at the LHC. The Higgs couplings are mostly driven by $M_A$ and $\tan\beta$, but can receive contributions from higher order corrections involving supersymmetric particles in loops. The lightest Higgs mass receives very large corrections from stops and to a lesser extent sbottoms, and the measured mass of 125 GeV imposes strong constraints on the stop sector \cite{Arbey:2011ab}. The heavy Higgs bosons $H^0,A^0,H^\pm$ have in general similar masses. 

Generally, to be detectable at the LHC the supersymmetric particle masses are expected to be in the range between 50 GeV and 2 TeV.

Three supersymmetric particles are candidates for dark matter in this scenario: the lightest neutralino, the lightest sneutrino and the gravitino, which in each case needs to be the lightest supersymmetric particle (LSP). The lightest sneutrino has been shown to be too strongly interacting to be a viable dark matter candidate \cite{Falk:1994es,Arina:2007tm}. There remain therefore two candidates: neutralino or gravitino, two cases that we will review in the following.

\subsection{Neutralino dark matter}

The neutralino dark matter scenario has been up-to-now the most studied new physics scenario, and is actively searched for at the LHC.

As discussed in Section~\ref{sec:collider}, the characteristic final states in the MSSM with neutralino dark matter include Standard Model particles along with two undetectable neutralinos which contribute to missing transverse energy. The largest cross sections are obtained in channels where gluinos or squarks are pair-produced, then decayed into two neutralinos and jets. With 13 TeV in the centre-of-mass, the LHC can probe the supersymmetric parameter space with masses of gluinos and squarks as large as 3~TeV. 

The LEP and Tevatron experiments have provided lower bounds on the superpartner masses in the neutralino LSP scenario, which are relatively independent from the choice of parameters \cite{Nakamura:2010zzi}. These limits are given in Table~\ref{tab:LEP_MSSM}, where the lightest neutralino is heavier than 46 GeV.\footnote{See for example Refs.~\cite{Arbey:2012na,Belanger:2013pna,Arbey:2013aba} for specific scenarios in which the neutralino can be lighter than 46 GeV.}

\begin{table}[!t]
\begin{center}
\begin{tabular}{|c|c|c|}
\hline
Particle & Limits & ~~~~~~Conditions~~~~~~\\
\hline\hline
$\tilde \chi^0_1$ & 46 &  \\
\hline
$\tilde \chi^0_2$ & 62.4 & $\tan\beta < 40$\\
\hline
$\tilde \chi^0_3$ & 99.9 & $\tan\beta < 40$\\
\hline
$\tilde \chi^0_4$ & 116 & $\tan\beta < 40$\\
\hline
$\tilde \chi^\pm_1$ & 94 & $\tan\beta < 40$, $m_{\tilde \chi^\pm_1} - m_{\tilde \chi^0_1} > 5$ GeV\\
\hline
$\tilde{e}_R$ & 73 & \\
\hline
$\tilde{e}_L$ & 107 & \\
\hline
$\tilde{\tau}_1$ & 81.9 & $m_{\tilde{\tau}_1} - m_{\tilde \chi^0_1} > 15$ GeV\\
\hline
$\tilde{u}_R$ & 100 & $m_{\tilde{u}_R} - m_{\tilde \chi^0_1} > 10$ GeV\\
\hline
$\tilde{u}_L$ & 100 & $m_{\tilde{u}_L} - m_{\tilde \chi^0_1} > 10$ GeV\\
\hline
$\tilde{t}_1$ & 95.7 & $m_{\tilde{t}_1} - m_{\tilde \chi^0_1} > 10$ GeV\\
\hline
$\tilde{d}_R$ & 100 & $m_{\tilde{d}_R} - m_{\tilde \chi^0_1} > 10$ GeV\\
\hline
$\tilde{d}_L$ & 100 & $m_{\tilde{d}_L} - m_{\tilde \chi^0_1} > 10$ GeV\\
\hline
& 248 & $m_{\tilde \chi^0_1} < 70$ GeV, $m_{\tilde{b}_1} - m_{\tilde \chi^0_1} > 30$ GeV\\
& 220 & $m_{\tilde \chi^0_1} < 80$ GeV, $m_{\tilde{b}_1} - m_{\tilde \chi^0_1} > 30$ GeV\\
$\tilde{b}_1$ & 210 & $m_{\tilde \chi^0_1} < 100$ GeV, $m_{\tilde{b}_1} - m_{\tilde \chi^0_1} > 30$ GeV\\
& 200 & $m_{\tilde \chi^0_1} < 105$ GeV, $m_{\tilde{b}_1} - m_{\tilde \chi^0_1} > 30$ GeV\\
& 100 & $m_{\tilde{b}_1} - m_{\tilde \chi^0_1} > 5$ GeV\\
\hline
$\tilde{g}$ & 195 & \\
\hline
\end{tabular}
\end{center}
\caption{Constraints on the supersymmetric particle masses (in GeV) from searches at LEP and the Tevatron. From~\cite{Tanabashi:2018oca,Arbey:2011un}.
\label{tab:LEP_MSSM}}
\end{table}

In the following the gravitino is assumed to be extremely heavy and to have no influence on the results. We perform random scans over the remaining 19 pMSSM parameters, with masses between 50 GeV and 3 TeV, trilinear couplings between $-15$ at 10 TeV, and $\tan\beta$ between 2 and 60, before applying constraints.

\begin{figure}[!t]
 \begin{center}
  \includegraphics[width=0.5\textwidth]{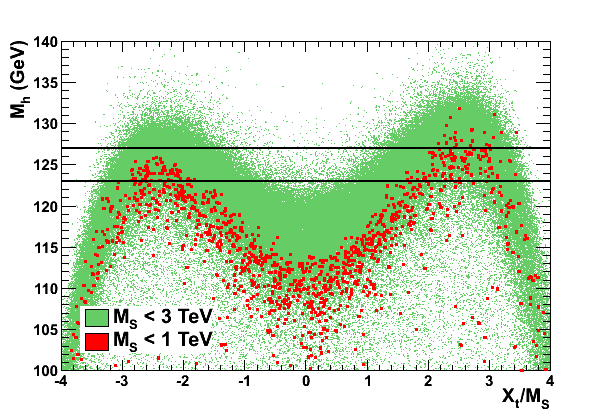}\includegraphics[width=0.5\textwidth]{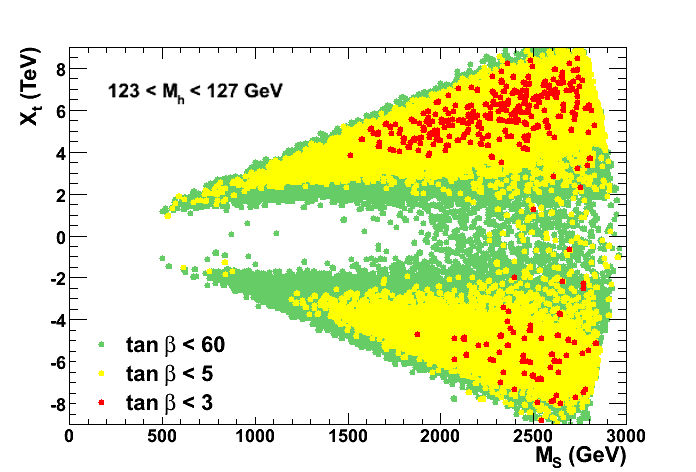}
\caption{Maximal value of the $h^0$ mass as a function of $X_t/M_S$ with $X_t (=A_t-\mu/\tan\beta)$ the stop mixing parameter and $M_S(=\sqrt{m_{\tilde{t}_1}m_{\tilde{t}_2}})$ the supersymmetry breaking scale (left) and regions in which $123<M_h<127$ GeV in the $(M_S,X_t)$ plane for some selected ranges of $\tan\beta$ (right). From \cite{Arbey:2011ab}.\label{fig:higgs}}
 \end{center}
\end{figure}

The measurement of the lightest Higgs boson mass of 125 GeV has set strong constraints on the stop sector, as can be seen in Fig.~\ref{fig:higgs}, where the Higgs mass is shown as a function of $X_t/M_S$, with $X_t = A_t-\mu/\tan\beta$ the stop mixing parameter and $M_S(=\sqrt{m_{\tilde{t}_1}m_{\tilde{t}_2}})$ the supersymmetry breaking scale. The model points in agreement with the Higgs mass constraints are also shown in the $(M_S,X_t)$ parameter planes in the right panel. Both plots show that scenarios with large stop masses and stop mixing parameters are favoured. 

\begin{figure}[!t]
\begin{center}
\hspace*{-0.5cm}\includegraphics[width=.35\textwidth]{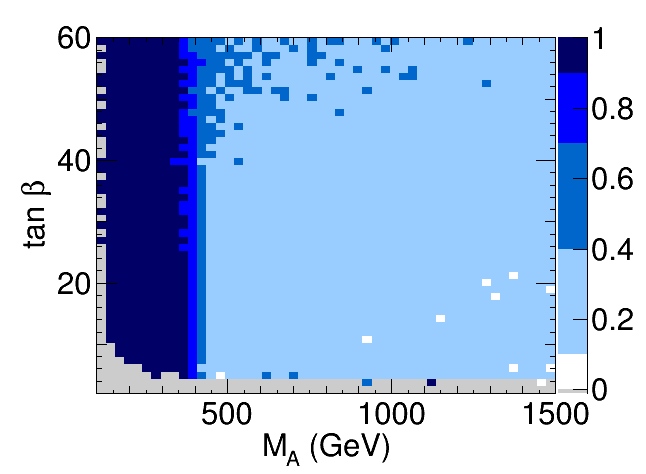}\includegraphics[width=.35\textwidth]{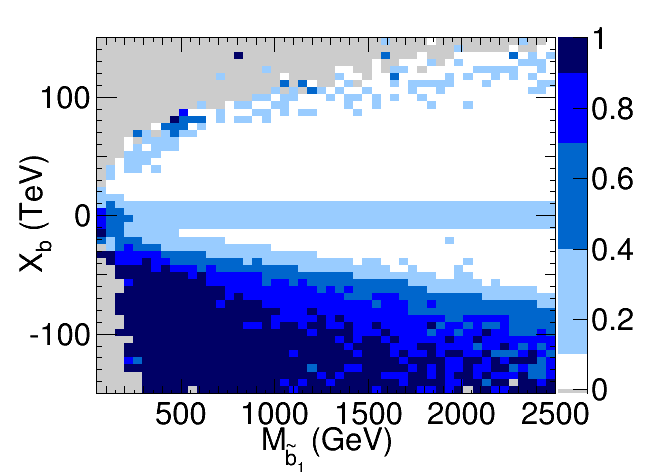}\includegraphics[width=.35\textwidth]{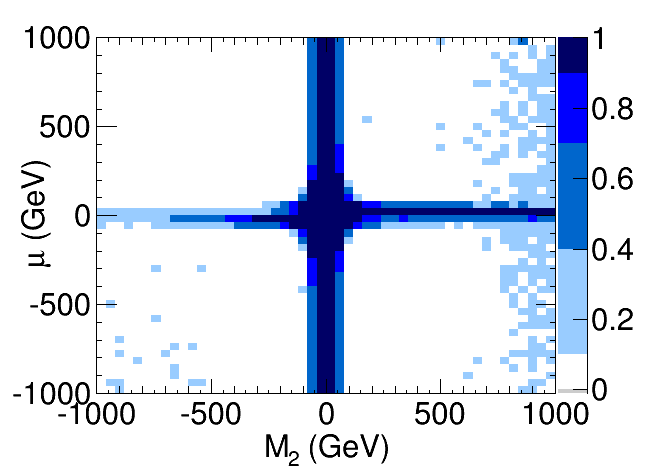}
\caption{Fraction of excluded points by Higgs coupling measurements in the $(M_A,\tan\beta)$ (left), $(M_{\tilde{b}_1},X_b=A_b-\mu\tan\beta)$ (centre) and $(M_2,\mu)$ (right) parameter planes. From~\cite{Arbey:2018wjb}.\label{fig:higgscoup}}
\end{center}
\end{figure}

The Higgs couplings have also been measured with a good precision, and they set additional constraints, as shown in Fig.~\ref{fig:higgscoup}, where the fraction of points excluded by the Higgs coupling constraints are shown in the $(M_A,\tan\beta)$, $(M_{\tilde{b}_1},X_b=A_b-\mu\tan\beta)$ and $(M_2,\mu)$ parameter planes. This shows that the Higgs coupling data favour large CP-odd Higgs mass, positive sbottom mixing parameter $X_b$ and not too small $M_2$ or $\mu$ parameters. Similarly searches for heavy Higgs bosons set limits on the same parameter planes \cite{Arbey:2012bp,Arbey:2013jla,Arbey:2015aca,Arbey:2018wjb}.

As discussed in the previous section the neutralinos are mixed states of zino, photino and Higgsinos. One refers to ``bino'' when the neutralino mass is mostly given by $M_1$, ``wino'' for $M_2$ and Higgsino for $\mu$, and mixed states otherwise. It is important to note that the binos have in general suppressed couplings as compared to the Higgsinos and winos. The charginos can similarly be winos or Higgsinos. Therefore a bino neutralino is relatively independent from the charginos, but a wino neutralino is accompanied by a wino chargino close-in-mass, and a Higgsino neutralino by a second Higgsino neutralino and a Higgsino chargino which are both close-in-mass. This has strong implications, in particular for the relic density, as we will see in the following.

\begin{figure}[!t]
\begin{center}
\includegraphics[width=.45\textwidth]{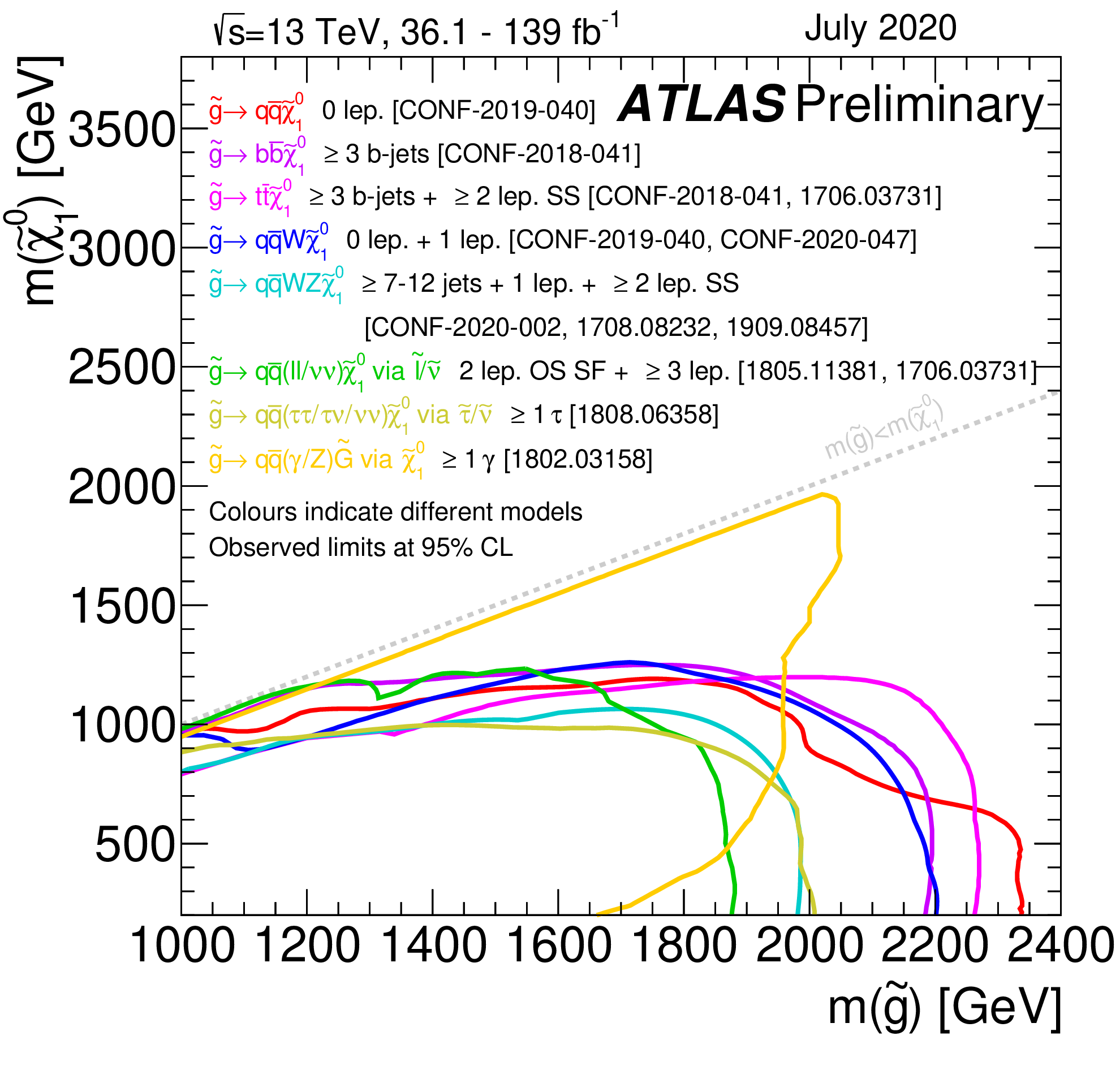}\includegraphics[width=.45\textwidth]{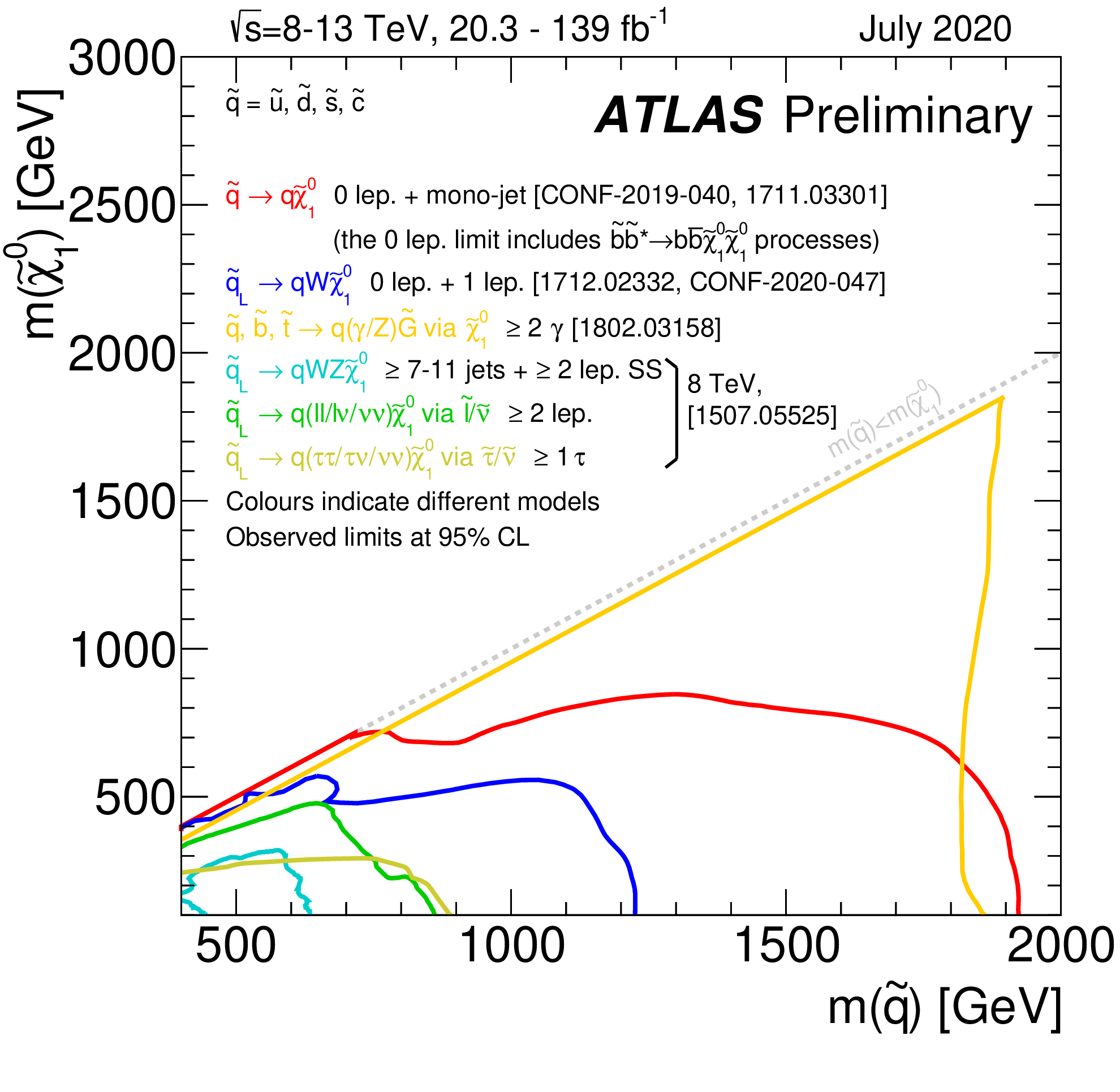}
\caption{Recent ATLAS exclusion contours in the gluino vs. lightest neutralino (left) and squark vs. lightest neutralino (right) mass in simplified MSSM scenarios, for different decay chains. From~\cite{ATL-PHYS-PUB-2020-020}.\label{fig:ATLAS_SUSY}}
\end{center}
\end{figure}

\begin{figure}[!t]
\begin{center}
\hspace*{-0.2cm}\includegraphics[height=6.5cm]{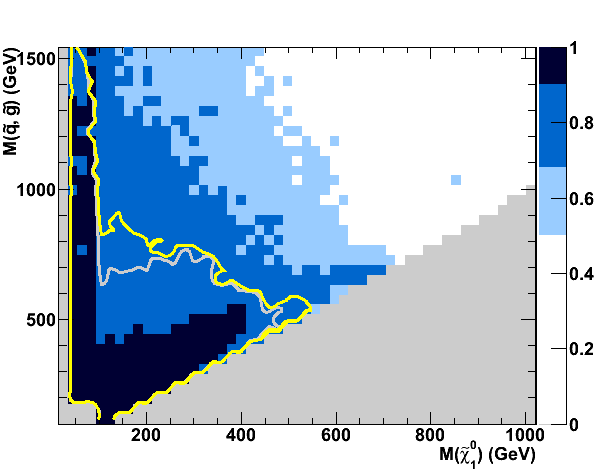}\includegraphics[height=6.5cm]{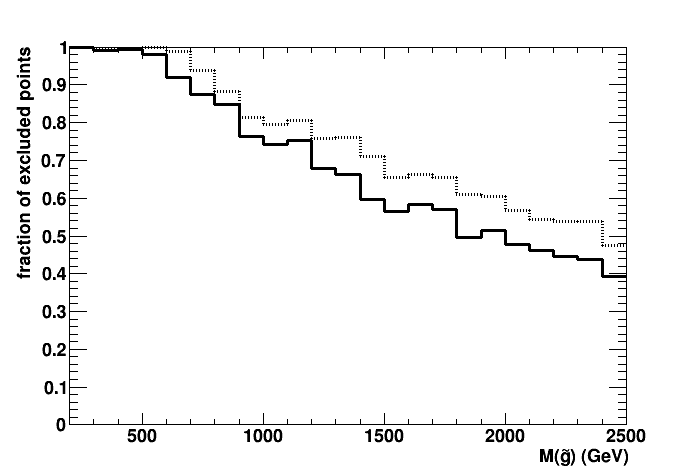}
\caption{Left: Fraction of excluded pMSSM points in the plane of the gluino/squark minimal mass vs. lightest neutralino mass. The yellow line corresponds to the limit below which 95\% of the points are excluded by supersymmetric and monojet searches, and the grey line to the exclusion by supersymmetric searches only.  Right: Fraction of pMSSM points excluded by supersymmetric searches (solid) and in addition monojet searches (dotted) as a function of the gluino mass. These results have been obtained using the ATLAS limits with a luminosity of about 20 fb$^{-1}$. From~\cite{Mahmoudi:2014cya}.\label{fig:pMSSM_constraints}}
\end{center}
\end{figure}

At colliders, on the one hand flavour physics and in particular $B$-meson decays are sensitive to heavy Higgs states, stops and charginos. The current limits favour large $M_A$, small $\tan\beta$, and large $|M_2|$ and $M_{\tilde{t}_{1,2}}$ \cite{Arbey:2012ax,Mahmoudi:2014mja}. On the other hand high transverse energy searches for new particles can probe directly masses up to the TeV scale. There exists a vast program of searches for supersymmetric particles at the LHC, which has provided severe constraints on the MSSM parameter space. The results are generally presented in simplified MSSM scenarios, where only one heavy superpartner and the lightest neutralino are considered, with the assumption of one decay chain for the heavy supersymmetric particle, as shown for example in Fig.~\ref{fig:ATLAS_SUSY}. The important parameter here is the mass splitting between the considered heavy superpartner and the lightest neutralino, which determines the energies of the decay products.
However because of the large number of pMSSM parameters the hierarchy of supersymmetric masses is highly model-dependent, and it is necessary to generate events and simulate the ATLAS and CMS detectors in order to recast the results in the pMSSM. Additionally monojet searches are known to be complementary to supersymmetric searches \cite{Arbey:2013iza,Arbey:2015hca}. In the 19-parameter pMSSM it is not possible to draw strict limits in 2-dimensional mass parameter planes, because there exist many possibilities to evade such constraints. This can be seen in Fig.~\ref{fig:pMSSM_constraints}, which shows the fraction of excluded points in the neutralino mass vs. lightest squark/gluino mass parameter plane, and as a function of the gluino mass. As can be seen, a non-negligible fraction of pMSSM points featuring squarks with masses below 1 TeV can still be consistent with the experimental data.

\begin{figure}[!t]
\begin{center}
\includegraphics[width=.5\textwidth]{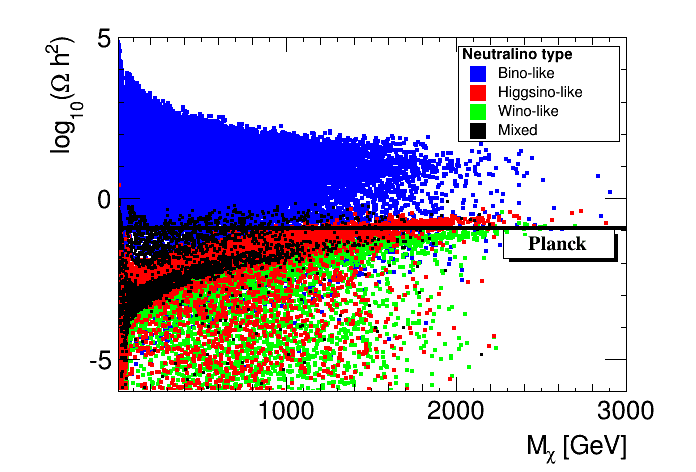}\includegraphics[width=.5\textwidth]{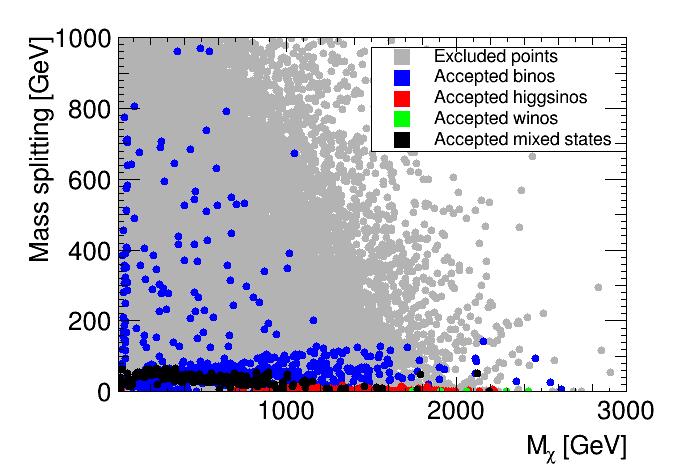}
\caption{Left: Relic density as a function of the lightest neutralino mass, for the different neutralino types. The black horizontal line corresponds to the dark matter value measured by Planck. The thickness of the line corresponds to the uncertainty. Right: pMSSM points in agreement with the Planck dark matter value in the plane of mass splitting between the next-to-lightest supersymmetric particle and the lightest neutralino vs. lightest neutralino mass. From~\cite{Arbey:2017eos}.\label{fig:pMSSM_relic}}
\end{center}
\end{figure}

We now turn to dark matter constraints in the pMSSM. The supersymmetric particle mass spectrum as well as the nature of the lightest neutralino strongly impact the reinterpretation of the dark matter constraints in terms of pMSSM parameters. This is particularly visible when considering the relic density constraints, as shown in Fig.~\ref{fig:pMSSM_relic} in the relic density vs. lightest neutralino mass plane, where it is shown that the relic density spans over about 10 orders of magnitude around the Planck dark matter density value. The right-hand plot shows the points in agreement with the upper and lower relic density bounds in the parameter plane of lightest neutralino mass vs. mass splitting with the next-to-lightest supersymmetric particle.
Winos and Higgsinos provide a relic density compatible with the observations if their masses are between 1.5 and 2.5 TeV. Binos have in general a too large relic density, which is due to the fact that they have suppressed interactions and annihilate less in the early Universe. They can nevertheless provide a correct relic density if they are accompanied by a next-to-lightest supersymmetric particle with a mass of about 100 GeV larger than the lightest neutralino, or less. Since winos and Higgsinos are always accompanied by a close-in-mass chargino and/or neutralino, the relic density can be in agreement with the observed dark matter density only when there exists at least one supersymmetric particle close-in-mass, which would increase the (co-)annihilation rate, whatever the type of the lightest neutralino is.

A rather usual hypothesis on the relic density constraint is to consider only the upper limit of the observed dark matter density. Indeed, if the relic density is too large it would overclose the Universe and be observationally inconsistent. On the other hand, if the relic density is smaller than the observed dark matter density it leaves room for another type of dark matter, for example axions or primordial black holes.

Dark matter direct and indirect detections have also strong implications on the pMSSM. Since only the neutralino enters directly into the scattering and annihilation cross sections the existence of a close-in-mass next-to-lightest supersymmetric particle has less consequences on the results and modifies the amplitudes only at loop level. Nevertheless the couplings of the neutralinos to SM particles still strongly depend on the choice of the pMSSM parameters.

\begin{figure}[!t]
\begin{center}
\includegraphics[width=.5\textwidth]{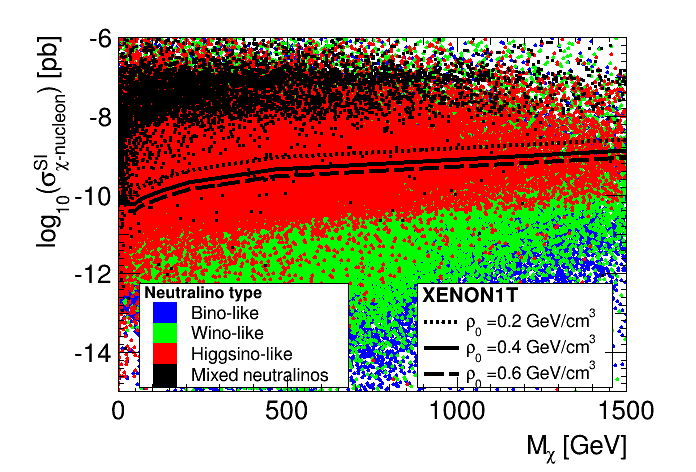}\includegraphics[width=.5\textwidth]{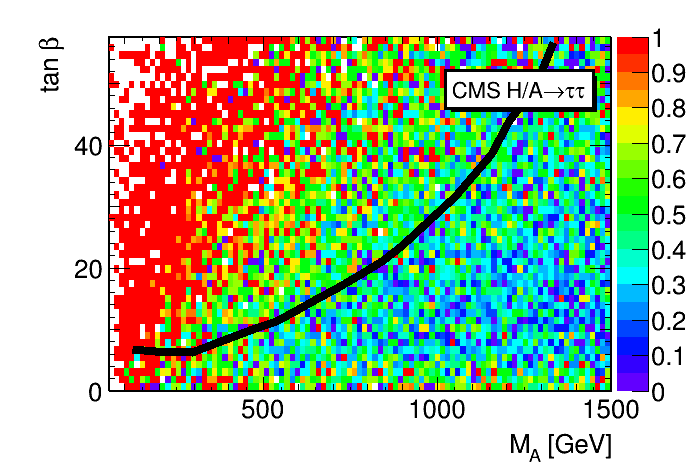}
\caption{Left: Scattering cross section of the lightest neutralino with nucleons as a function of the lightest neutralino mass. The black line corresponds to the upper limit set by XENON1T for different generally admitted local dark matter densities. Right: Fraction of points excluded by the XENON1T limit in the $(M_A,\tan\beta)$ parameter plane. The black line shows the limit of searches for heavy Higgs bosons in the $\tau^+\tau^-$ channel by the CMS experiment. From~\cite{Arbey:2017eos}.\label{fig:pMSSM_DD}}
\end{center}
\end{figure}

Dark matter direct detection experiments set limits on the scattering cross section of the lightest neutralino with nucleons, i.e. protons or neutrons. In Fig.~\ref{fig:pMSSM_DD} the limits set by the XENON1T experiment \cite{Aprile:2017iyp} are shown in the scattering cross section vs. lightest neutralino mass plane. Since the binos have smaller couplings they have very small cross sections and are well below the limit. The scatterings are mainly mediated by Higgs bosons, so that the Higgsinos and the mixed-state neutralinos have larger cross sections and are more likely to be probed by dark matter direct detection. There is a rather strong dependence on the heavy Higgs masses, which is shown on the right-hand panel of the figure, where the exclusion fraction of pMSSM points is presented in the heavy Higgs mass vs. $\tan\beta$ parameter plane. This exclusion is comparable to the ones obtained by heavy Higgs searches at the LHC.

\begin{figure}[!t]
\begin{center}
\includegraphics[width=.5\textwidth]{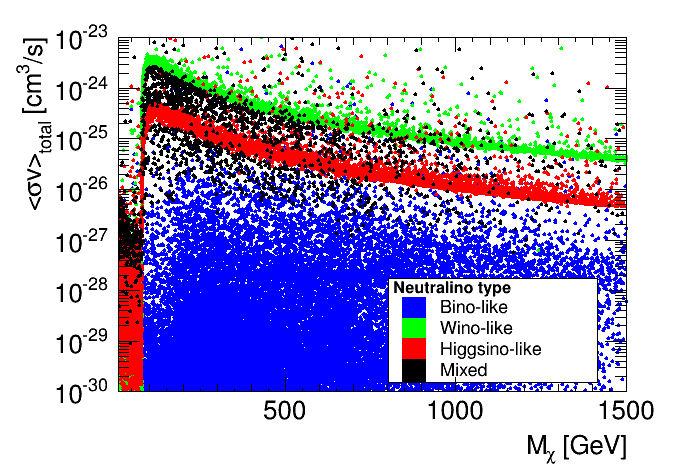}\includegraphics[width=.5\textwidth]{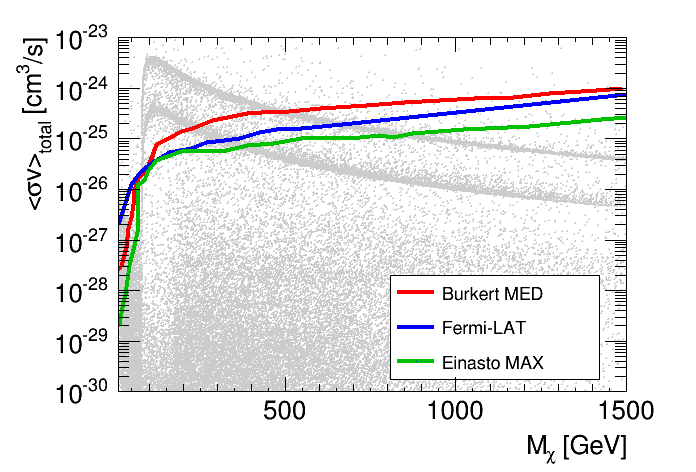}
\caption{Total annihilation cross section as a function of the lightest neutralino mass, for the different types of neutralinos (left) and upper limits by AMS-02 with Burkert profile and MED propagation model, with Einasto profile and MAX propagation model, and by Fermi-LAT. From~\cite{Arbey:2017eos}.\label{fig:pMSSM_ID}}
\end{center}
\end{figure}

Dark matter indirect searches set complementary constraints on the annihilation cross sections of the neutralinos into SM particles. We consider here the Fermi-LAT limits \cite{Fermi-LAT:2016uux} on the annihilation into gamma-rays, and the AMS-02 antiproton results \cite{Aguilar:2016kjl}, which are related to the annihilation into charged or coloured particles hadronizing into protons and antiprotons. In Fig.~\ref{fig:pMSSM_ID} the total annihilation cross sections for the pMSSM points are shown as a function of the lightest neutralino mass, for the different types of neutralinos. Similarly to the scattering cross sections the binos have small annihilation cross sections well below the detection limits. The largest cross sections correspond to mixed-state neutralinos and to winos. This highlights the complementarity with direct detection, which constrains mostly Higgsinos. Three upper limits are shown: the Fermi-LAT gamma-ray constraints which are unaffected by propagation, the AMS-02 limits which are recast into a conservative case, based on the choice of the Burkert dark matter density and MED antiproton propagation model, and a stringent case corresponding to the choice of the Einasto profile and MAX propagation model.\footnote{See Ref.~\cite{Arbey:2017eos} for more details and definitions.} As can be seen, the limits can be changed by one order of magnitude. In all cases winos are probed up to masses of about 1 TeV and Higgsinos up to about 400 GeV.

\begin{figure}[!t]
\begin{center}
\includegraphics[width=.5\textwidth]{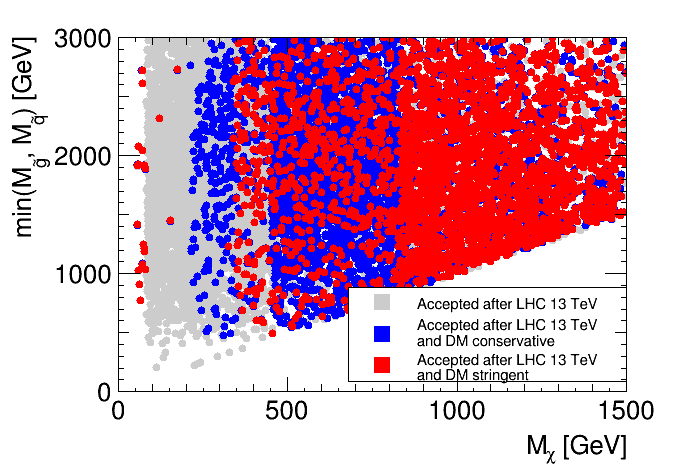}\includegraphics[width=.5\textwidth]{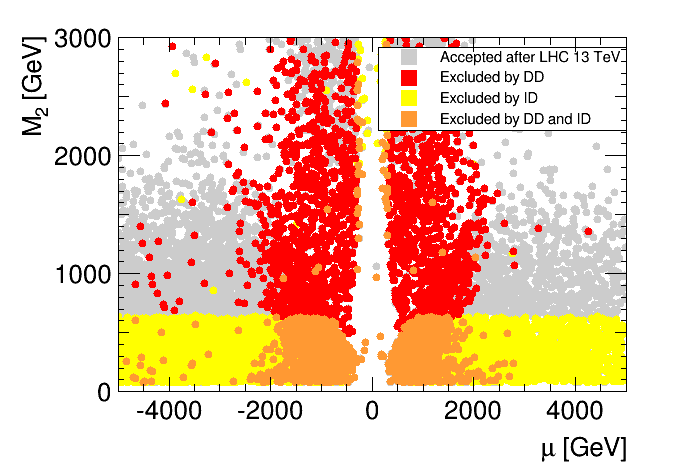}
\caption{pMSSM points in agreement with the LHC, dark matter direct and indirect detection constraints, in the lightest squark/gluino mass vs. lightest neutralino mass (left) and $(\mu,M_2)$ (right) parameter planes. From~\cite{Arbey:2017eos}.\label{fig:pMSSM_combined}}
\end{center}
\end{figure}

\begin{figure}[!t]
\begin{center}
\includegraphics[width=.33\textwidth]{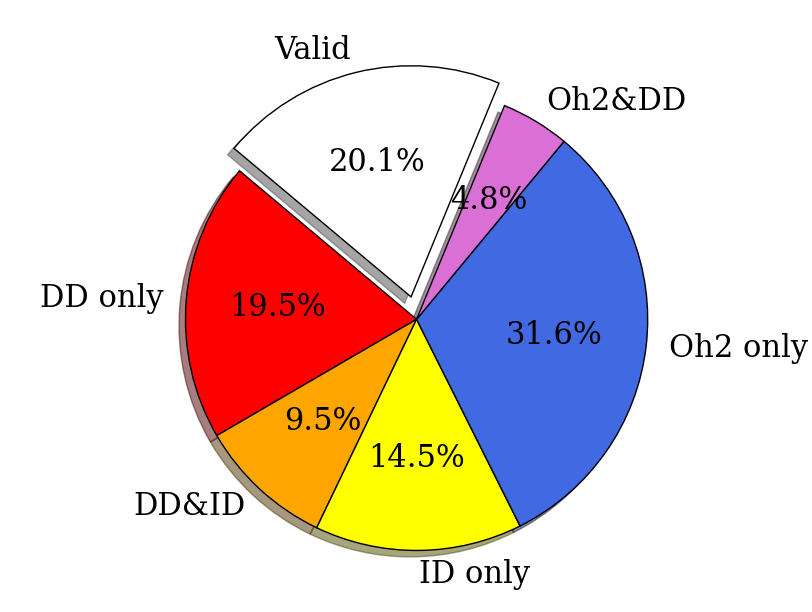}\hspace*{0.5cm}\includegraphics[width=.33\textwidth]{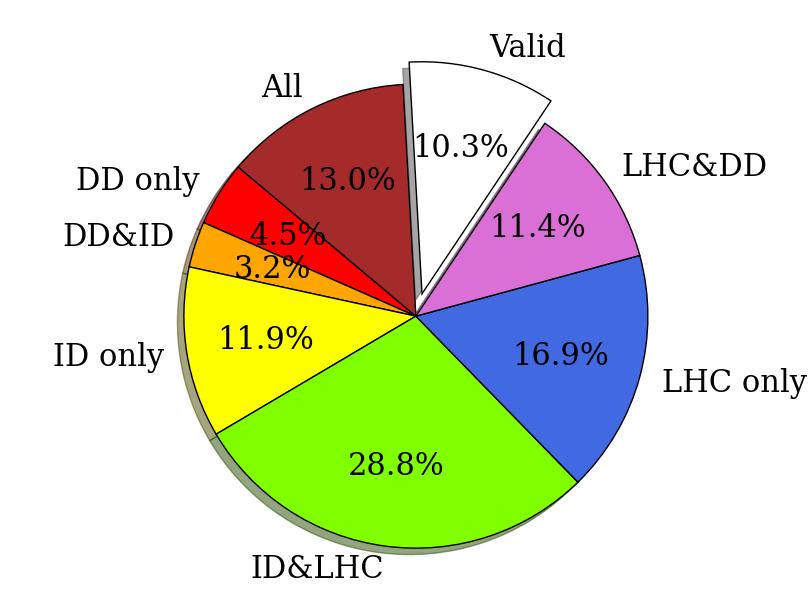}
\caption{Fraction of pMSSM points excluded by the different searches, except LHC searches (left) and except relic density (right). From~\cite{Arbey:2017eos}.\label{fig:pMSSM_pie}}
\end{center}
\end{figure}

Comparing Figs. \ref{fig:pMSSM_relic}, \ref{fig:pMSSM_DD} and \ref{fig:pMSSM_ID} clearly shows that there exists an interplay between direct and indirect searches for constraining the pMSSM parameter space, as well as with relic density and LHC searches. This is exemplified in Fig.~\ref{fig:pMSSM_combined}, where the points compatible or excluded by the different searches are shown in the lightest squark/gluino mass vs. lightest neutralino mass, and $\mu$ vs. $M_2$ parameter planes. It is clear that dark matter detection limits provide complementary constraints to LHC searches.
In particular searches for supersymmetry at the LHC are only mildly sensitive to the neutralino sector, whereas dark matter detection is especially sensitive to the neutralino properties, revealing the interplay between different new physics searches. The left panel shows that dark matter searches exclude particularly well small lightest neutralino masses, beyond LHC searches. In the right panel it is shown that dark matter direct detection excludes small $|\mu|$ values and indirect detection small $M_2$ values, which shows the complementary role of both kinds of dark matter detection limits. In Fig.~\ref{fig:pMSSM_pie} the proportion of pMSSM points excluded by the different searches (including the upper bound of the relic density constraint) is shown, demonstrating how strongly the pMSSM parameter plane is constrained by imposing simultaneously all constraints.

\begin{figure}[!t]
\begin{center}
\includegraphics[width=.33\textwidth]{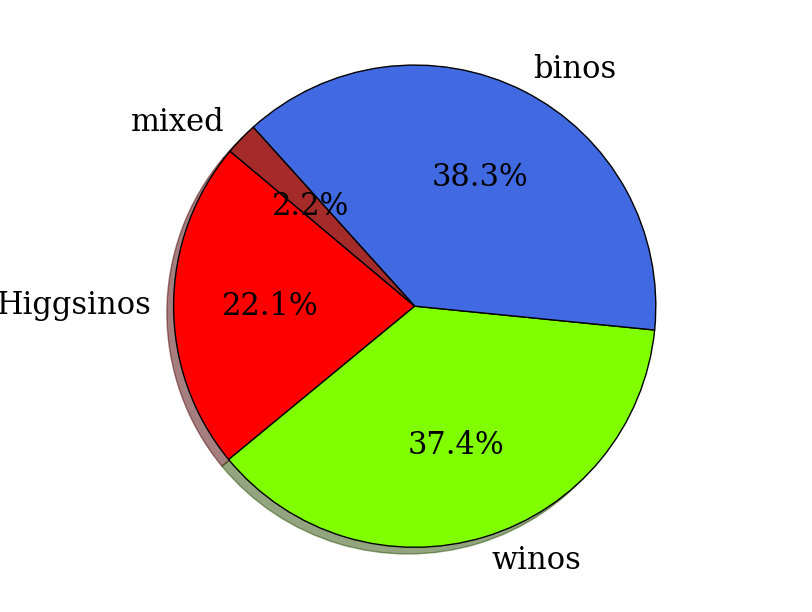}\includegraphics[width=.33\textwidth]{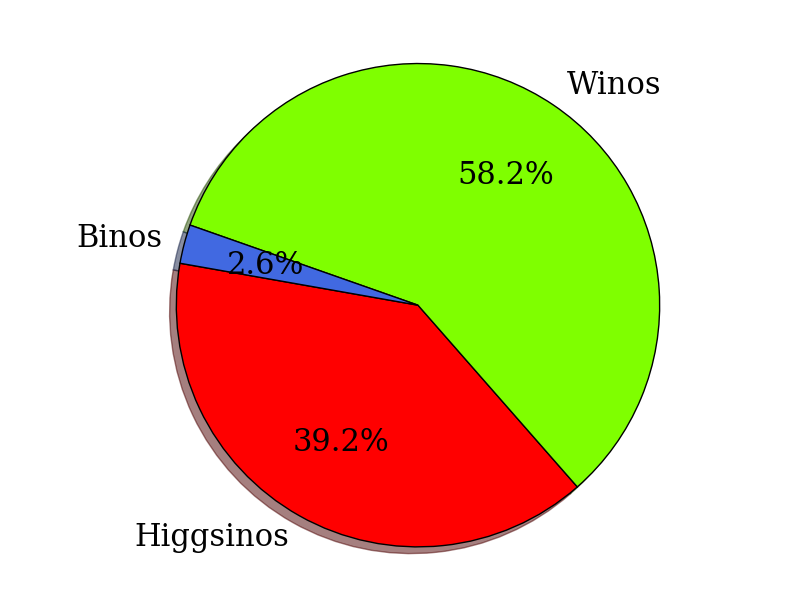}
\includegraphics[width=.33\textwidth]{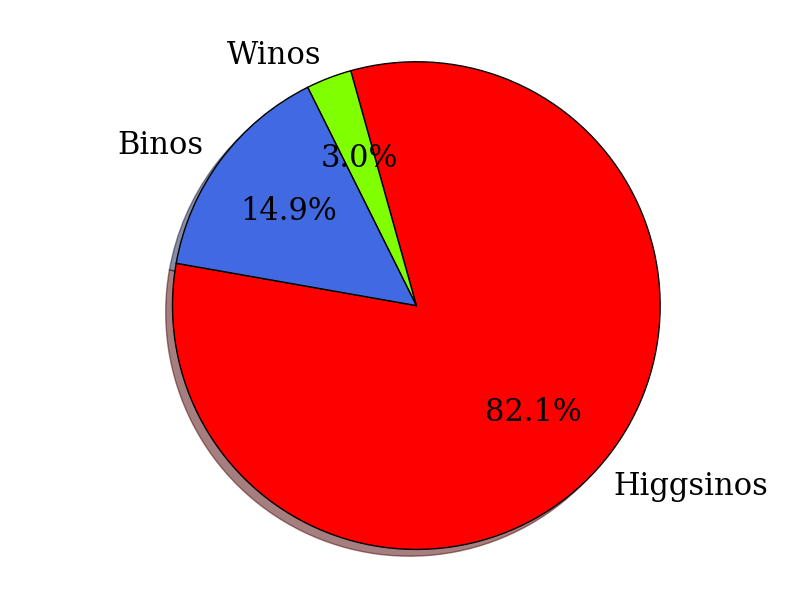}
\caption{Fraction of the different lightest neutralino types of the pMSSM points, before imposing the dark matter constraints (left), after imposing direct detection, indirect detection and upper relic density constraints (centre) and after imposing in addition the lower relic density limit (right). From~\cite{Arbey:2017eos}.\label{fig:pMSSM_neutralino_relic}}
\end{center}
\end{figure}

If one assumes the MSSM to provide a candidate to explain the whole dark matter it is then necessary to impose in addition the lower bound on the relic density. As discussed before this imposes that the lightest neutralino is accompanied by a close-in-mass next-to-lightest supersymmetric particle, which has a big impact on the pMSSM parameter space: first the fraction of points passing all the constraints becomes extremely small, second the types of surviving lightest neutralinos are particularly affected, as demonstrated in Fig.~\ref{fig:pMSSM_neutralino_relic}.
The role of the relic density constraint becomes therefore particularly important.

In the case new physics is discovered at colliders or in dark matter detection experiments, it will be essential to determine the underlying model and its parameters. While collider constraints are determined in an environment under control, this is not the case of dark matter constraints. As we have seen direct detection relies in particular on the local density and velocity of dark matter, and indirect detection on the dark matter galactic profile, and in the case of charged particles on propagation models. We have seen that these assumptions generate uncertainties on the dark matter detection constraints, resulting in uncertainties of up to one order of magnitude on the dark matter annihilation or scattering cross sections. It is therefore important to take these uncertainties into consideration when trying to determine the parameters of the models. On the other hand relic density sets extremely strict constraints when both upper and lower limits are imposed, but it is important to keep in mind that the calculation of the relic density relies on simple assumptions on the early Universe properties, which may prove to be inexact, and would generate strong biases when determining the underlying model.

%%%

\subsection{Gravitino dark matter}
\label{sec:gravitino}

We now consider the case of gravitino dark matter. In terms of pMSSM parameters in this scenario the gravitino mass parameter is smaller than the other superparticle masses. Being the superpartner of the graviton which mediates gravity, the gravitino is coupled only very weakly to other particles. As a consequence, contrary to the neutralino case the gravitino is not a thermal relic, because it interacts too weakly to be in thermal equilibrium in the early Universe. Since the gravitino is the lightest supersymmetric particle the next-to-lightest supersymmetric particle has a very long lifetime, and the heaviest supersymmetric particles decay into the next-to-lightest supersymmetric particle which subsequently decays into one gravitino and SM particles. In addition, gravitinos can be produced substantially right after inflation in the primordial thermal bath from scattering off the other supersymmetric particles, in particular off the coloured or charged supersymmetric particles.

Turning to the phenomenology of gravitino dark matter scenarios, for dark matter direct and indirect detections the gravitino is so weakly coupled that any detection through its scattering or annihilation is impossible. Therefore, no constraint can be set from dark matter detection experiments. At colliders, the gravitino is so weakly coupled to the other supersymmetric particles that it cannot appear in the final states. On the contrary, the next-to-lightest supersymmetric particle will always be present in the final states, and will decay much later outside the detectors. If this next-to-lightest supersymmetric particle is charged it can leave tracks in the detector, making the detection easier. In the case the next-to-lightest supersymmetric particle is not a neutralino, most of the collider supersymmetric search results are not valid anymore and specific searches may have to be designed, even if recasting the results is still possible \cite{CahillRowley:2012cb}.

To simplify the discussion, let us consider here only the case in which the next-to-lightest supersymmetric particle is a neutralino, following Ref.~\cite{Arvey:2015nra}. The immediate consequence is that all the collider constraints presented in the neutralino dark matter section are unchanged in the gravitino dark matter scenario. However, contrary to the neutralino dark matter case, dark matter direct and indirect detection experiments cannot set constraints on the pMSSM, because the gravitino is too weakly interacting to be detected in such experiments.

Relic density on the other hand can still impose constraints on the pMSSM gravitino dark matter scenario. The lifetime of the lightest neutralino is an important parameter here and is given by \cite{Covi:2009bk}:
\begin{equation}
\tau_{{\tilde\chi}_1^0} \approx (57\, \mbox{s} ) \left(\frac{M_{{\tilde\chi}_1^0}}{1\;\mbox{TeV}}\right)^{-5}   
 \left(\frac{M_{\tilde{G}}}{10\;\mbox{GeV}}\right)^{2}\,.
 \label{eq:NLSPlifetime}
\end{equation}
The main decay of the neutralino is into a gravitino and a photon or a $Z$ boson. The only requirement in the above formula is that the gravitino has to be lighter than the neutralino, which has a lower limit of 46 GeV set by LEP. One can therefore expect the neutralino lifetime to be in the ballpark of the second, which confirms that the collider phenomenology is unchanged since the neutralino is quasi-stable at colliders. In addition this implies that the neutralinos in the thermal bath would not decay before the freeze-out. The standard calculation of relic density can therefore be applied to the lightest neutralino, and the obtained value can be rescaled by number conservation when the neutralinos will decay into gravitinos, so that the contribution to the gravitino relic abundance of the neutralino decays is:
\begin{equation}
 \Omega_{\tilde{G},\,\mbox{$\chi$-decay}} \,h^2 = \frac{M_{\tilde{G}}}{M_{{\tilde\chi}_1^0}}\, \Omega_{\chi,\mbox{standard}} \,h^2\,,
\end{equation}
where $\Omega_{\chi,\,\mbox{standard}}$ denotes the neutralino relic abundance computed as if it were the lightest supersymmetric particle. Therefore, one can expect for small gravitino masses relic densities much smaller than the neutralino relic densities. There however exists another contribution to the gravitino relic density which comes from the scattering at high energy of the other supersymmetric particles in the early Universe thermal bath after inflation, which amounts to \cite{Bolz:2000fu,Pradler:2006qh}
\begin{equation}
\Omega_{\tilde{G},\mbox{reheating}}\,h^2=0.83\; \frac{T_{RH}}{10^9\,\mbox{GeV}} 
\left(\frac{M_{\tilde{G}}}{1\,\mbox{GeV}}\right)^{-1}\sum_{i=1}^3 \gamma_{i} (T_{RH})
\left(\frac{M_{i}}{300\,\mbox{GeV} }\right)^2 \,,
\end{equation}
where $\gamma_{i} (T_{RH})$ are numerical factors of order 1 showing the dependence on the gauge coupling for each SM gauge group and the evolution of the gauge couplings and gaugino masses $M_{i}$ to the scale of the reheating temperature $T_{RH}$. Practically, the higher the reheating temperature is, the higher the value of this contribution. The gravitino relic density is therefore given by the sum of both contributions:
\begin{equation}
\Omega_{\tilde{G}}\,h^2 \;=\;\Omega_{\tilde{G},\,\mbox{$\chi$-decay}} \,h^2 \;+\; \Omega_{\tilde{G},\mbox{reheating}}\,h^2\,,
\end{equation}
which can be compared directly to the observed dark matter density. In addition to the 20 pMSSM parameters the gravitino relic density is thus driven by the reheating temperature $T_{RH}$, which can be seen in this context as the energy scale of leptogenesis. Leptogenesis models give $T_{RH} \sim 10^9$ GeV \cite{Barbieri:1999ma,Fujii:2003nr,Buchmuller:2004nz,Fong:2013wr}, and setting constraints on the relic density can lead to upper limits on the reheating temperature \cite{Arvey:2015nra}.

\begin{figure}[!t]
\begin{center}
\includegraphics[width=.45\textwidth]{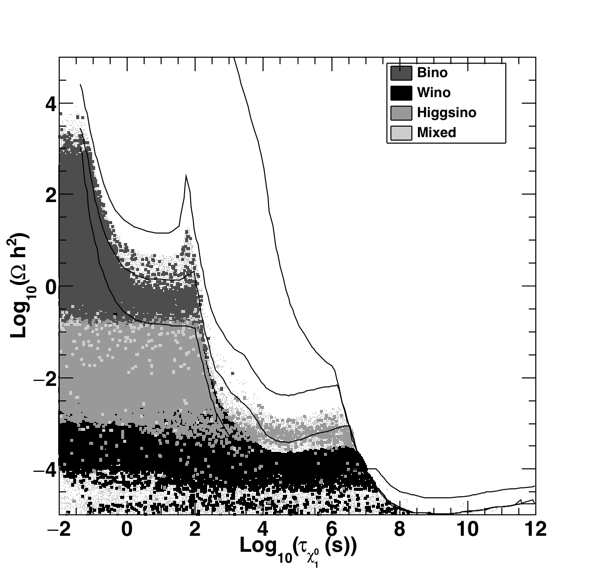}
\includegraphics[width=.45\textwidth]{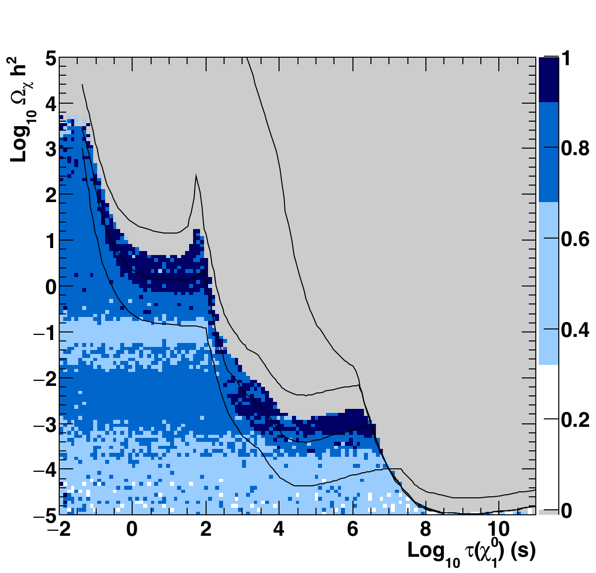}
\caption{Big-Bang nucleosynthesis constraints from Ref.~\cite{Jedamzik:2006xz} in the neutralino relic density vs. neutralino lifetime parameter plane, for the different types of lightest neutralino in the pMSSM (left) and with the fraction of pMSSM points excluded by the LHC constraints with 300 fb$ ^{-1}$ encoded by the colour scale. The different lines correspond to upper limits of lightest neutralino hadronic branching fractions of 100\%, 10\%, 1\% and 0\% from bottom to top. From~\cite{Arvey:2015nra}.\label{fig:BBN_gravitino}}
\end{center}
\end{figure}

\begin{figure}[!t]
\begin{center}
\includegraphics[width=.5\textwidth]{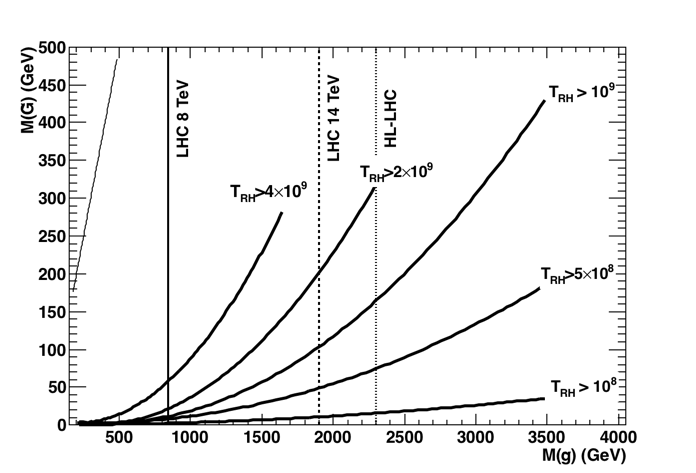}
\caption{Minimum values of the gravitino mass as a function of the gluino mass for
different choices of reheating temperature (in GeV). The vertical lines correspond to the gluino masses below which at least 95\% of the pMSSM points are excluded by the LHC supersymmetric searches at 7+8 TeV (continuous), and the projections at 14 TeV for a luminosity of 300 fb$^{-1}$ (dashed) and during the high luminosity (HL) run with 3000 fb$^{-1}$ (dotted). From~\cite{Arvey:2015nra}.\label{fig:gravitino_reheating}}
\end{center}
\end{figure}

In terms of phenomenology, collider searches can help setting constraints on the neutralino decay part of relic density. In addition, if the neutralino lifetime is of the order of the second, neutralinos of the thermal bath decay during Big-Bang nucleosynthesis, which can be modified or disrupted by the injection of particles resulting from the decays of the neutralino \cite{Jedamzik:2006xz,Covi:2009bk}. BBN can thus set constraints on the gravitino scenario, related to the neutralino relic abundance and lifetime, which can also be combined with the LHC constraints, as shown in Fig.~\ref{fig:BBN_gravitino} in the relic density vs. neutralino lifetime plane: large relic abundances and long lifetimes are excluded by BBN constraints relatively independently from the hadronic branching fractions.
One of the main parameters which drive the neutralino lifetime is the gluino mass term $M_3$. Gluino searches at the LHC therefore set constraints on the neutralino lifetime. By combining all constraints from colliders and cosmology, it has been shown that constraints can be obtained on the reheating temperature \cite{Arvey:2015nra}, as presented in Fig.~\ref{fig:gravitino_reheating} in the gravitino mass vs. gluino mass plane, where reheating temperature contours are also shown: with 300 fb$^{-1}$ of data and in absence of gluino discovery, an upper limit on the reheating temperature of $4\times10^9$ GeV can be set.

This example shows how particle physics and dark matter searches are connected, and that collider searches may allow us to set constraints on the primordial Universe properties (here reheating temperature) if new physics/new particles were detected in collider experiments.

%%%%%%%%%%%%%%%%%%%%%%

\section{Dark matter and early Universe}

As we have seen in the previous sections, dark matter relic density is strongly related to the early Universe properties. This was shown in particular in Section~\ref{sec:gravitino} in the case of gravitino dark matter, in which the reheating temperature -- or leptogenesis scale -- plays an important role. We will consider here two additional examples of interplay between primordial Universe and relic density, in the context of the pMSSM neutralino dark matter scenario, where the neutralino is a thermal relic: quintessence model and decaying scalar field scenario.

\subsection{Relic density and quintessence}

We first assume that dark energy is made of a quintessence scalar field. In the present Universe the scalar field potential dominates its density, so that the scalar field equation-of-state mimics a cosmological constant. In the early Universe on the other hand the scalar field density could be dominated by its kinetic term, leading to a kination behaviour. In such a case, the scalar field may dominate the expansion in the very early Universe, as opposed to radiation domination in the standard cosmological model. Such a modified expansion rate can affect the relic density since accelerating the expansion would increase the distance between dark matter particles, making the freeze-out occur earlier, which lowers the annihilation probability and increases the relic density \cite{Salati:2002md,Arbey:2008kv}.

\begin{figure}[!t]
\begin{center}
\includegraphics[width=.48\textwidth]{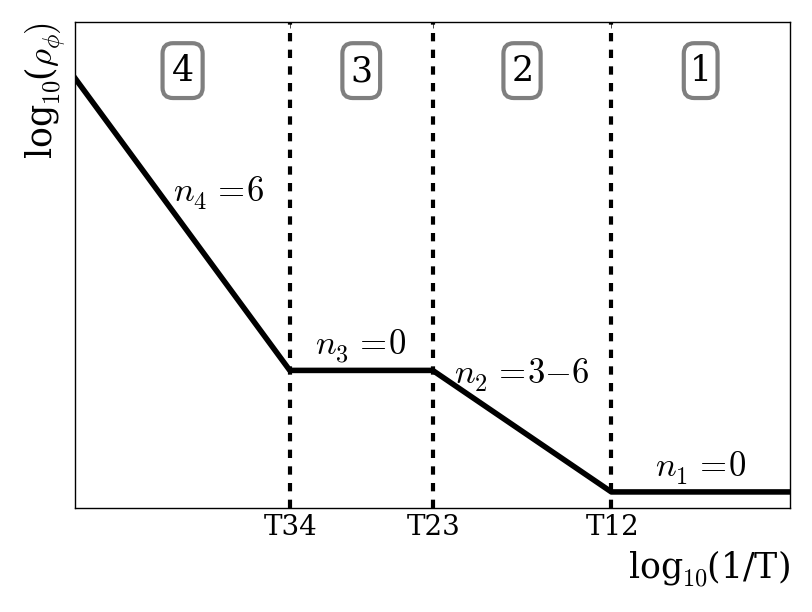}
\caption{Scalar field density as a function of the temperature in the 4-phase simplified quintessence model. From~\cite{Arbey:2018uho}.\label{fig:quint_model}}
\end{center}
\end{figure}

\begin{figure}[!t]
\begin{center}
\includegraphics[width=.45\textwidth]{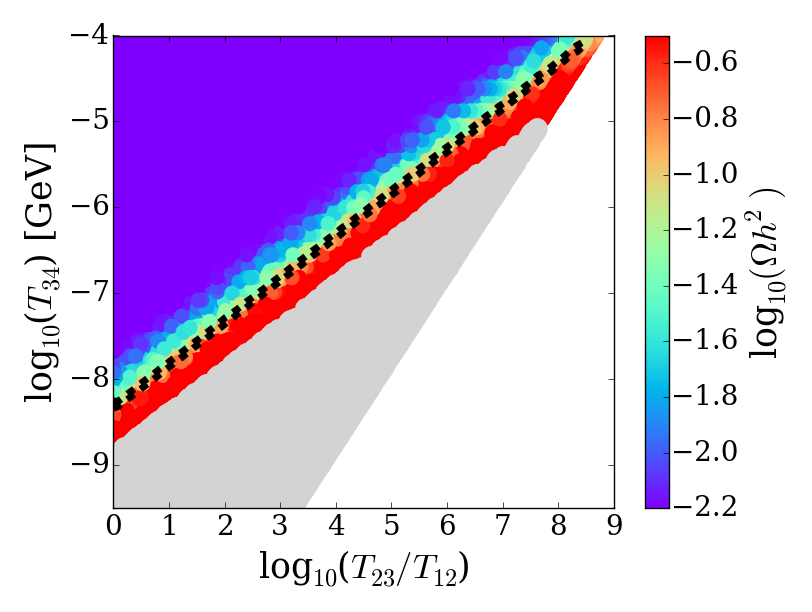}\includegraphics[width=.45\textwidth]{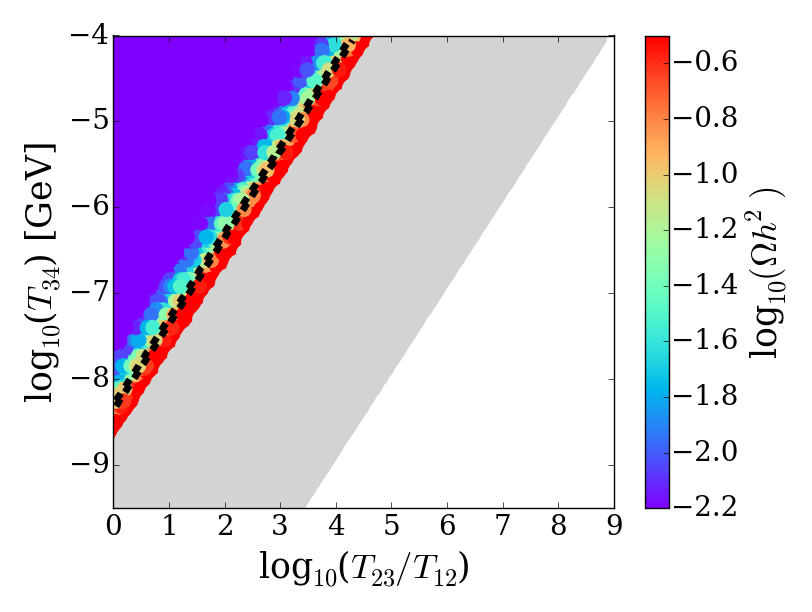}
\caption{Relic density value in the $T_{23}/T_{12}$ vs. $T_{34}$ parameter plane, for $n_2=3$ (left) and $n_2=6$ (right). The dashed lines correspond to the observed dark matter density and the grey zone to the parameter regions excluded by Big-Bang nucleosynthesis constraints. From~\cite{Arbey:2018uho}.\label{fig:quint_CMSSM}}
\end{center}
\end{figure}

\begin{figure}[!t]
\begin{center}
\includegraphics[width=.5\textwidth]{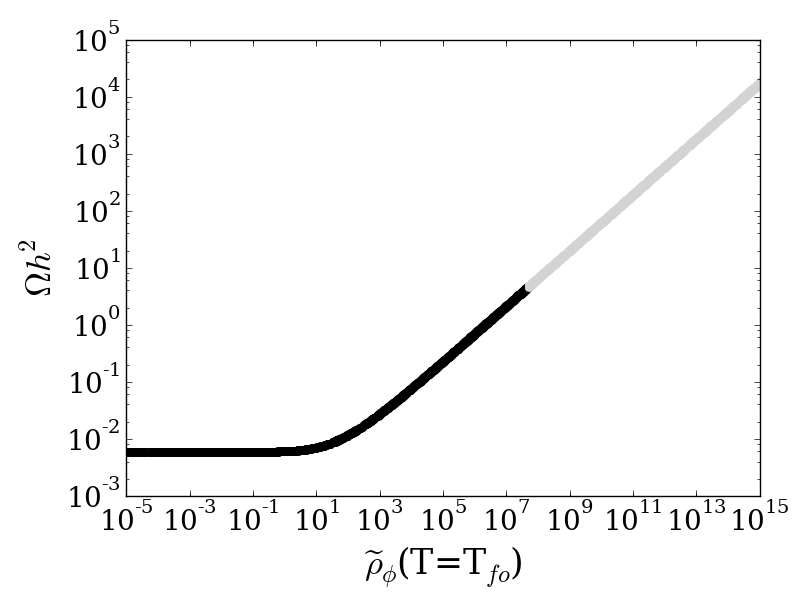}
\caption{Relic abundance as a function of the ratio of scalar field density over radiation density at the freeze-out temperature, for an initial relic density of $5.6 \times 10^{-3}$ in absence of a scalar field. The grey part of the curve corresponds to the region excluded by Big-Bang nucleosynthesis. From~\cite{Arbey:2018uho}.\label{fig:quint_CMSSM2}}
\end{center}
\end{figure}

In practice the Friedmann equation (\ref{eq:friedmann_stand}) is modified as
\begin{equation}
H^2=\frac{8 \pi G}{3} (\rho_{rad} + \rho_\phi)\,, \label{eq:friedmann_mod}
\end{equation}
where $\rho_\phi$ is the scalar field density, which follows the Klein-Gordon equation:
\begin{equation}
\frac{d\rho_\phi}{dt} = -3 H ( \rho_\phi + P_\phi )\,, 
\end{equation}
in which $\rho_\phi = \dot{\phi}^2/2 + V(\phi)$ and $P_\phi = \dot{\phi}^2/2 - V(\phi)$. The choice of the potential is crucial, and can strongly affect the equation-of-state of quintessence \cite{Tsujikawa:2013fta,Arbey:2019cpf}. Concerning relic density, the freeze-out is expected to occur at a temperature around $M_\chi/10$, typically about $10-100$ GeV for standard thermal relics. This corresponds to a scale factor $a/a_0 \sim 10^{-15}-10^{-14}$. At this epoch we have no constraint on the expansion rate, and quintessence scenarios have behaviours which can be easily modelled as described in Fig.~\ref{fig:quint_model}. There exist four main phases \cite{Arbey:2018uho}:\vspace*{-0.1cm}
\begin{enumerate}
 \item The present Universe with the scalar field density nearly constant, and a density equal to the cosmological constant density.\vspace*{-0.1cm}
 \item Above the temperature $T_{12}$ and below $T_{23}$ when the scalar field has an equation-of-state $w_\phi\in[0,1]$, with a density evolving as $\rho\propto a^{-n_2}$, for $n_2 \in [3,6]$.\vspace*{-0.1cm}
 \item A period of constant density between $T_{23}$ and $T_{34}$, as in the double exponential potential scenario \cite{Barreiro:1999zs}.\vspace*{-0.1cm}
 \item Above $T_{34}$ when the density is dominated by the kinetic term and evolves as $a^{-6}$.\vspace*{-0.1cm}
\end{enumerate}
The value of the scalar field density is fixed to the cosmological constant density today. Hence, the quintessence density is completely fixed at any time by the choice of $T_{12}$, $T_{23}$, $T_{34}$ and $n_2$. At $T_{12}$ the scalar field has a negligible density in comparison to the other cosmological densities, so that the value of $T_{12}$ has no effect on the relic density. Therefore we are left with three parameters: $T_{23}/T_{12}$, $T_{34}$ and $n_2$. Since the relic density is expected to increase we consider a pMSSM point with a Higgsino lightest neutralino and a relic density $\Omega h^2 = 5.6 \times 10^{-3}$ \cite{Arbey:2018uho}, i.e. two orders of magnitude below the observed dark matter density. We now study whether quintessence scenarios can modify the relic density and make it consistent with the observations. Scanning over the three quintessence parameters it is shown in Fig.~\ref{fig:quint_CMSSM} that the relic density can be increased by two orders of magnitude and be consistent with the observations without disrupting Big-Bang nucleosynthesis.
A more careful analysis shows that the main parameter of the model which determines the value of the relic density is the scalar field density at the freeze-out temperature, as can be seen in Fig.~\ref{fig:quint_CMSSM2}.


The fact that the presence of quintessence in the early Universe can increase the relic density by several orders of magnitude is particularly relevant for the interpretation of the relic density constraint in particle physics. Indeed if directly applied, the dark matter density bound would exclude too low relic densities \emph{within the standard cosmological model}. It is however important to study whether cosmological phenomena can modify the relic density and make it compatible with dark matter observations, thus allowing pMSSM scenarios normally excluded to be consistent with cosmological observations.
\begin{figure}[!t]
\begin{center}
\includegraphics[width=.5\textwidth]{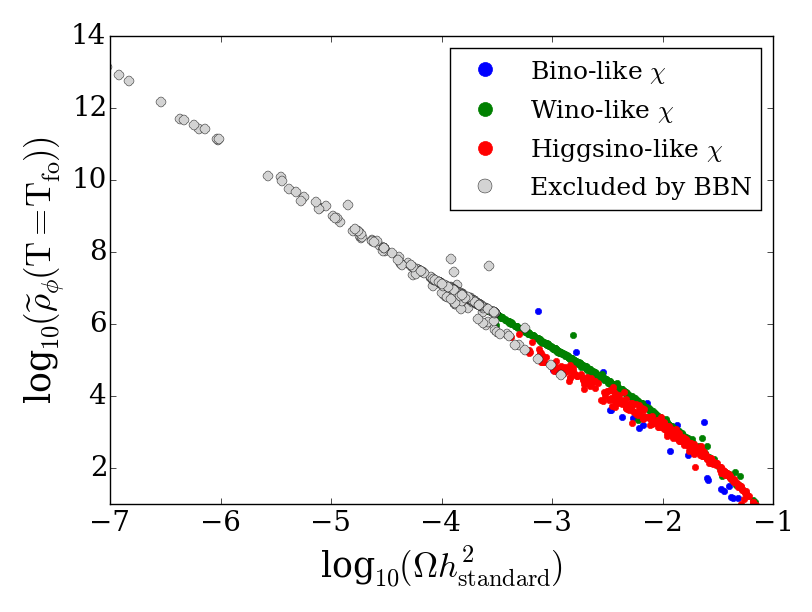}
\caption{Value of the ratio of scalar field density over radiation density required to increase the relic density obtained in the standard cosmological scenario up to the observed dark matter density, for a sample of pMSSM points. The grey points have no solution since too large scalar field densities are excluded by Big-Bang nucleosynthesis. From~\cite{Arbey:2018uho}.\label{fig:quint_pMSSM}}
\end{center}
\end{figure}
Figure~\ref{fig:quint_pMSSM} shows for a large sample of pMSSM points the required scalar field density at the freeze-out time to increase the relic density so that it reaches the observed dark matter density, for the different types of lightest neutralinos. An important result is that relic densities smaller than $3\times10^{-4}$ in the standard cosmological scenario cannot be increased sufficiently to be consistent with the observed dark matter constraint because of Big-Bang nucleosynthesis constraints.

The results presented in this section make it clear that the upper and lower limits of the relic density values obtained following Section~\ref{sec:relic_standard} cannot be used to constrain particle physics scenarios, because the cosmological properties of the early Universe may be different from the ones of the standard cosmological model.

\subsection{Relic density and decaying scalar field}

We now consider another scenario, in which a scalar field decays in the early Universe. We do not make any assumption on the nature of the scalar field, it may be an inflaton, a scalar field participating in a phase transition, in leptogenesis or baryogenesis, etc. One can assume that the scalar field decays before Big-Bang nucleosynthesis in order to leave the abundance of the elements unaffected. The scalar field has therefore disappeared in the very early Universe, and have no effect on the present Universe. To simplify, we consider that the field is pressureless. The Klein-Gordon equation reads:
\begin{equation}
\frac{d\rho_\phi}{dt} = -3 H \rho_\phi - \Gamma_\phi \rho_\phi\,, 
\end{equation}
where $\Gamma_\phi$ is the decay width of the scalar field. One considers that the field decays mostly into radiation, i.e. relativistic particles, with a tiny branching fraction $b$ into relics. The number density of relic particles is therefore given by:
\begin{equation}
dn/dt=-3Hn-\langle \sigma_{\mbox{eff}} v\rangle (n^2 - n_{\mbox{eq}}^2) + \frac{b}{M_\chi} \Gamma_\phi \rho_\phi \,. \label{eq:evol_eq_mod}
\end{equation}
The expansion rate can receive in addition to the scalar field density a contribution from the dark matter density if it is produced non-thermally in large proportions:
\begin{equation}
H^2=\frac{8 \pi G}{3} (\rho_{rad} + \rho_\phi + \rho_\chi)\,, \label{eq:friedmann_mod2}
\end{equation}
where $\rho_\chi$ is the dark matter density. Finally the radiation entropy density may also be modified by radiation injection:
\begin{equation}
\frac{ds_{\rm rad}}{dt} = -3 H s_{\rm rad}  + (1-b) \, \frac{\Gamma_\phi \rho_\phi}{T}\,. 
\end{equation}
The reheating temperature $T_{RH}$ is defined implicitly following Eq.~(\ref{eq:TRH}). We also define $\tilde\rho_\phi = \rho_\phi / \rho_{\rm rad}$ and
\begin{equation}
 \tilde\Sigma^* = \frac{\Sigma_\phi \rho_\phi}{3 H T S_{\rm rad}} \,,
\end{equation}
so that
\begin{equation}
\frac{ds_{\rm rad}}{dt} = -3 H \bigl(1-\tilde\Sigma^*\bigr) s_{\rm rad}\,. 
\end{equation}
If $\tilde\Sigma^* > 1$ the entropy injection is stronger than entropy dilution generated by the expansion, and inversely. The scalar field evolution depends therefore on three parameters: the reheating temperature $T_{RH}$ (or equivalently the scalar field decay width), the branching fraction into relic particles $b$ (or equivalently $\eta = b \times (\mbox{1\,GeV})/ m_\phi$) and the initial value of the scalar field, that we parametrize with $\kappa_\phi = \rho_\phi(T_{\rm init})/\rho_{\rm rad}(T_{\rm init})$.

The decaying scalar field has therefore three consequences:\vspace*{-0.1cm}
\begin{itemize}
 \item It can accelerate the expansion by increasing the Hubble parameter, leading to an increased relic density as in the case of quintessence.\vspace*{-0.1cm}
 \item Its decay can produce relic particles non-thermally, increasing the final relic density.\vspace*{-0.1cm}
 \item Its decay injects radiation entropy, which alters the relation between time and temperature, leading to a decrease of the relic density \cite{Gelmini:2006pw,Arbey:2009gt}.\vspace*{-0.1cm}
\end{itemize}
A small reheating temperature corresponds to a late decay, which can affect Big-Bang nucleosynthesis. It was shown in Ref.~\cite{Arbey:2018uho} that Big-Bang nucleosynthesis sets in this context a lower limit of $T_{RH} \gtrsim 6$ MeV.

We consider that the branching ratio into relic particles is negligibly small. The scalar field decay thus injects radiation entropy, which locally increases the temperature, leading to increased annihilation rates and decreased relic density. To show the effect of the scalar field decay we first consider a point with a large relic density of 1.3, which is two orders of magnitude above the observed dark matter density. In general, too large relic densities are systematically excluded by the imposed constraints, since they overclose the Universe.
\begin{figure}[!t]
\begin{center}
$\begin{array}{cc}
\includegraphics[width=.45\textwidth]{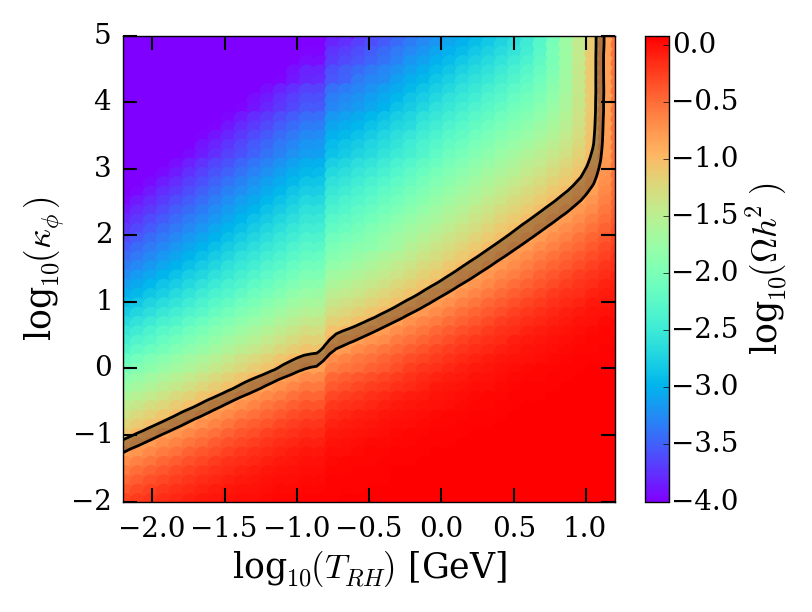}&\includegraphics[width=.45\textwidth]{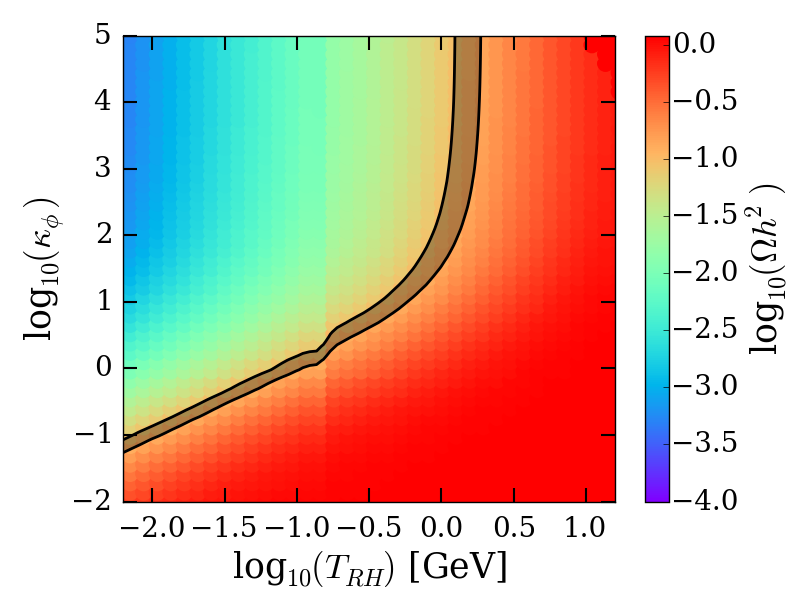}\\
{\rm (a)}\; \eta=0 & {\rm (b)}\; \eta=10^{-12}\\
\includegraphics[width=.45\textwidth]{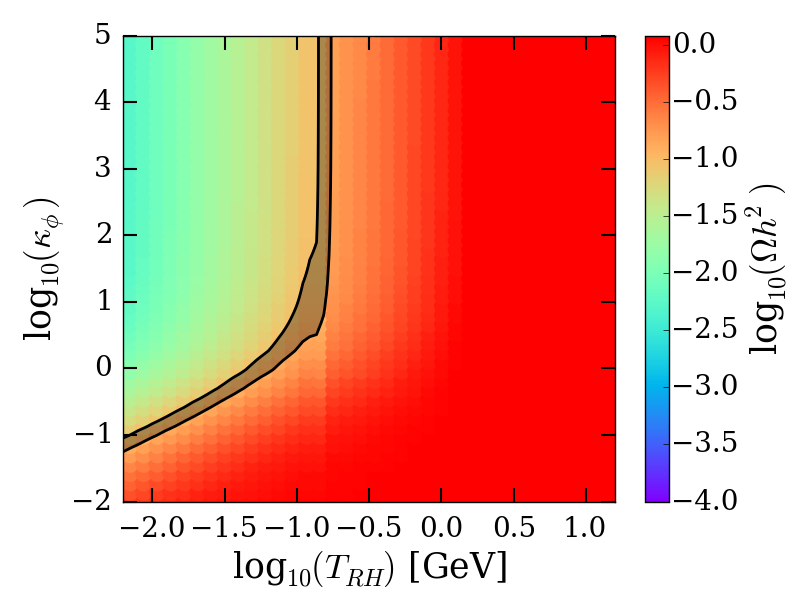}&\includegraphics[width=.45\textwidth]{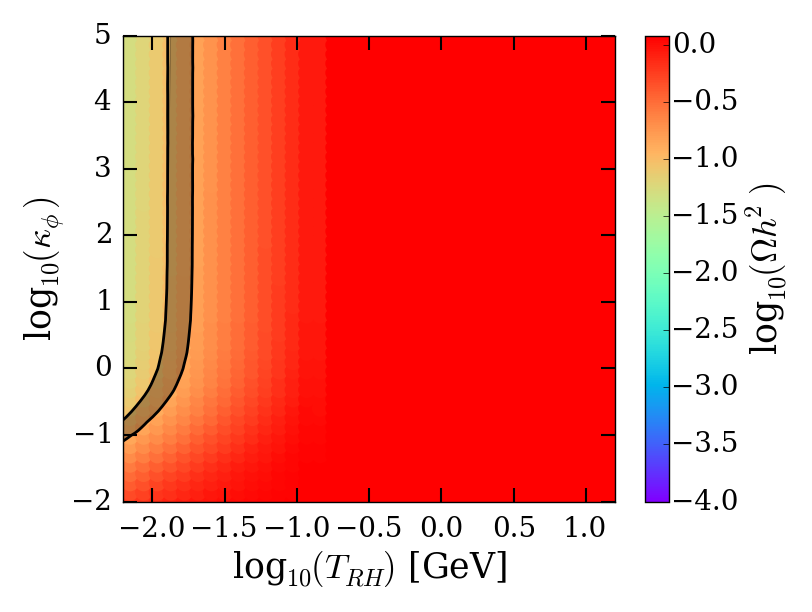}\\
{\rm (c)}\; \eta=10^{-11} & {\rm (d)}\; \eta=10^{-10}
\end{array}$
\caption{Relic density as a function of the reheating temperature $T_{RH}$ and the initial scalar field density to radiation density ratio $\kappa_\phi$, for different choices of the branching ratio into relic particles $b=\eta m_\phi/(\mbox{1\,GeV})$, for a relic density of 1.3 in the standard cosmological model. In the grey regions the relic density is in agreement with the observed dark matter density. From~\cite{Arbey:2018uho}.\label{fig:pMSSM_decaying}}
\end{center}
\end{figure}
In Fig.~\ref{fig:pMSSM_decaying} the relic density of this pMSSM point is shown as a function of the reheating temperature and the initial density of scalar field, for different values of the scalar field branching ratio into relic particles, for reheating temperatures compatible with Big-Bang nucleosynthesis constraints. As expected in absence of decay into relic particle ($b=0$) one sees that the relic density can decrease by several orders of magnitude when the initial scalar field density and the reheating temperature increase. One can also notice that the effect of a nonzero branching fraction into relics is to increase the relic density. The two effects go therefore in opposite directions, and larger values of $b$ would alleviate the possibility to decrease the relic density. For the four plots presented above however there exist solutions to make the too small standard relic density compatible with the observed dark matter density.

\begin{figure}[!t]
\begin{center}
$\begin{array}{cc}
\includegraphics[width=.45\textwidth]{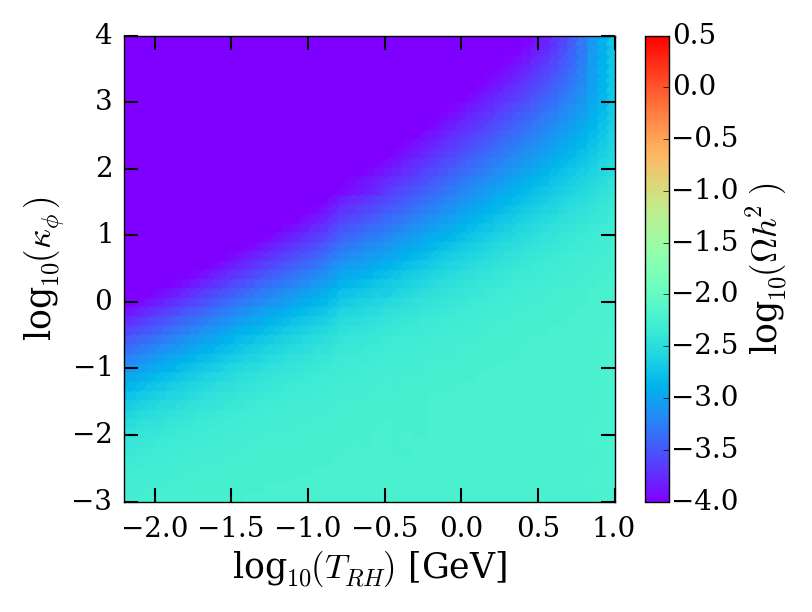}&\includegraphics[width=.45\textwidth]{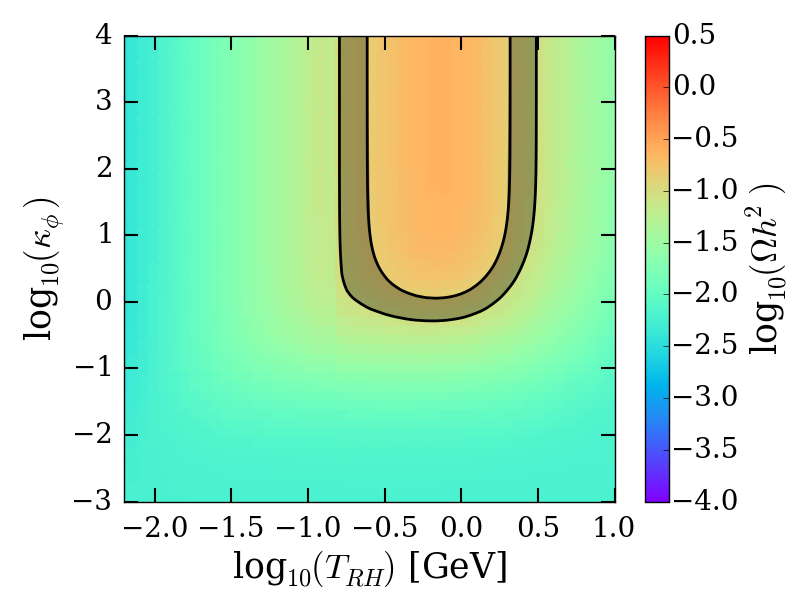}\\
{\rm (a)}\; \eta=0 & {\rm (b)}\; \eta=10^{-11}\\
\includegraphics[width=.45\textwidth]{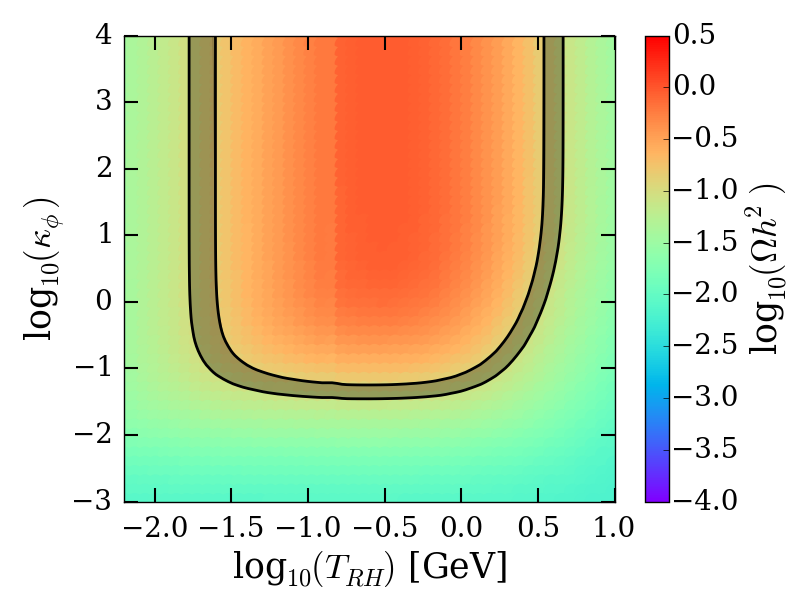}&\includegraphics[width=.45\textwidth]{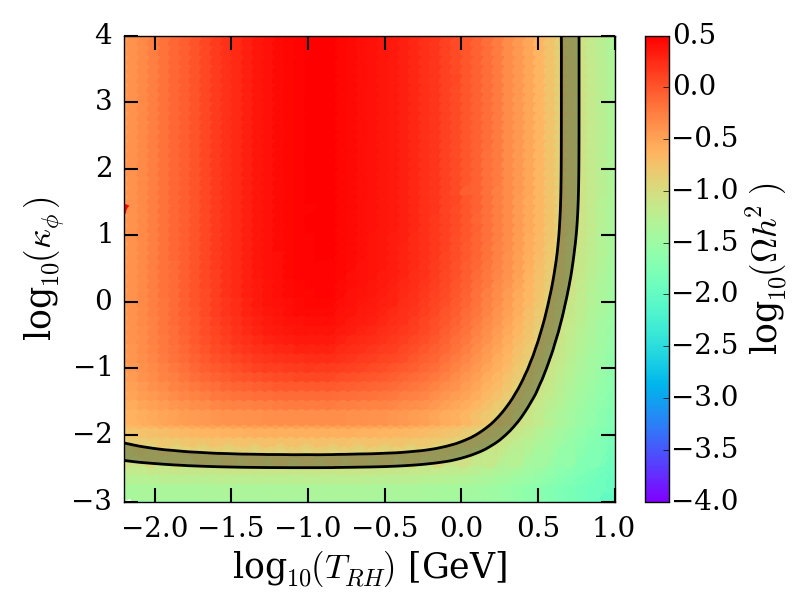}\\
{\rm (c)}\; \eta=10^{-10} & {\rm (d)}\; \eta=10^{-9}
\end{array}$
\caption{Relic density as a function of the reheating temperature $T_{RH}$ and the initial scalar field density to radiation density ratio $\kappa_\phi$, for different choices of the branching ratio into relic particles $b=\eta m_\phi/(\mbox{1\,GeV})$, for a relic density of $5.6 \times 10^{-3}$ in the standard cosmological model. In the grey regions the relic density is compatible with the observed dark matter density. From~\cite{Arbey:2018uho}.\label{fig:pMSSM2_decaying}}
\end{center}
\end{figure}
We now consider the case of a pMSSM point with a too small relic density of $5.6 \times 10^{-3}$. In Fig.~\ref{fig:pMSSM2_decaying} the relic density of this point is shown as a function of the reheating temperature and the initial density of scalar field, for different values of the scalar field branching ratio into relic particles, for reheating temperatures compatible with Big-Bang nucleosynthesis constraints. As expected, in absence of decay into relic particles the relic density is decreased by the presence of the scalar field, but with a branching ratio into relic particles as low as $\eta=10^{-11}$ the relic abundance can be increased by two orders of magnitude and reach the observed dark matter abundance.

A conclusion of this analysis is that the presence of a decaying scalar field in the early Universe can modify the relic density by orders of magnitude, making new physics scenarios originally incompatible with the dark matter constraints within the standard cosmological model in agreement with the observational data. However there is sufficient flexibility in the parameters to make \emph{any} standard relic density compatible with the observed dark matter scenario. This shows again that the relic density cannot be used to set robust constraints on new physics scenarios.

\begin{figure}[!t]
\begin{center}
\includegraphics[width=.5\textwidth]{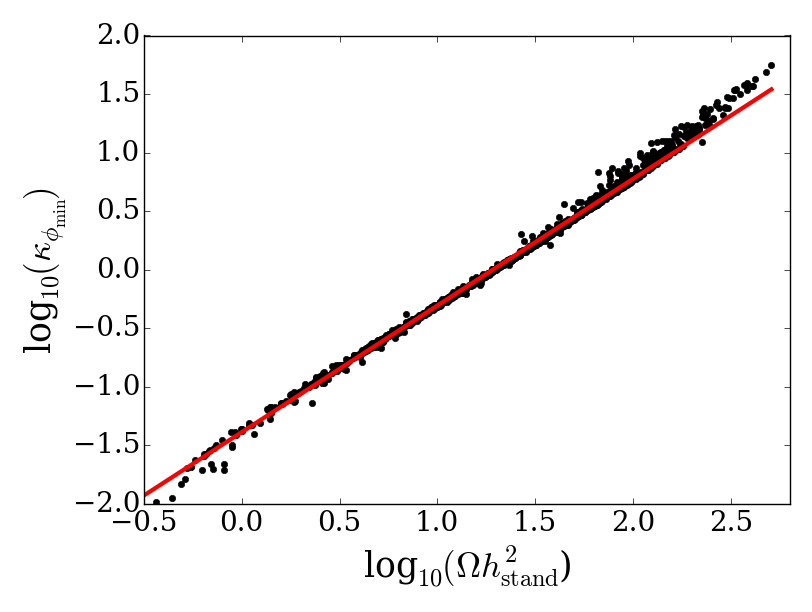}
\caption{Minimum scalar field density $\kappa_\phi$ as a function of the relic density in the standard cosmological model $\kappa_\phi$, in order for the relic density to be compatible with the observed dark matter constraints, for a reheating temperature of 6 MeV. From~\cite{Arbey:2018uho}.\label{fig:pMSSM3_decaying}}
\end{center}
\end{figure}

\begin{figure}[!t]
\begin{center}
\includegraphics[width=.6\textwidth]{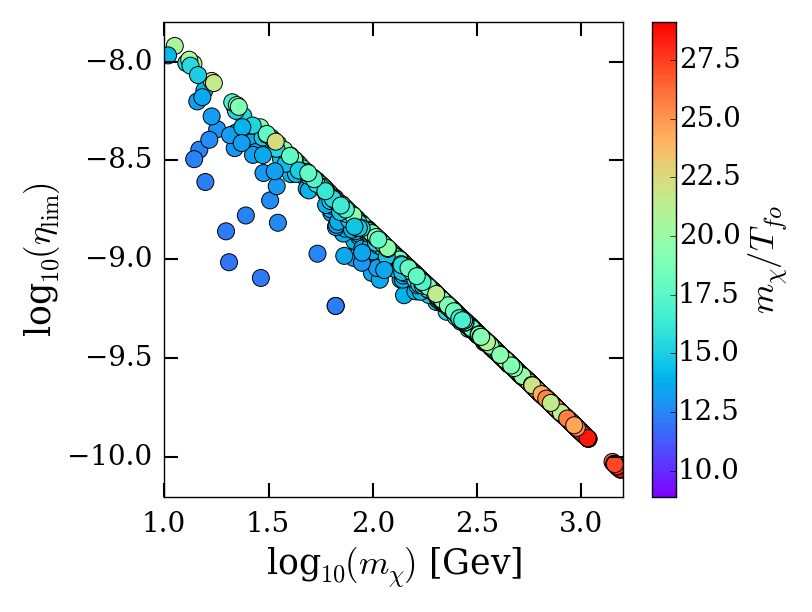}
\caption{Maximum decay branching fraction into relic particles $\eta$ as a function of the relic mass in order for the relic density to be compatible with the observed dark matter constraints. The colour scale gives the value of the freeze-out temperature for these points. From~\cite{Arbey:2018uho}.\label{fig:pMSSM4_decaying}}
\end{center}
\end{figure}

On the other hand for a given new physics scenario it may be possible to set constraints on the decaying scalar field model. For example, for scenarios with too small relic densities, by choosing a reheating temperature as small as what is allowed by the Big-Bang nucleosynthesis constraints, one can determine the minimum scalar field density to make the relic density compatible with the observed dark matter density. This is shown in Fig.~\ref{fig:pMSSM3_decaying}, for a set of pMSSM parameter points.
Similarly in Fig.~\ref{fig:pMSSM4_decaying} one can see the maximum value of the decay branching fraction into relic particles $\eta$ for a set of pMSSM parameters for a large range of lightest neutralino masses.

These results show a clear interplay between the underlying dark matter model and the cosmological properties of the early Universe.

%%%%%%%%%%%%%%%%%%%%%%

\section{Conclusion}

The nature of dark matter remains one of the most actively debated question in cosmology as well as in particle physics. Moreover, in spite of the unprecedented progress in observational cosmology and the determination of many parameters of the standard cosmological model, many questions are still unanswered, such as the nature of dark energy, the origin of the particle-antiparticle asymmetry in the Universe, and the reasons of the inflationary period. As we have seen in this review, dark matter originates from the early Universe, and may have been affected by many unknown phenomena at the period before Big-Bang nucleosynthesis.

Often considered as made of new particles, dark matter is actively searched for at colliders and in dark matter detection experiments. In absence of new physics signals, scenarios of new physics beyond the Standard Model can be severely constrained by the experimental limits, as well as by the cosmological and astrophysical observations, and they can connect the early Universe properties to the properties of new physics particles. We have seen that in this context the relic density, when compared to the observed dark matter density, can set extremely strong constraints on new physics parameters, but is also very much dependent on the properties of the early Universe, which makes it an important observable for both new physics and cosmological scenarios. The example of supersymmetry demonstrates in particular that by combining particle, astroparticle and cosmological data, it becomes possible not only to set strong constraints on the new physics parameters, but also to get limits on cosmological models, such as quintessence or decaying scalar field scenarios. Beyond the example of supersymmetry, such approaches can apply to any new physics scenarios that provide one or several candidates for dark matter.

If new physics particles or phenomena are discovered at colliders or in dark matter experiments in the future, it will be of utmost importance to determine the underlying model and to eventually identify the nature of dark matter. It will be important to remember then that dark matter observables are dependent on cosmology and astrophysics, and that the identification of dark matter and the underlying new physics model will only be possible under assumptions on the primordial properties of the Universe, or via the simultaneous determination of the particle physics model and cosmological scenario. As a consequence, the discovery of dark matter at colliders will provide new ways to access the earlier time of the cosmological history.

%%%%%%%%%%%%%%%%%%%%%%

\bibliography{biblio}

\end{document}